\documentclass [12pt]{article}
\usepackage{lscape}                                           %
\usepackage{amsmath}
\usepackage{amsfonts} 
\usepackage{color}
\usepackage{graphicx}
\usepackage{lscape}
\newcommand{\ul}{{u_{loc}}}
\newcommand{\bul}{{{\bar u}_{loc}}}
\newcommand{\vl}{{v_{loc}}}
\newcommand{\bvl}{{{\bar v}_{loc}}}
\newcommand{\zl}{{z_{loc}}}

\newcommand{\xl}{{x_{loc}}}

\newcommand{\xii}[1]{{x_{#1}}}
\newcommand{\loc}{_{loc}}
\newcommand{\bou}{_{bou}}

\newcommand{\ttt}[1]{{t_{#1}}}

\newcommand{\mdda} { \frac{-2}{\alpha'} }
\newcommand{\madd} { \frac{-\alpha'}{2} }

\newcommand{\DDelta}{D}

\newcommand{\barX}{{\bar X}}

\newcommand{\bep}{{\bar \epsilon}}

\newcommand{\bc}{{\bar c}}
\newcommand{\bk}{{\bar k}}
\newcommand{\bt}{{\bar t}}
\newcommand{\bu}{{\bar u}}
\newcommand{\bv}{{\bar v}}
\newcommand{\bz}{{\bar z}}

\newcommand{\bX}{{\bar X}}

\newcommand{\cD}{{\cal D}}

\newcommand{\cM}{{\cal M}}
\newcommand{\cN}{{\cal N}}

\newcommand{\cS}{{\cal S}}
\newcommand{\cT}{{\cal T}}

\newcommand{\cX}{{\cal X}}
\newcommand{\cbX}{{\bar{ \cal X}}}
\newcommand{\cXX}{{\mathfrak X}}
\newcommand{\bcXX}{{\mathfrak X}}

\newcommand{\htt}{{\hat t}}

\newcommand{\Z}{\mathbb{Z}}

\newcommand{\R}{\mathbb{R}}

\newcommand{\C}{\mathbb{C}}

\newcommand{\du}{\partial_u}
\newcommand{\dub}{{\bar\partial_{\bar u}}}
\newcommand{\dz}{\partial}
\newcommand{\dzb}{{\bar\partial}}
\newcommand{\dw}{\partial}

\newcommand{\dy}{{\partial_y}}

\newcommand{\ke}{{k_\epsilon}}
\newcommand{\kbe}{k_{\bar\epsilon}}
\newcommand{\kei}[1]{{k_{\epsilon_{#1}}}}
\newcommand{\kbei}[1]{{k_{\bar\epsilon_{#1}}}}

\newcommand{\comb}[2]{\left( \begin{array}{c} #1 \\ #2 \end{array} \right) }

\newcommand{\oh}{\frac{1}{2}}

\newcommand{\COMMENTO}[1]{}
\newcommand{\COMMENTOOK}[1]{}
\newcommand{\COMMENTONO}[1]{}
\newcommand{\COMMENTOO}[1]{}

\begin{document}
\title{
Correlators of arbitrary untwisted operators and excited twist operators 
for $N$ branes at angles.
}

\author{
{Igor Pesando$^1$}
\\
~\\
~\\
$^1$Dipartimento di Fisica, Universit\`a di Torino\\
and I.N.F.N. - sezione di Torino \\
Via P. Giuria 1, I-10125 Torino, Italy\\
\vspace{0.3cm}
\\{ipesando@to.infn.it}
}

\maketitle
\thispagestyle{empty}

\abstract{
We compute the generic correlator  with $L$ untwisted
operators and $N$ (excited) twist fields
for branes at angles on $T^2$ 
and show that it is given by a generalization of the Wick theorem.
We give also the recipe to compute efficiently the 
generic OPE between an untwisted operator and an
excited twisted state.
}
\\
\\
keywords: {D-branes, Conformal Field Theory}
\\
\\

\newpage

\section{Introduction and conclusions}

Since their introduction, D-branes have been very important in the formal
development of string theory as well as in attempts to apply string
theory to particle phenomenology and cosmology. 
However, the requirement of chirality in any physically realistic
model  leads to a somewhat restricted number of possible D-brane
set-ups.  An important class is intersecting brane models where
chiral fermions can arise at the intersection of two branes at angles.
An important issue for these models is the computation of Yukawa
couplings and flavour changing neutral currents.

Besides the previous computations many  other computations 
often involve correlators of twist fields and excited
twist fields. 
It is therefore important and interesting in its own to be able to
compute these correlators also because it is annoying to be able to
compute, at least theoretically, all possible correlators involving
all kinds of excited spin fields while not being able to do so with
twist fields.
As known in the literature \cite{Dixon:1986qv} 
and explicitly shown in \cite{Pesando:2011ce}
for the case of magnetized branes these computations boil down to the
knowledge of the Green function in presence of twist fields and of the
correlators of the plain twist fields.
In many previous papers correlators with excited twisted fields have been 
computed on a case by case basis without a clear global picture, see
for example
(\cite{Burwick:1990tu},
\cite{Erler:1992gt}, \cite{Anastasopoulos:2013sta})

In this technical paper we have analyzed the $N$ excited twist fields
amplitudes with $L$ boundary vertices at tree
level for open strings localized at $D$-branes intersections on $R^2$
(or $T^2$)
using the classical path integral approach
(\cite{Dixon:1986qv},\cite{Atick:1987kd}) which is more
efficient than the also classical sewing approach
(\cite{Corrigan:1975sn}, \cite{DiBartolomeo:1990fw}).
This approach has been explored in many papers in the branes at
angles setup as well as the T dual magnetic branes setup see for example
(\cite{Bianchi:1991rd}, \cite{Antoniadis:1993jp},
\cite{Gava:1997jt}, \cite{David:2000um},
\cite{Abel:2003vv}, \cite{Cvetic:2003ch}, 
\cite{Lust:2004cx},
\cite{Bertolini:2005qh}, \cite{Lawrence:2007bk},
\cite{Duo:2007he},\cite{Choi:2007nb},
\cite{Conlon:2011jq},
\cite{Pesando:2012cx}).
We will nevertheless follow a slightly different approach, the so
called Reggeon vertex \cite{SDS}, 
which allows to compute the generating function
of all correlators, in particular we will use the formulation put
forward in \cite{Petersen:1988cf}.

This paper is organized as follows. In section 2 we review the
geometrical framework of branes at angles and fix our conventions.
In the same section we discuss carefully how to make use of the doubling
trick in presence of multiple cuts and the existence of local and
global constraints.
In section 3 we compute the OPE of chiral and boundary vertex
operators with an arbitrary excited twist operator by relying on the
operator to state correspondence. We propose also a better notation
for excited twist operators than that usually used which requires a
new symbol for any excited twist operator.
In this same section and for use in the  fourth
 we establish also which chiral operators are best suited
to obtain excited twist operators in the easiest way.
Finally in section 4 we compute the generating function of correlators
of $N$ excited twist operators with $L$ boundary operators.
We do this in steps by first computing the interaction of boundary and
chiral vertices with twist field operators and then computing the
desired correlators by letting appropriate combinations of chiral
vertex operators collide with twist fields.
Our main result is the generating function given in
eq. (\ref{reggeon-excited-twists+bou}) which shows that all
correlators can be computed once the $N$ plain twist  operators
correlator  together with the Green function in presence of
these $N$ twists are known.
This expression remains nevertheless quite formal since 
it requires the precise knowledge of
the Green function \footnote{
Note that the Green functions used in this paper are dimensionful and
normalized as $\partial_u \bar\partial_\bu G^{I J}(u,\bu; v, \bv;\{
\epsilon_t\}) = -\frac{\alpha'}{2} \delta^{I J} \delta^2(u-v)$.}
 and its regularized versions.
Therefore
somewhat explicit expressions of these quantities are given in appendix  
\ref{app:functions-for_excited}
(see also appendixes \ref{app:Green_functions} and
\ref{app:boundray_green})
and completely explicit expressions for all involved quantities in the $N=3$
case are given in appendix \ref{app:explicit_expressions_N=3}.
From these expressions it is clear that the computation of amplitudes,
i.e. moduli integrated correlators, with
(untwisted) states carrying momenta are very unwieldy because Green
functions can at best be expressed as sum of product of type D
Lauricella functions. This should however not be a complete surprise
since in \cite{Hamidi:1986vh} it was shown that twist fields
correlators in orbifold setup are connected to loop amplitudes which,
up to now, have not been expressed in term of simpler functions.

\section{Review of branes at angles}
\begin{figure}[hbt]
\begin{center}
\def\svgwidth{300px}
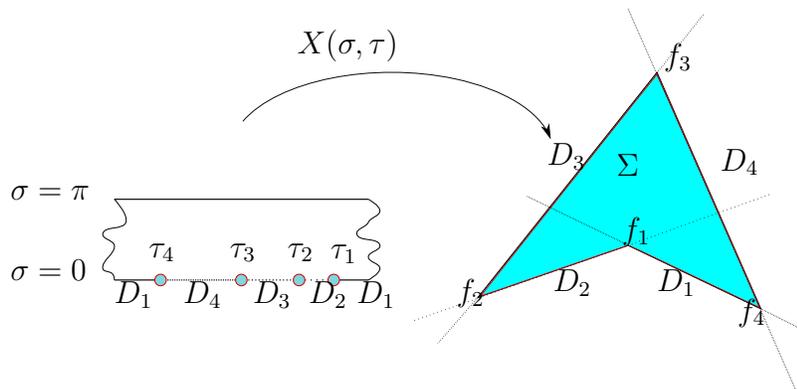
\end{center}
\vskip -0.5cm
\caption{Map from the Minkowskian worldsheet to the target polygon $\Sigma$.}
\label{fig:stripe2polygon}
\end{figure}

\begin{figure}[hbt]
\begin{center}
\def\svgwidth{250px}
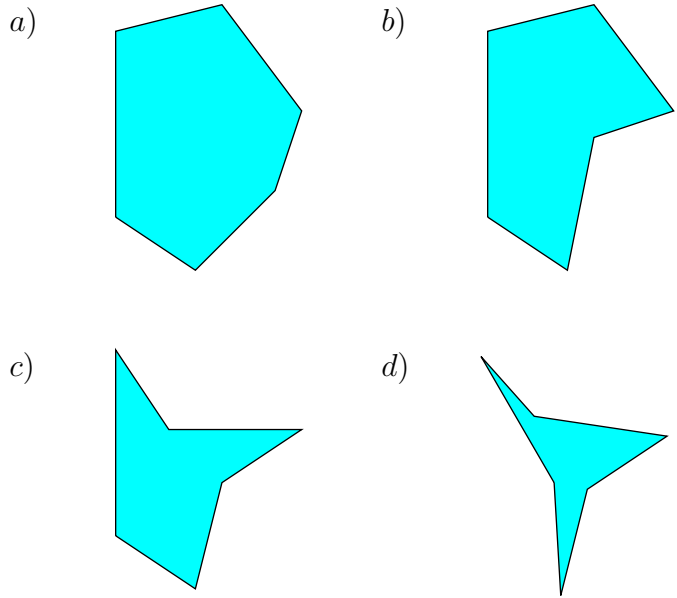
\end{center}
\vskip -0.5cm
\caption{The four different cases with $N=6$. 
$a)$ $M_{ccw}=2$ and $M=4$ where $M_{ccw}$ is measured
  counterclockwise and $M_{cw}$ clockwise.
$b)$ $M_{ccw}=3$ and $M=3$.
$c)$ $M_{ccw}=4$ and $M=2$.
$d)$ $M_{ccw}=5$ and $M=1$.
}
\label{fig:6gons}
\end{figure}

The Euclidean action for a string configuration is given by
\begin{equation}
S
=
\frac{1}{4\pi \alpha'} \int d\tau_E \int_0^\pi d\sigma~ (\partial_\alpha X^I)^2
=
\frac{1}{4\pi \alpha'} \int_H d^2 u~ 
(\du X \dub \barX + \dub X \du \barX )
\end{equation}
where $u\in H$, the upper half plane, 
$d^2 u= e^{2 \tau_E} d\tau_E d\sigma= \frac{d u ~d\bar  u}{2 i}$
and $I=1,2$ or $z, \bz$ so that $X= X^z= \frac{1}{\sqrt{2}}(X^1+i X^2)$, 
$\barX= X^\bz= X^*$.
The complex string coordinate is a map from the upper half plane to a
closed polygon $\Sigma$ in $\C$, i.e. $X:H \rightarrow \Sigma\subset \C$.
For example in fig. \ref{fig:stripe2polygon} we have pictured the interaction
of $N=4$ branes at angles $D_t$ with $t=1,\dots N$. The interaction
between brane $D_t$ and $D_{t+1}$ is in $f_t\in\C$. We use the rule
that index $t$ is defined modulo $N$.
As shown in \cite{Pesando:2011ce} given the number of twist fields
$N$ there are $N-2$ different sectors which correspond to the number
of reflex angles (the interior angles bigger than $\pi$) as shown in
figure \ref{fig:6gons} in the case $N=6$.
The intuitive reason why they are different is that we need go through
the straight line, i.e. no twist,  if we want to go
from a reflex angles to a more usual convex one.

\begin{figure}[hbt]
  \begin{minipage}[t]{0.45\linewidth}
     \def\svgwidth{200px}
     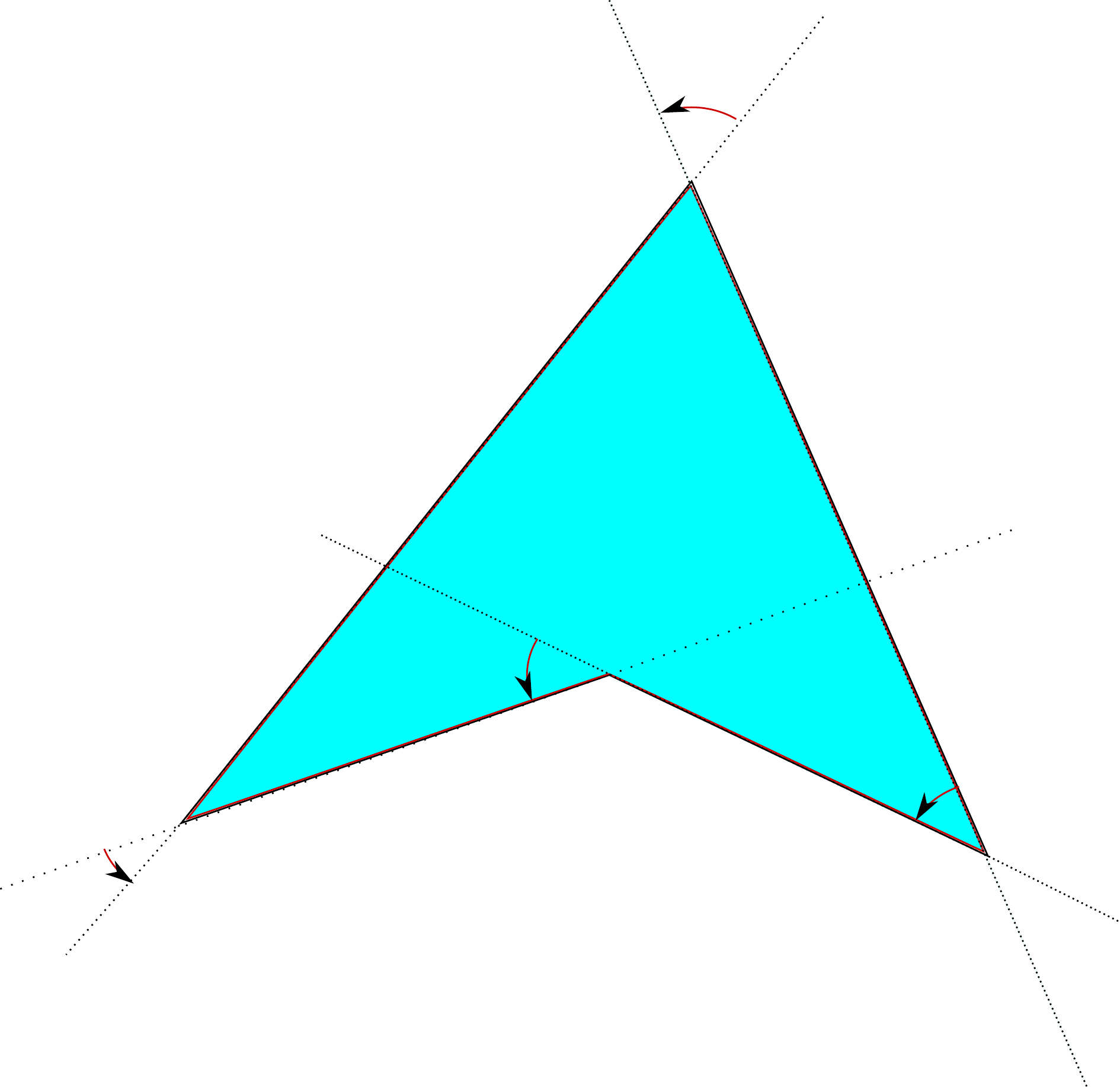
     \vskip -0.5cm
     \caption{A polygon $\Sigma$ with an reflex angle and branes
       labeled clockwise with $N=4$ and $M=1$. }
     \label{fig:N4Sigma_clockwise}
  \end{minipage}
  \begin{minipage}[t]{0.45\linewidth}
     \def\svgwidth{200px}
     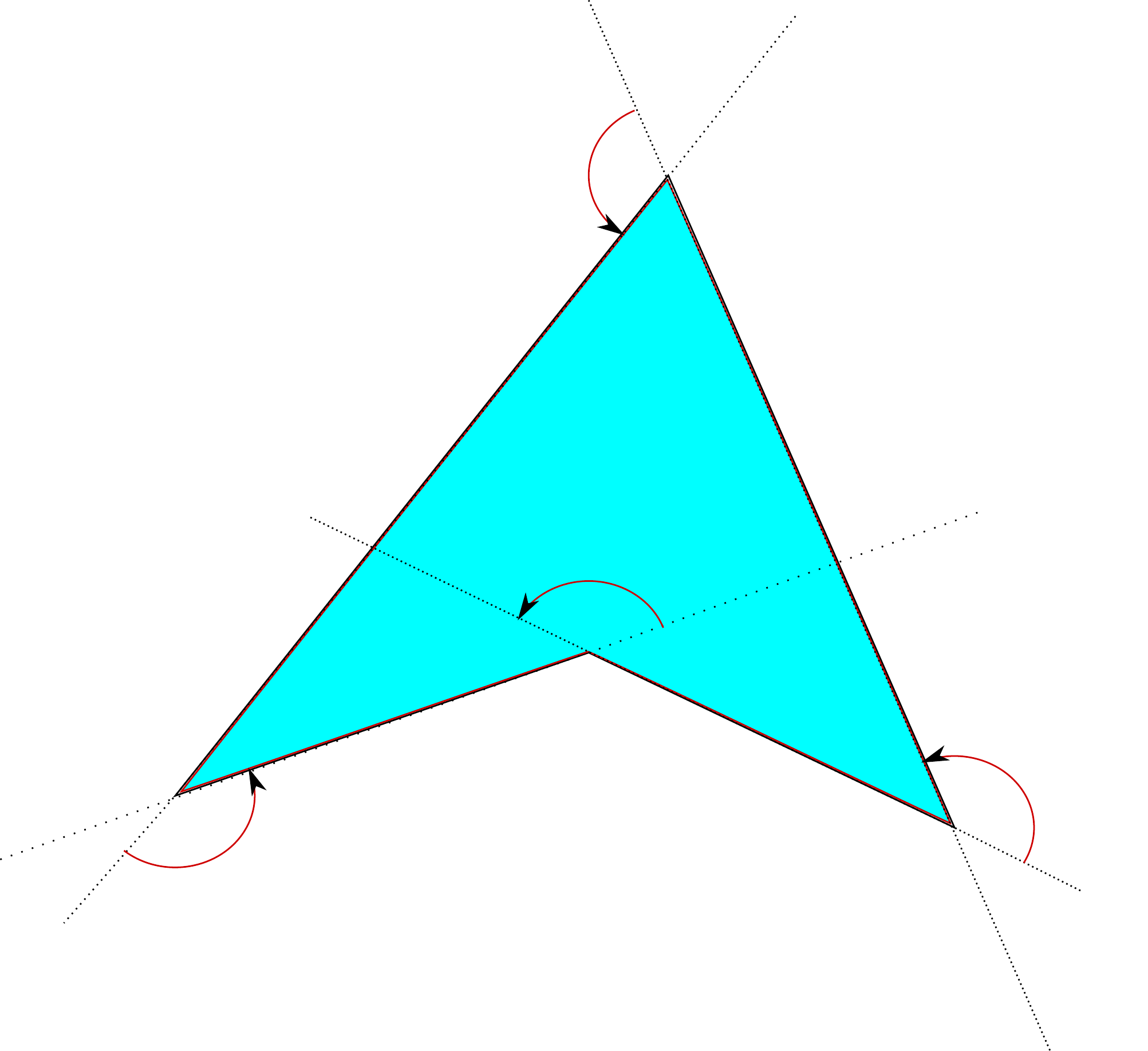
     \vskip -0.5cm
     \caption{A polygon $\Sigma$ with an reflex angle and branes labeled
       counterclockwise with $N=4$ and $M_{ccw}=3$. }
     \label{fig:N4Sigma}
  \end{minipage}
\end{figure}

\begin{figure}[hbt]
\begin{center}
\def\svgwidth{300px}
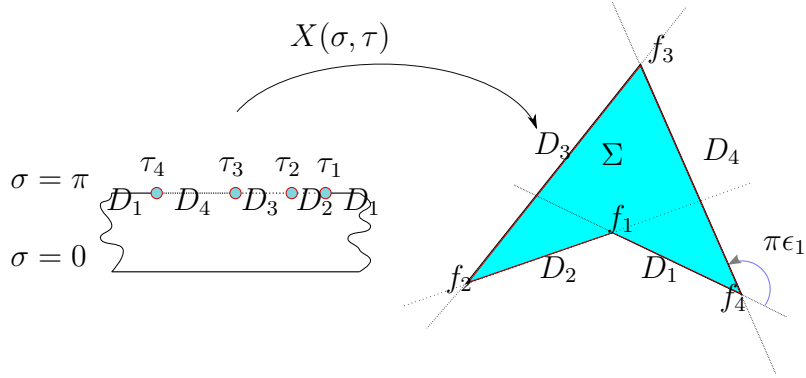
\end{center}
\vskip -0.5cm
\caption{Map from the worldsheet to the target polygon $\Sigma$ with
  complementary angles.}
\label{fig:stripe2polygon_other_boundary}
\end{figure}

They are labeled by an integer $1\le M \le N-2$ given by
\begin{equation}
M=\sum_{t=1}^N \epsilon_t
\end{equation}
where $0<\epsilon_t<1$ (we define also $\bar \epsilon_t = 1-\epsilon_t$
for simplifying the expressions) are the twists defined as in
eq. (\ref{eps-alf-alf}) and correspond to the angles measured from
brane $D_t$ to brane $D_{t+1}$
 when they are labeled clockwise as in figure 
\ref{fig:N4Sigma_clockwise}.
\COMMENTOOK{As written below it is wrong we keep the same branes
  ordering and measure angles in the complementary way then we get the
  time reverse amplitude
}
It is also possible to have the very same geometrical configuration 
where branes are simply labeled 
counterclockwise as shown in figure \ref{fig:N4Sigma} for which we have
$M_{ccw}= N-M$ when we still measure angles from brane $D_t$ to brane $D_{t+1}$ .
To understand the meaning we notice that
if we relabel the branes as in the clockwise case, i.e. we consider
the branes labeled as in figure \ref{fig:N4Sigma_clockwise} but with
the angles measured as in figure \ref{fig:N4Sigma} and we use the
conventions we use for the local description, described in the next section, 
the interactions on worldsheet would take
place in the reversed world sheet time order or in the same order but
on the other boundary as shown in figure
\ref{fig:stripe2polygon_other_boundary}.
\COMMENTOO{write it correctly}
Since the physics must be connected and actually the two correlators are 
connected by complex conjugation and exchange $\epsilon \leftrightarrow
1- \epsilon$ we
have chosen to measure them clockwise.

\subsection{The local description}
Locally at the interaction point $f_t\in\partial \Sigma \subset \C$
the boundary conditions for the brane $D_{t}$ are given by
\begin{align}
Re( e^{-i \pi \alpha_{t}} X'_{loc} |_{\sigma=0} ) 
=
Im( e^{-i \pi \alpha_{t}} X_{loc} |_{\sigma=0}) -g_{t} 
=0
\label{local-boundary-condition-s=0}
\end{align}
where $g_t\in\R$ is the distance from the line parallel to the brane
going through the origin.
Similarly the boundary conditions for the brane $D_{t+1}$ is given by
\begin{align}
Re( e^{-i \pi \alpha_{t+1}} X'_{loc} |_{\sigma=\pi} ) 
=
Im( e^{-i \pi \alpha_{t+1}} X_{loc} |_{\sigma=\pi}) -g _{t+1}
=0
\label{local-boundary-condition-s=pi}
\end{align}
The interaction point is then
\begin{equation}
f_t= \frac{e^{i \pi \alpha_{t+1}} g_t - e^{i \pi \alpha_{t}} g_{t+1} }
{\sin~ \pi(\alpha_{t+1}-\alpha_{t})}
\end{equation}

\begin{figure}[hbt]
\begin{center}
\def\svgwidth{300px}
\input{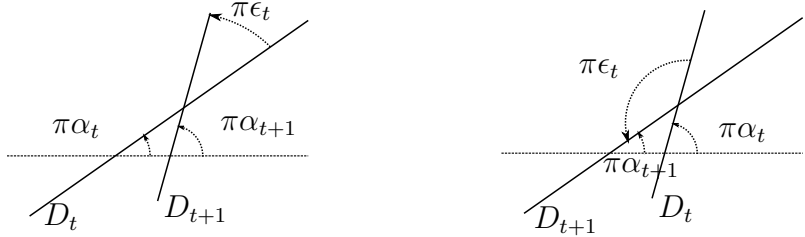}
\end{center}
\vskip -0.5cm
\caption{The connection between $\epsilon$ and the geometrical angles
  $\alpha$s defining the branes.}
\label{fig:angles}
\end{figure}

When we write the Minkowskian string expansion as 
$X(\sigma,\tau)=X_L(\tau+\sigma)+X_R(\tau-\sigma)$
the previous boundary conditions imply (and not become since they are
not completely equivalent because of zero modes)
\begin{align}
X'_{L~loc}(\xi)= e^{i 2\pi \alpha_t} X'_{R~loc}(\xi)
,~~~~
X'_{L~loc}(\xi+\pi)= e^{i 2\pi \alpha_{t+1}} X'_{R~loc}(\xi-\pi)
\label{local-boundary-condition-upper}
\end{align}
or in a more useful way in order to explicitly compute the mode expansion
\begin{align}
X'_{L~loc}(\xi+2\pi)= e^{i 2\pi \epsilon_t} X'_{L~loc}(\xi)
,~~~~
X'_{R~loc}(\xi+2\pi)= e^{-i 2\pi \epsilon_t} X'_{R~loc}(\xi)
\label{local-boundary-condition}
\end{align}
where we have defined
\begin{equation}
\epsilon_t=
\left\{\begin{array}{c c}
(\alpha_{t+1}-\alpha_t) & \alpha_{t+1}>\alpha_t
\\
1+(\alpha_{t+1}-\alpha_t) & \alpha_{t+1}<\alpha_t
\end{array}
\right.
\label{eps-alf-alf}
\end{equation}
so that $0<\epsilon_t<1$ and there is no ambiguity in the phase 
$e^{i  2\pi \epsilon_t}$ entering the boundary conditions.
The quantity $\pi \epsilon_t$ is the angle between the two branes
$D_t$ and $D_{t+1}$ measured counterclockwise as shown in
fig. \ref{fig:angles}. 
A consequence of this definition is that $\epsilon$ becomes
$1-\epsilon$  when we flip the order of two branes.
For example the angles in fig. \ref{fig:N4Sigma} become those in
fig. \ref{fig:N4Sigma_clockwise} when we reverse the order we count
the branes, i.e. when we follow the boundary clockwise instead of
counterclockwise.

We introduce as usual the Euclidean fields 
$X_{loc}(\ul,\bul)$, $\bar X_{loc}(\ul,\bul)$ 
by a worldsheet Wick rotation in such a way they are defined on the 
upper half plane by $\ul=e^{\tau_E+i\sigma} \in H$. 
The previous choice of having brane $D_t$ at $\sigma=0$
(\ref{local-boundary-condition-s=0}) and brane $D_{t+1}$ at
$\sigma=\pi$ (\ref{local-boundary-condition-s=pi}) implies that 
in the local description where the interaction point is at $x=0$ $D_t$
is mapped into $x>0$ and $D_{t+1}$ into $x<0$.
The boundary conditions (\ref{local-boundary-condition-upper}) can
then immediately be written as
\begin{align}
 \partial X_{loc}(x\loc+ i 0^+ ) 
&= 
e^{i 2 \pi \alpha_t}
 \bar \partial \bar X_{loc}( x\loc- i 0^+)
~~~~~~
0<x\loc,
\nonumber\\
 \partial X_{loc}(x\loc+ i 0^+ ) 
&= 
e^{i 2 \pi \alpha_{t+1}}
\bar  \partial \bar X_{loc}(x\loc - i 0^+)
~~~~
x\loc<0
\label{local-boundary-upper}
\end{align}
and similarly relations for $\bX$ which can be obtained by complex conjugation. 
When we add to the previous conditions the further constraints
\begin{equation}
Im(e^{-i\pi \alpha_t} X(x,x))=g_t~~~x>0
\label{local-boundary-points}
\end{equation}
we obtain a system of conditions which are equivalent to the original
ones (\ref{local-boundary-condition-s=0}, \ref{local-boundary-condition-s=pi}).

In order to express the boundary conditions 
(\ref{local-boundary-condition}) in the Euclidean formulation
it is better to introduce the local fields 
defined on the whole complex plane by the doubling trick as
\begin{align}
\dz\cX_{loc}(\zl)
&=
\left\{
\begin{array}{cc}
  \du X_{loc}(\ul) 
  & \zl=\ul\mbox{ with }{Im~} \zl >0\mbox{ or } \zl\in\R^+ 
  \\
  e^{i 2\pi \alpha_t} \dub \bX_{loc}(\bul) 
  & \zl=\bul\mbox{ with }{Im~} \zl <0\mbox{ or } \zl\in\R^+ 
\end{array}
\right.
\nonumber\\
\dz \cbX_{loc}(\zl)
&=
\left\{
\begin{array}{cc}
  \du \bX_{loc}(\ul) 
  & 
  \zl= \ul\mbox{ with }{Im~} \zl >0\mbox{ or } \zl\in\R^+ 
  \\
  e^{-i 2\pi \alpha_t} \dub X_{loc}(\bul) 
  & 
  \zl=\bul\mbox{ with }{Im~} \zl <0\mbox{ or } \zl\in\R^+ 
\end{array}
\right.
\label{loc-calX-calXbar}
\end{align}
In this way we can write  eq.s (\ref{local-boundary-condition}) as
\begin{align}
 \partial\cX_{loc}(e^{i 2\pi} \delta ) 
= 
e^{i 2 \pi \epsilon_t}
 \partial\cX_{loc}(\delta)
,~~~~
 \partial\cbX_{loc}(e^{i 2\pi} \delta ) 
= 
e^{-i 2 \pi \epsilon_t}
 \partial\cbX_{loc}(\delta)
\label{loc-Euc-bc}
\end{align}
Notice that the two Minkowskian boundary conditions in eq.s
(\ref{local-boundary-condition}) can be mapped one into the other by
complex conjugation 
while the corresponding ones in the Euclidean version
given in eq.s  (\ref{loc-Euc-bc}) are independent and each is
mapped into itself by complex conjugation therefore the Euclidean
classical solutions for $\cX\loc$ and $\cbX\loc$ are independent.

The quantization of the string with given boundary conditions yields
\begin{align}
X_{loc}(\ul,\bul)
&=
f_t
&+
i\oh \sqrt{2\alpha'}
e^{i \pi \alpha_{t}}
\sum_{n=0}^\infty
\left[ 
\frac{\bar \alpha_{(t)n+\bar\epsilon_t}}{ n+\bar\epsilon_t} \ul^{-(n+\bar\epsilon_t)}
-
\frac{\alpha_{(t)n+\epsilon_t}^\dagger}{ n+\epsilon_t} \ul^{n+\epsilon_t}
\right]
\nonumber\\
&
&+
i \oh \sqrt{2\alpha'}
e^{i \pi \alpha_{t}}
\sum_{n=0}^\infty 
\left[
-
\frac{\bar \alpha_{(t)n+\bar\epsilon_t}^\dagger}{ n+\bar\epsilon_t} \bul^{n+\bar\epsilon_t}
+
\frac{\alpha_{(t)n+\epsilon_t}}{ n+\epsilon_t} \bul^{-(n+\epsilon_t)}
\right]
\nonumber\\
\bar X_{loc}(\ul,\bul)
&=
f_t^*
&+
i \oh \sqrt{2\alpha'}
e^{-i \pi \alpha_{t}}
\sum_{n=0}^\infty 
\left[
-
\frac{\bar \alpha_{(t)n+\bar\epsilon_t}^\dagger}{ n+\bar\epsilon_t} \ul^{n+\bar\epsilon_t}
+
\frac{\alpha_{(t)n+\epsilon_t}}{ n+\epsilon_t} \ul^{-(n+\epsilon_t)}
\right]
\nonumber\\
&
&+
i \oh \sqrt{2\alpha'}
e^{-i \pi \alpha_{t}}
\sum_{n=0}^\infty
\left[ 
\frac{\bar \alpha_{(t)n+\bar\epsilon_t}}{ n+\bar\epsilon_t} \bul^{-(n+\bar\epsilon_t)}
-
\frac{\alpha_{(t)n+\epsilon_t}^\dagger}{ n+\epsilon_i} \bul^{n+\epsilon_t}
\right]
\label{loc-exp-X}
\end{align}
with $\bar\epsilon_t =1 -\epsilon_t$ and we can interpret $f_t$ as the
classical solution
\begin{equation}
X_{loc,cl}=f_t,~~~~
\bar X_{loc,cl}=\bar f_t
\end{equation}
since it is the only solution to the equations of motion with finite
euclidean action.

We find also the non trivial commutation relations ($n,m \ge 0$)
\begin{equation}
[ \alpha_{(t) n+\epsilon_t}, \alpha_{(t) m+\epsilon_t}^\dagger]
=( n+\epsilon_t) \delta_{m,n}
,~~~~
[ \bar\alpha_{(t) n+\bar\epsilon_t}, \bar\alpha_{(t) m+\bar\epsilon_t}^\dagger]
=( n+\bar\epsilon_t) \delta_{m,n}
\end{equation}
and the vacuum is defined in the usual way by
\begin{equation}
\alpha_{(t) n+\epsilon_t} |T_t\rangle =\bar \alpha_{(t) n+\bar\epsilon_t }|T_t\rangle =0
~~~~ n\ge 0
\label{twisted-vacuum}
\end{equation}
At first sight it may seem that the vacuum encodes the $\epsilon_t$
information only but as we show in eq. (\ref{eikX-sigma+})
it contains also information about $f_t$ 
\COMMENTOOK{CHECK}
and $\alpha_t$ and $\alpha_{t+1}$ which can be extracted
from the proper OPEs.

\subsection{Global description}
In the local description, where the interaction point is at $\xl=0$, $D_t$
is mapped into $\xl>0$ and $D_{t+1}$ into $\xl<0$
this means that in the global description the
world sheet interaction points are mapped on the boundary of the upper
half plane  so that $x_{t+1}< x_t$.
The global equivalent of the local boundary conditions given in
eq.s (\ref{local-boundary-upper}) which is useful
for the path integral formulation where we cannot use the equations
of motion is
\begin{align}
 \partial_u X(u,\bu)|_{u=x+ i 0^+ } 
&= 
e^{i 2 \pi \alpha_t}
 \bar \partial_\bu \bar X(u,\bu)|_{u=x+ i 0^+ } 
~~~~
x_{t}<x<x_{t-1}
\nonumber\\
 \partial_u \bX(u,\bu)|_{u=x+ i 0^+ } 
&= 
e^{-i 2 \pi \alpha_t}
 \bar \partial_\bu X(u,\bu)|_{u=x+ i 0^+ } 
~~~~
x_{t}<x<x_{t-1}
\label{path-integral-global-boundary-upper}
\end{align}
To the previous constraints one must also add the global equivalent to
eq. (\ref{local-boundary-points})
\begin{equation}
X(x_t,\bar x_t)= f_t,~~~~
\bar X(x_t ,\bar x_t)= f_t^*
\label{global-boundary-points}
\end{equation}
in order to get a system of boundary conditions equivalent to the
original ones.
%
When using an operatorial approach the global equivalent of eq.s 
(\ref{local-boundary-upper}) become
\begin{align}
 \partial X_L(x+ i 0^+ ) 
&= 
e^{i 2 \pi \alpha_t}
 \bar \partial \bar X_R(x - i 0^+)
~~~~
x_{t}<x<x_{t-1}
\nonumber\\
 \partial \bX_L(x+ i 0^+ ) 
&= 
e^{-i 2 \pi \alpha_t}
 \bar \partial X_R(x - i 0^+)
~~~~
x_{t}<x<x_{t-1}
\label{global-boundary-upper}
\end{align} 
If we introduce the global fields 
defined on the whole complex plane by the doubling trick as
\begin{align}
\dz\cX(z)
&=
\left\{
\begin{array}{cc}
  \du X(u) 
  & z=u\mbox{ with }{Im~} z >0\mbox{ or } z\in\R-[x_N,x_1] 
  \\
  e^{i 2\pi \alpha_1} \dub \bX(\bar u) 
  & z=\bar u\mbox{ with }{Im~} z <0\mbox{ or } z\in\R-[x_N,x_1] 
\end{array}
\right.
\nonumber\\
\dz \cbX(z)
&=
\left\{
\begin{array}{cc}
  \du \bX(u) 
  & 
  z= u\mbox{ with }{Im~} z >0\mbox{ or } z\in\R-[x_N,x_1] 
  \\
  e^{-i 2\pi \alpha_1} \dub X(\bar u) 
  & 
  z=\bar u\mbox{ with }{Im~} z <0\mbox{ or } z\in\R-[x_N,x_1] 
\end{array}
\right.
\label{calX-calXbar}
\end{align}
the local boundary conditions (\ref{local-boundary-condition}) 
can be written in the global formulation as
\begin{align}
 \partial\cX(x_t+ e^{i 2\pi} \delta ) 
&= 
e^{i 2 \pi \epsilon_t}
 \partial\cX(x_t +\delta)
\nonumber\\
 \partial\cbX(x_t+ e^{i 2\pi} \delta ) 
&= 
e^{-i 2 \pi \epsilon_t}
 \partial\cbX(x_t +\delta)
.
\end{align}
Finally it is worth noticing the behavior
of the previously introduced fields under complex conjugation when $z$
is restricted to $z\in \C-[-\infty,x_1]$
\begin{align}
[ \dz \cX(z)]^*
&=
e^{-i 2\pi \alpha _1} \dz \cX(z\rightarrow \bar z) 
=
\dzb \cbX(\bar z)
=
\left\{
\begin{array}{c c}
  \dub \bX(\bu) 
  & \bz=\bu
  \\
  e^{-i 2\pi \alpha_1} \du X(u) 
  & \bz= u
\end{array}
\right.
\nonumber\\
[ \dz \cbX(z)]^*
&=
e^{-i 2\pi \alpha _1} \dz \cbX(z\rightarrow \bar z) 
=
\dzb \cX(\bar z)
=
\left\{
\begin{array}{c c}
  \dub X(\bu) 
  & \bz=\bu
  \\
  e^{i 2\pi \alpha_1} \du \bX(u) 
  & \bz= u
\end{array}
\right.
\label{dcX-dbcX}
\end{align}
where $ \dz \cX(z\rightarrow \bar z)$ means that the holomorphic $\dz
\cX(z)$ is evaluated at $\bz$.
The previous expressions also show that it is not necessary to introduce the
antiholomorphic fields $\dzb \cX(\bar z)$ and
 $\dzb \cbX(\bar z)$ which it is possible to construct applying the doubling
trick on  $\dub X(\bar u)$ and $\dub \bX(\bar u)$ respectively.

\section{Twisted Fock Space and OPEs}
Given the vacuum $|T\rangle$~~ \footnote{
In this section we have dropped the dependence on $t$ as much as
possible to simplify the notation  e.g.  $\epsilon_t\rightarrow \epsilon$.
}
defined as usual in eq. (\ref{twisted-vacuum})
and the expansions (\ref{loc-exp-X}) we can immediately write a
normalized basis element of the Fock space as
\begin{equation}
\prod_{n=0}^\infty 
\left[
\frac{1}{N_n!}\left( \frac{\alpha^\dagger_{n+\epsilon}}
                     {\sqrt{n+\epsilon}}\right)^{N_n}
\frac{1}{\bar N_n!}\left( 
\frac{\bar\alpha^\dagger_{n+\bar\epsilon}}{\sqrt{n+\bar\epsilon}} \right)^{\bar N_n}
\right]
|T\rangle
\end{equation}
We want now to explore the state to operator correspondence.
To the vacuum $|T\rangle$ we associate the twist field
$\sigma_{\epsilon, f}(x)$ which depends both on the twist $\epsilon$
and on the position $f\in\C$ so that
\begin{equation}
|T\rangle= \lim_{x\rightarrow0}\sigma_{\epsilon, f}(x) |0\rangle_{SL(2)}
\end{equation}
with normalization
\begin{equation}
\langle T | T \rangle =1
\end{equation}
For the other states in Fock space it is however better to
introduce the non normalized states
\begin{equation}
\prod_{n=0}^\infty 
\left(\ke n! \alpha^\dagger_{n+\epsilon}\right)^{N_n}
\left(\kbe n! \bar\alpha^\dagger_{n+\bar\epsilon} \right)^{\bar N_n}
|T\rangle
\label{twisted-sts}
\end{equation}
with $\ke= -i\oh\sqrt{2\alpha'} e^{i \pi \alpha_{t}}$ and
$\kbe= -i\oh\sqrt{2\alpha'} e^{-i \pi \alpha_{t}}$ 
to which we make correspond the (generically non primary) {\sl chiral}
operators
\begin{equation}
\left[\prod_{n=0}^\infty 
\left(\partial^{n+1}_u X  \right)^{ N_n}
\left( \partial^{n+1}_u \bar X  \right)^{\bar N_n}
\sigma_{\epsilon, f}\right](x)
\label{twisted-ops}
\end{equation}
This notation can be partially misleading since, for example,
it is {\sl not} true that (see eq. (\ref{d2x-sigma}) for the right OPE)
\footnote{
Notice that on shell $\partial^2_u X(u,\bu)= \oh \partial^2_u X_L(u)$
},
\begin{equation}
\partial^2_u X(u,\bu) ~\sigma_{\epsilon, f}(x) \sim
\frac{1}{(u-x)^\#} (\partial^2_u X \sigma_{\epsilon, f}) (x) + \dots
\label{wrong-OPE}
\end{equation}
and therefore the operator (\ref{twisted-ops}) is just a way to
write the operator which corresponds to the state (\ref{twisted-sts})
under the operator to state correspondence. 
The advantage of this notation is that it clearly shows which state
corresponds to which operator and that it is consistent with the usual
untwisted state to operator correspondence for which
\begin{equation}
\prod_{n=1}^\infty 
\left[
\left(\ke n! \alpha^\dagger_{n}\right)^{N_n}
\left(\kbe n! \bar\alpha^\dagger_{n} \right)^{\bar N_n}
\right]
|0\rangle_{SL(2)}
\leftrightarrow
\left[\prod_{n=1}^\infty 
\left( \partial^n_u X  \right)^{N_n}
\left( \partial^n_u \bar X \right)^{\bar N_n}
\right](u,\bu)
\end{equation}

We have in particular, as we soon show, that 
\begin{align}
\du X(u,\bu) \sigma_{\epsilon,f}(x) 
&\sim (u-x)^{\epsilon-1} (\du X \sigma_{\epsilon,f})(x)
\nonumber\\
\du \bX(u,\bu) \sigma_{\epsilon,f}(x) &\sim (u-x)^{\bar\epsilon -1} 
(\du \bX \sigma_{\epsilon,f})(x)
\label{OPEs-epsilon}
\end{align}
so that the excited twist operators 
$(\du X \sigma_{\epsilon,f})(x)$ and 
$(\du \bX \sigma_{\epsilon,f})(x)$ are what usually written as
$\tau(x)$ and $\bar \tau(x)$.
This observation also hints to another reason why to use the
notation (\ref{twisted-ops}) which is to avoid the proliferation of
new  symbols, one for each excited twist operator.

Finally notice that almost all boundary operators can be recovered from chiral
ones, for example
when $ x_t<x< x_{t-1}$ and with the help of the boundary conditions
(\ref{path-integral-global-boundary-upper}) we get
\begin{equation}
\partial_x( X(x,x)) 
=
\partial_u X(u,\bu) |_{u=x}+\bar\partial_\bu X(u,\bu) |_{u=x}
=
\partial_u X(u,\bu) |_{u=x}+ e^{i 2\pi \alpha_t} \partial_u \bar X(u,\bu) |_{u=x}
\label{dxX=duX+dubX}
\end{equation}
where both $\partial_u X(u,\bu)|_{u=x}$ and $\partial_u \bar
X(u,\bu)|_{u=x}$ are chiral operators computed on the boundary.
A notable exception is however $e^{i \bk X(x,x) + i k \bar X(x,x)}$
which is intrinsically non chiral since it cannot be expressed off
shell using chiral operators.
In particular from the previous eq. (\ref{dxX=duX+dubX}) it follows 
that boundary operators have more complex OPEs, such as
\begin{align}
\partial_x X(u,\bu)|_{u=\xii1} \sigma_{\epsilon,f}(\xii2) 
\sim 
(\xii1-\xii2)^{\epsilon-1} (\du X \sigma_{\epsilon,f})(\xii2)
+
(\xii1-\xii2)^{\bar\epsilon -1} 
e^{i 2\pi \alpha_t}
(\du \bX \sigma_{\epsilon,f})(\xii2)
.
\end{align}

\subsection{Chiral OPEs from local Fock space}
\label{sect:chiral_OPE}
Let us now see how to compute OPEs of a chiral operator with an excited twist
field using the local operatorial formalism.
We will make several examples to make clear the way to proceed.

We start considering the simplest example, i.e. the OPE 
$\du X(u,\bu) \sigma_{\epsilon,f}(x)$.   
In order to find this OPE we compute\footnote{
\COMMENTOOK{RETHINK}
One could wonder whether eq. (\ref{dX-sigma-ope}) should actually
written using the quantum fluctuation, i.e.
$\lim_{x\rightarrow0} \du X_q(u,\bu)\sigma_{\epsilon, f}(x) |0\rangle_{SL(2)}
= \du X_{loc,q}(\ul) |T\rangle$.
The answer is no because CFT has no meaningful way of
splitting $X$ into the classical and quantum part, moreover if we
assume the previous expression we naively get
$
\lim_{x\rightarrow 0} \du X(u,\bu)\sigma_{\epsilon, f}(x) |0\rangle_{SL(2)}
=\lim_{x\rightarrow 0} \du (X_{cl}+X_q)(u,\bu)\sigma_{\epsilon, f}(x) |0\rangle_{SL(2)}
= \du (X_{cl}+X_{loc,q})(\ul) |T\rangle$
and while $\du X_{cl}$ has the expected singularity for $u \rightarrow
x_t$ it contains also
information on the location of the other twist fields, explicitly
$\du X_{cl}\sim (u-x_t)^{\epsilon-1} R(u,\{x_{\htt\ne t}\})$ which would imply
that full excited twists would contain information on these locations,
e.g.
$ (\du X \sigma_{\epsilon_t,f_t})(x_t)=  
\left( R(x_t,\{x_{\htt\ne t}\}) \sigma_{\epsilon_t,f_t} \right)(x_t)
+(\du X_q \sigma_{\epsilon_t,f_t})(x_t)
$.
 
}
\begin{equation}
\lim_{x\rightarrow0} \du X(u,\bu)\sigma_{\epsilon, f}(x) |0\rangle_{SL(2)}
= \partial_\ul X_{loc}(\ul) |T\rangle
\label{dX-sigma-ope}
\end{equation}
where $ \du X_{loc}(\ul)$ is the operator in the twisted Fock space
which corresponds to the abstract operator $ \du X(u,\bu)$ and we can
identify $\ul= u-x$.
Now using the explicit expansion (\ref{loc-exp-X}) we get
\begin{equation}
\partial_\ul X_{loc}(\ul) |T\rangle
=
\ul^{\epsilon-1}~
\ke \alpha_{\epsilon}^\dagger |T\rangle 
+
\ul^{\epsilon}~
\ke \alpha_{\epsilon+1}^\dagger |T\rangle
+\dots
\end{equation}
from which we deduce not only the leading order of the OPE (\ref{OPEs-epsilon})
but also the higher order terms 
\begin{equation}
\du X(u,\bu)\sigma_{\epsilon, f}(x)
=
(u-x)^{\epsilon-1} ~(\du X \sigma_{\epsilon,f})(x)
+
(u-x)^{\epsilon} ~(\du^2 X \sigma_{\epsilon,f})(x)
+\dots
\end{equation}

As a second example we consider the OPE
$\du^2 X(u,\bu)\sigma_{\epsilon, f}(x)$. Proceeding as before and using
the fact that the local Fock space operator which corresponds to
$\du^2 X(u,\bu)$ is $\partial^2_\ul X_{loc}(\ul)$ we can find
\begin{equation}
\partial^2_\ul X_{loc}(\ul) |T\rangle
=
\ul^{\epsilon-2}~(\epsilon-1)~
\ke \alpha_{\epsilon}^\dagger |T\rangle 
+
\ul^{\epsilon-1}~\epsilon~
\ke \alpha_{\epsilon+1}^\dagger |T\rangle
+\dots
\label{d1x-sigma}
\end{equation} 
so that we can deduce the OPE
\begin{equation}
\du^2 X(u,\bu)\sigma_{\epsilon, f}(x)
=
(u-x)^{\epsilon-2} ~(\epsilon-1)~(\du X \sigma_{\epsilon,f})(x)
+
(u-x)^{\epsilon-1} ~\epsilon~ (\du^2 X \sigma_{\epsilon,f})(x)
+\dots
\label{d2x-sigma}
\end{equation}
which shows clearly what stated before about the wrongness of
eq. (\ref{wrong-OPE}).  
Obviously eq. (\ref{d2x-sigma}) is compatible with
eq. (\ref{d1x-sigma}) since the former can be obtained from the latter
by taking the derivative $\partial_u$.

In the previous examples the local Fock space operator has the same
functional form of the abstract one but as discussed in
\cite{Pesando:2011yd} for the T dual configuration of branes with
magnetic field
this is not always the case.
The correct mapping, which is rederived in section
\ref{sect:deriv_gen_fun}, is given by
\begin{align}
\left[
\prod_{n=1}^\infty 
\left(\partial^n_u X \right)^{N_n}
\left(\partial^n_u \bar X \right)^{\bar N_n}
\right](u,\bu)
\leftrightarrow
\left.
\prod_{n=1}^\infty 
\frac{\partial^{N_n}}{\partial \bar c_n^{N_n}}
\frac{\partial^{\bar N_n}}{\partial c_n^{\bar N_n}}
\cS_c(c,\bc) \right|_{\bar c_{n\ge1}=c_{n\ge1}=0}
\end{align}
where the chiral generating function $\cS_c(c,\bc)$ is given by
\begin{align}
\cS_c(c,\bc)
=
&:
\exp\left\{
\sum_{n=1}^\infty  
\left[ \bar c_n \partial^n_\ul X_{loc}(\ul,\bul)
+ c_n \partial^n_\ul \bar X_{loc}(\ul,\bul) \right]
\right\}
:
\nonumber\\
&
\exp\left\{
\sum_{n,m=1}^\infty  
\bar c_n~c_m ~\partial^n_\ul ~\partial^m_\vl|_{\vl=\ul} 
\Delta^{z\bz}_c(\ul,\bul; \vl,\bvl; \epsilon)
\right\}
\label{S_chiral}
\end{align}
where $:\dots:$ is the normal ordering and 
\begin{equation}
\Delta^{z\bz}_c(\ul,\bul; \vl,\bvl; \epsilon)=
G^{z \bz}_{N=2}(\ul,\bul; \vl,\bvl;\{ 0,\epsilon; \infty, \bar\epsilon\}) 
-
G^{z \bz}_{U(t)}(\ul,\bul; \vl,\bvl) 
\end{equation}
is the regularized Green function with the twist $\sigma_{\epsilon,f}$ 
at $\xl=0$ and the anti-twist $ \sigma_{\bar \epsilon,f}$ in
$\xl=\infty$ and $G^{z \bz}_{U(t)}(\ul,\bul,\vl,\bvl)$ is the Green function for
the untwisted string with boundary conditions corresponding to
$D_t${} (see appendix \ref{app:Delta} for more details)\footnote{
For the chiral correlators there is no difference in using the
untwisted Green function for $D_t$ or $D_{t+1}$ since the 
$G^{z  \bz}_{U(t)}$ is the only piece of the untwisted Green function $G^{I J}_U$
which contributes and it is insensitive to the angle at which the brane is
rotated, this is not anymore true for the  boundary correlators as we
discuss later.
}.
It is worth discussing how $\Delta_c$ has to be interpreted either as a 
 kind of generating function or as a difference of Green
 functions, one of which associated to a couple of twist fields.
This point of view is important in order to avoid confusions 
which could arise 
when considering the role of $\Delta_c$ in  correlators with many twist fields.
The answer is given in the derivation of eq. (\ref{V_N+M}) which shows
that $\Delta_c$ is a difference of Green functions.
%

As an example of the consequences of the previous expression (\ref{S_chiral}) we can compute the local operator which corresponds to
the energy-momentum tensor 
$T(u) =- \frac{2}{\alpha'} \partial_u X \partial_u \bar X$.
We find the local operator
\begin{align}
T\loc(\ul) 
&=
- \frac{2}{\alpha'} :\partial_\ul X\loc \partial_\ul \bar X\loc :
- \frac{2}{\alpha'} \partial_\ul\partial_\vl \Delta^{z\bz}_c|_{\vl=\ul}
\nonumber\\
&=
- \frac{2}{\alpha'} :\partial_\ul X\loc \partial_\ul \bar X\loc:
- \frac{\epsilon \bar\epsilon}{2} \frac{1}{\ul^2}
\label{T-loc}
\end{align}
Then we can compute the OPE 
$T(u) \sigma_{\epsilon, f}(x)$ from
\begin{equation}
\left[- \frac{2}{\alpha'} :\partial_\ul X\loc \partial_\ul \bar X\loc:
- \frac{\epsilon \bar\epsilon}{2} \frac{1}{\ul^2}\right]
| T \rangle
=
- \frac{\epsilon \bar\epsilon}{2} \frac{1}{\ul^2} | T \rangle
+ \frac{1}{\ul} \alpha^\dagger_\epsilon  \bar\alpha^\dagger_{\bar\epsilon}
|T\rangle
+O(1)
\end{equation}
to be
\begin{equation}
T(u) \sigma_{\epsilon, f}(x)
\sim
\frac{\epsilon \bar\epsilon}{2} \frac{1}{(u-x)^2} \sigma_{\epsilon,  f}(x)
+
\frac{1}{(u-x)^2} 
\frac{1}{\ke\kbe}
\left(\partial_u X \partial_u \bar X \sigma_{\epsilon,  f}\right)(x)
+O(1)
\end{equation}
from which we read both the conformal dimension of $ \sigma_{\epsilon,
  f}$ and that $\ke\kbe \partial_x \sigma_{\epsilon,  f}=
\left(\partial_u X \partial_u \bar X \sigma_{\epsilon,  f}\right)(x)$.
It is noteworthy that the double pole comes from the extra piece 
$\partial\partial\Delta_c$ in eq. (\ref{T-loc}) which would not be
present in a naive state to operator correspondence.

As a last example to make clear the algorithm we consider the more
complex OPE
$(\partial^2_u X\, \partial_u X\, \partial_u \bar X)(u) 
\left(\partial^2_u \bar X \sigma_{\epsilon, f}\right)(x)$
at the leading order. First we compute the operatorial realization of
$(\partial^2_u X\, \partial_u X\, \partial_u \bar X)(u)$ to be
\begin{align}
&:(\partial^2_\ul X\loc \partial_\ul X\loc \partial_\ul \bar X\loc)(u):
\nonumber\\
&+
\partial^2_\ul\partial_\vl \Delta^{z\bz}_c|_{\vl=\ul}  \partial_\ul X\loc
+
\partial_\ul\partial_\vl \Delta^{z\bz}_c|_{\vl=\ul}  \partial^2_\ul X\loc
\nonumber\\
&=
:(\partial^2_\ul X\loc \partial_\ul X\loc \partial_\ul \bar X\loc)(u):
-\frac{\ke\kbe \epsilon(1-\epsilon)(2-\epsilon)}{2 \ul^3} \partial_\ul X\loc
-\frac{\ke\kbe \epsilon(1-\epsilon)}{2 \ul^2} \partial^2_\ul X\loc
\end{align}
then we associate the state 
$\kbe \bar\alpha^\dagger_{\bar\epsilon+2} |T\rangle$ to the excited
twist $\left(\partial^2_u \bar X \sigma_{\epsilon, f}\right)(x)$ then
an easy computation gives
\begin{equation}
(\partial^2_u X\, \partial_u X\, \partial_u \bar X)(u) 
\left(\partial^2_u \bar X \sigma_{\epsilon, f}\right)(x)
\sim
(\ke\kbe)^2 \frac{ \epsilon(1-\epsilon)^2}{2 (u-x)^{4-\epsilon}} \sigma_{\epsilon, f}(x)
\end{equation}
where the leading order contribution comes from the terms linear in
$X\loc$.

\subsection{Boundary OPEs from local Fock space}
In the computation of the interaction of twisted states with untwisted
ones quite often we are not interested in chiral operators but in
boundary operators such as $e^{i \bk X(x,x) + i k \bar X(x,x)}$.
For this case we must extend the analysis given in the previous section.
The correct mapping is then given by
\begin{align}
\left[
e^{i \bk X + i k \bar X}
\prod_{n=1}^\infty 
\left(\partial^n_x X \right)^{N_n}
\left(\partial^n_x \bar X \right)^{\bar N_n}
\right](x+i0^+, x-i 0^+)
\leftrightarrow
\left.
\prod_{n=1}^\infty 
\frac{\partial^{N_n}}{\partial \bar c_n^{N_n}}
\frac{\partial^{\bar N_n}}{\partial c_n^{\bar N_n}}
\cS(c,\bc) \right|_{c_0=i\,k,c_{n\ge1}=0}
\end{align}
where the generating function $\cS(c,\bc)$ is given by
\begin{align}
\cS(c,\bc)
=
&:
\exp\left\{
\sum_{n=0}^\infty  
\left[ \bar c_n \partial^n_\xl X_{loc}(\xl+i0^+,\xl-i0^+)
+ c_n \partial^n_\xl \bar X_{loc}(\xl+i0^+,\xl-i0^+) \right]
\right\}
:
\nonumber\\
&
\exp\left\{
\sum_{n,m=0}^\infty  
\bar c_n~c_m ~\partial^n_\xii1 ~\partial^m_\xii2
\Delta\bou^{z \bz}(\xii1; \xii2; \epsilon)
|_{\xii1=\xii2=\xl} 
\right\}
\nonumber\\
&
\exp\left\{
\oh
\sum_{n,m=0}^\infty  
c_n~c_m ~\partial^n_\xii1 ~\partial^m_\xii2
\Delta\bou^{\bz \bz}(\xii1; \xii2; \epsilon) 
|_{\xii1=\xii2=\xl} 
\right\}
\nonumber\\
&
\exp\left\{
\oh
\sum_{n,m=0}^\infty  
\bar c_n~\bar c_m ~\partial^n_\xii1 ~\partial^m_\xii2 
\Delta\bou^{z z}(\xii1; \xii2; \epsilon)
|_{\xii1=\xii2=\xl} 
\right\}
\label{S_bou}
\end{align}
where $:\dots:$ is the normal ordering and we defined 
\begin{equation}
\Delta\bou^{I J}(\xii1; \xii2; \epsilon)
=
\left\{\begin{array}{r l}
\scriptstyle
G^{I J}(\xii1+i0^+,\xii1-i0^+; \xii2+i0^+,\xii2-i0^+; 
\{ 0,\epsilon; \infty, \bar\epsilon\}) 
& 
\scriptstyle
- G^{I J}_{U(t)}(\xii1+i0^+,\xii1-i0^+; \xii2+i0^+,\xii2-i0^+)
\\
& \xii1,\xii2>0
\\
\scriptstyle
G^{I J}(\xii1+i0^+,\xii1-i0^+; \xii2+i0^+,\xii2-i0^+; \{ 0,\epsilon;
\infty, \bar\epsilon\})  
&
\scriptstyle
- G^{I J}_{U(t+1)}(\xii1+i0^+,\xii1-i0^+; \xii2+i0^+,\xii2-i0^+)
\\
& \xii1,\xii2<0
\end{array}\right. 
\end{equation}
with $G^{I J}(\ul,\bul,\vl,\bvl; \{ 0,\epsilon; \infty, \bar\epsilon\})$ 
the Green function\footnote{We normalize the Green function such that
$\partial \bar\partial G(\ul,\bul,\vl,\bvl) = \madd \delta^2(\ul-\vl)$.
} with 
the twist $\sigma_{\epsilon,f}$  at $\xl=0$ and the anti-twist $
\sigma_{\bar \epsilon,f}$ in $\xl=\infty$  and
$ G^{I J}_{U(t)}(\ul,\bul,\vl,\bvl)$ is the Green function of the
untwisted string with both ends on $D_t$
and the need to distinguish $x>0$ from $x<0$ is due to the different
boundary conditions of the twisted string in this ranges
\COMMENTOOK{
Result changes if $x<0$ or $x>0$ since there are two different branes
for which the allowed momentum is different since it must be tangential
}
As shown in eq. (\ref{expl-Delta-expr}) in appendix \ref{app:Delta}
all $\Delta^{I J}\bou$ are equal to a real symmetric function
$\Delta\bou(x_1,x_2)$  up to phases which combine to allow the previous
generating function to be written as
\begin{align}
\cS(c,\bc)
=
&:
\exp\left\{
\sum_{n=0}^\infty  
\left[ \bar c_n \partial^n_\xl X_{loc}(\xl+i0^+,\xl-i0^+)
+ c_n \partial^n_\xl \bar X_{loc}(\xl+i0^+,\xl-i0^+) \right]
\right\}
:
\nonumber\\
&
\exp\left\{
\sum_{n,m=0}^\infty  
c_{n \parallel D_t} ~c_{m \parallel D_t} ~\partial^n_\xii1 ~\partial^m_\xii2
\Delta\bou(\xii1; \xii2; \epsilon)
|_{\xii1=\xii2=\xl} 
\right\}
\end{align}
when $x\loc>0$ with 
\begin{equation}
c_{n \parallel D_t}= 
\frac{ e^{-i \pi \alpha_t} c_n + e^{i \pi \alpha_t} \bar c_n }{\sqrt{2}}.
\label{c-parallel}
\end{equation} 
When $x\loc<0$ we get a very similar expression but with the substitution
$c_{\parallel D_t}\rightarrow c_{\parallel D_{t+1}}$ since in this
case the vertex is on the brane $D_{t+1}$.

Given the previous results we can now compute the operatorial
realization of the the vertex $e^{i k\cdot X(x, x)}$ to be,
similarly to the results (\cite{Pesando:2011yd},\cite{DiVecchia:2011mf})
\begin{equation}
\left\{
\begin{array}{c c}
|\xl|^{ - \alpha' k^2_{\parallel D_t}} 
e^{-\frac{1}{2} R^2(\epsilon_t)  \alpha' k^2_{\parallel D_t} }
~:e^{i ( \bar k X\loc(\xl, \xl)+ k \bar X\loc(\xl, \xl))}:
&
\xl>0 ~[\xii t< x <\xii {t-1}]
\\
|\xl|^{ - \alpha' k^2_{\parallel D_{t+1} }} 
e^{-\frac{1}{2} R^2(\epsilon_t)  \alpha' k^2_{\parallel D_{t+1}} }
~:e^{i ( \bar k X\loc(\xl, \xl)+ k \bar X\loc(\xl, \xl))}:
&
\xl<0 ~[\xii {t+1}< x <\xii {t}]
\end{array}
\right.
\end{equation}
where
$R^2(\epsilon_t)=2 \psi(1)- \psi(\epsilon_t) -\psi(\bar\epsilon_t)>0$,
$\psi(z)= \frac{d \ln \Gamma(z)}{d z}$ being the digamma function.
Notice that we have not required the momentum $k$ to be tangent to the
brane since the normal part gives simply a phase due to the boundary
conditions nevertheless, as commented before, 
the explicit expression for $\Delta^{I J}$
implies that only the momentum parallel to the $D_t$ brane
$ k_{\parallel D_t}= \frac{ e^{-i \pi \alpha_t} k + e^{i \pi \alpha_t}
  \bar k}{\sqrt{2}}$
enters the form factor  
$e^{-\frac{1}{2} R^2(\epsilon_t)  \alpha'  k^2_{\parallel D_t} }$.

Using the previous operatorial realization we can then obtain the OPEs
\begin{align}
e^{i ( \bar k X(x, x)+ k \bar X(x, x))}  \sigma_{\epsilon_t,f_t}(\xii t)  
\sim& 
(x-\xii t)^{ - \alpha' k^2_{\parallel D_t}} 
e^{-\oh R^2(\epsilon_t)  \alpha' k^2_{\parallel D_t} } 
e^{i ( k  \bar f_t+ \bar  k   f_t ) }  
\nonumber\\
&\Big[
\sigma_{\epsilon_t,f_t}(\xii t)
-
(x-\xii t)^{\epsilon_t} 
\frac{ \sqrt{2} e^{-i \pi \alpha_t} k_{\parallel D_t}  }{ \epsilon_t }
( \partial_u X \sigma_{\epsilon_t,f_t} )(\xii t)
\nonumber\\
&~
-
(x-\xii t)^{\bar\epsilon_t} 
\frac{ \sqrt{2} e^{-i \pi \alpha_t} k_{\parallel D_t}  }{
  \bar\epsilon_t }
( \partial_u \bar X \sigma_{\epsilon_t,f_t} )(\xii t)
+ \dots
\Big]
~~~~
\xii t< x <\xii {t-1}
\label{eikX-sigma+}
\end{align}
and similarly for the $x\loc<0$ case which corresponds to 
$\xii {t+1}< x <\xii {t} $ with the substitution  
$k_{\parallel D_t}\rightarrow k_{\parallel D_{t+1}}$
and $(x-\xii t)^{\epsilon_t} \rightarrow (\xii t - x)^{\epsilon_t}$ 
\COMMENTO{
\begin{align}
e^{i ( \bar k X(x, x)+ k \bar X(x, x))}  \sigma_{\epsilon_t,f_t}(\xii t)  
\sim& 
|x-\xii t|^{ - \alpha' k^2_{\parallel D_{t+1}}} 
e^{-\oh R^2(\epsilon_t)  \alpha' k^2_{\parallel D_{t+1} } } 
e^{i ( k  \bar f_t+ \bar  k   f_t ) }  
\nonumber\\
&\Big[
\sigma_{\epsilon_t,f_t}(\xii t)
-
(x-\xii t)^{\epsilon_t} 
\frac{ \sqrt{2} e^{-i \pi \alpha_t} k_{\parallel D_{t+1} }  }{ \epsilon_t }
( \partial_u X \sigma_{\epsilon_t,f_t} )(\xii t)
\nonumber\\
&~
-
(x-\xii t)^{\bar\epsilon_t} 
\frac{ \sqrt{2} e^{-i \pi \alpha_t} k_{\parallel D_{t+1} }  }{
  \bar\epsilon_t }
( \partial_u \bar X \sigma_{\epsilon_t,f_t} )(\xii t)
+ \dots
\Big]
~~~~
\xii {t+1}< x <\xii {t}
\label{eikX-sigma-}
\end{align}
}
The previous OPE justify writing not only $\sigma_{\epsilon_t}(\xii t)$
but $\sigma_{\epsilon_t,f_t}(\xii t)$ since the $f_t$ and $\bar f_t$
can be computed using the phases of the leading order terms 
with  different momenta. 
Moreover the phases $e^{-i \pi \alpha_t}$ and
$e^{-i \pi \alpha_{t+1}}$ can be extracted from the projections
$k_{\parallel D_{t}}, k_{\parallel D_{t+1}} $ of
momentum $k$.
\COMMENTOOK{Check! Devo usare le due direzioni $\xl <>0$ per trovare
le due componenti di $f$}

In a similar way we can compute the operator associated to
 $ \partial_x X(x,x)~ e^{i k\cdot X(x, x)}$ to be
\begin{equation}
|\xl|^{ - \alpha' k^2_{\parallel D_t}} 
e^{-\frac{1}{2} R^2(\epsilon_t)  \alpha' k^2_{\parallel D_t} }
:\left( \partial_\xl X\loc(\xl,\xl)
-\frac{i e^{i \pi \alpha_t} \, \alpha'  k_{\parallel D_t} }{\sqrt{2}
\xl}\right)
e^{i ( \bar k X\loc(\xl, \xl)+ k \bar X\loc(\xl, \xl))}:
\label{boson-vertex-twisted-carrying-string}
\end{equation}
when $ \xl>0 ~[\xii t< x <\xii {t-1}] $ and similarly 
for $\xl<0~[\xii {t+1}< x <\xii {t}]$.
It is worth stressing that the term proportional to 
$\frac{i  \alpha'  k_{\parallel D_t} }{\xl}$ is fundamental in getting
the right interaction among a gluon and twisted matter 
(\cite{Billo:2002hm},\cite{Pesando:2011yd}, \cite{DiVecchia:2011mf})
when the amplitude is computed in operatorial formalism.

\subsection{Short derivation of the generating function $\cS$}
\label{sect:deriv_gen_fun}
We will now quickly review the motivations to write down the
generating vertex (\ref{S_bou}) and why it works.

Our aim is to get some hints on how to regularize the contact
divergences that appear in the path integral computation of amplitudes
in presence of twist fields.
We consider the boundary case since all the others can be treated in
an analogous manner.
In the untwisted case the operatorial generating function is simply the
Sciuto-Della Selva- Saito vertex \cite{SDS} \footnote{
Because of the boundary conditions not all terms at the exponent are
independent but this does not change the arguments in this section.
}
\begin{align}
\cS_{U}(c,\bc)
=
&:
e^{
\sum_{n=0}^\infty  
\left[ \bar c_n \partial^n_\xl X_{U, loc}(\xl+i0^+,\xl-i0^+)
+ c_n \partial^n_\xl \bar X_{U,loc}(\xl+i0^+,\xl-i0^+) \right]
}
:
\end{align}
Dropping the $\loc$ indication we can rewrite it as
\begin{equation}
\cS_U[J]
=
: e^{\int d x_a J_{I}(x_a) X^I_U(x_a,x_a)} :
\label{S-untwisted-bou-gen}
\end{equation}
where we have introduced the current $J_{I}(x_a)$ which must be set to
$
J_{I}(x_a)= \sum_{n=0}^\infty c_{I n} \partial^n_{x} \delta(x_a-x)
$
when we want to reproduce the original vertex but can also be taken
more general as we do in the following.

We can now compute the OPE of two such generating functions (vertices)
\begin{align}
\cS_U[J_1] \cS_U[J_2]
&=
e^{\int d\xii a \int d\xii b J_{1 I}(\xii a)  J_{2 J}(\xii b)
  G^{I J}_{U,bou}(\xii a,\xii b)}
: \cS_U[J_1] \cS_U[J_2] :
\nonumber\\
&=
e^{\int d\xii a \int d\xii b J_{1 I}(\xii a)  J_{2 J}(\xii b)
  G^{I J}_{U,bou}(\xii a,\xii b)}
\cS_U[J_1+J_2]
\end{align} 
which is valid for generic currents as long as they have compact
support and the points in the support of $J_1$ have bigger
absolute values than those  in the support of $J_2$ so that the
operatorial product is radial ordered.
Now the non operatorial term 
$e^{\int d\xii a \int d\xii b J_{1 I}(\xii a)  J_{2 J}(\xii b)
  G^{I J}_{U,bou}(\xii a,\xii b)}$ can be roughly understood as the generating
function for the OPE coefficients and we want to check that it is
reproduced when using the generating function (\ref{S_bou}) 
which is defined in the twisted Fock space.
In particular we consider the generating function 
with a more general current given by
\begin{equation}
\cS_T[J]= 
e^{\oh \int d\xii a \int d\xii b J_I(\xii a)  J_J(\xii b) \Delta^{I J}\bou(\xii a,\xii b)}
:   e^{\int d x_a J_I(x_a) X^I(x_a,x_a)}:
\label{S-bou-gen-J}
\end{equation}
It is then immediate to compute the product
\begin{align}
\cS_T[J_1] \cS_T[J_2]
&=
e^{\int d\xii a \int d\xii b J_{1 I}(\xii a)  J_{2 J}(\xii b)
  G^{I J}_{N=2,bou}(\xii a,\xii b;\{0,\epsilon; \infty; \bar\epsilon \})}
: \cS_T[J_1] \cS_T[J_2] :
\nonumber\\
&=
e^{\int d\xii a \int d\xii b J_{1 I}(\xii a)  J_{2 J}(\xii b)
  G^{I J}_{U,bou}(\xii a,\xii b)}
\cS_T[J_1+J_2]
\end{align}
where we have added and subtracted 
$ \int d\xii a \int d\xii b J_{1 I}(\xii a)  J_{2 J}(\xii b)
  G^{I J}_{U,bou}(\xii a,\xii b)$ 
to the exponent in order to complete the prefactor in
eq. (\ref{S-bou-gen-J}) in the case where $J=J_1+J_2$
and used the symmetry
$\Delta^{I J}{\bou}(\xii a,\xii b)=\Delta^{J I}{\bou}(\xii b,\xii a)$.
Having verified that the generating function (\ref{S_bou}) gives
vertices with the same OPEs as the untwisted ones we can exam the
reason which leads to it (\cite{Hamidi:1986vh}, \cite{Berkooz:2004yy}, \cite{Pesando:2011yd}).
In operatorial formalism the normal ordered vertex
(\ref{S-untwisted-bou-gen}) for the untwisted case 
can be obtained from a regularized non
normal ordered generating function by a multiplicative renormalization as
\begin{equation}
\cS_U[J]
=
\lim_{\eta\rightarrow0^+} \cN(\eta) ~\cS_{U,reg}[J, \eta]
\end{equation}
where the regularized generating function is defined by a point splitting as
\begin{equation}
\cS_{U,reg}[J, \eta]
=
e^{\int d x_a J_{I}(x_a) [ X^{I(+)}_U(x_a,x_a) 
+ X^{I(-)}_U(x_a e^{-\eta} ,x_a e^{-\eta}) ] }
\end{equation}
without normal ordering and the multiplicative renormalization  is
given by the inverse of the factor we get by normal ordering the
regularized generating function
\begin{equation}
 \cN(\eta)
=
e^{-\oh \int d\xii a \int d\xii b J_{I}(\xii a)  J_{J}(\xii b)
  G^{I J}_{U ,bou}(\xii a,\xii b e^{-\eta} )}
\label{calN}
\end{equation}
Now the generating function for the twisted string is defined in an
analogous way 
by regularizing the non normal ordered generating function 
and then renormalizing
in a minimal way using the same renormalization factor as in the
untwisted case (\ref{calN}), explicitly
\begin{equation}
\cS_T[J]
=
\lim_{\eta\rightarrow0^+} \cN(\eta) ~\cS_{T, reg}[J, \eta]
\end{equation}
with
\begin{equation}
\cS_{T, reg}[J, \eta]
=
e^{\int d x_a J_{I}(x_a) [ X^{I(+)}(x_a,x_a) 
+ X^{I(-)}(x_a e^{-\eta} ,x_a e^{-\eta}) ] }
\end{equation}


\subsection{Getting excited twists}
We are interested in excited twist states hence we would now  
write a kind of SDS vertex which generates these states
(\ref{twisted-sts}).
The main observation is then that
\begin{equation}
\partial_\ul^{n-1} \left[ \ul^{\bar \epsilon} \partial_\ul X\loc(\ul,\bul)\right]
= (n-1)! ~\ke \alpha_{n-1+\epsilon}^\dagger +O(\ul)
\label{simple-loc-op-for-excited}
\end{equation}
therefore a normal ordered products of these operators gives directly an
excited twist state, e.g.
\begin{align}
\lim_{\ul\rightarrow 0} 
&
:
\partial_\ul^{n-1} \left[ \ul^{\bar \epsilon} \partial_\ul X\loc(\ul,\bul)\right]
\partial_\ul^{m-1} \left[ \ul^{\epsilon} \partial_\ul \bX\loc(\ul,\bul)\right]
:
|T\rangle
\nonumber\\
&
=
\ke \kbe (n-1)! (m-1)! \alpha_{n-1+\epsilon}^\dagger \bar \alpha_{m-1+\epsilon}^\dagger |T\rangle
=
\left( 
\partial^n X \partial^m \bX
\sigma_{\epsilon,f}
\right)(0) |0\rangle_{SL(2)}
\label{example-dX-dbX}
\end{align}
The generating function for products of these operators  is obviously
\begin{align}
&\cT(d,\bar d)
=
\nonumber\\
& \lim_{\ul\rightarrow 0}
:
\exp\left\{
\sum_{n=1}^\infty  
\left[ \bar d_n 
\partial_\ul^{n-1} \left[ \ul^{\bar \epsilon} \partial_\ul X\loc(\ul,\bul)\right]
+ d_n 
\partial_\ul^{n-1} \left[ \ul^{\epsilon} \partial_\ul \bX\loc(\ul,\bul)\right]
\right]
\right\}
:
\label{T_chiral}
\end{align}
since
\begin{align}
\left[
\prod_{n=1}^\infty 
\left(\partial^n_u X \right)^{N_n}
\left(\partial^n_u \bar X \right)^{\bar N_n}
\sigma_{\epsilon,f}
\right](0) |0\rangle_{SL(2)}
\leftrightarrow
\lim_{\ul\rightarrow 0}
\left.
\prod_{n=1}^\infty 
\frac{\partial^{N_n}}{\partial \bar d_n^{N_n}}
\frac{\partial^{\bar N_n}}{\partial d_n^{\bar N_n}}
\cT(d,\bar d) \right|_{\bar d=d=0}
|T\rangle
\end{align}

Comparing with eq. (\ref{S_chiral}) we realize that there is not the exponent
quadratic in $d$, this means that the abstract operator corresponding
to e.g. eq. (\ref{example-dX-dbX}) is not simply 
$\lim_{u\rightarrow x}
\partial_u^{n-1} \left[ (u-x)^{\bar \epsilon} \partial_u  X \right]
\partial_u^{m-1} \left[ (u-x)^{\epsilon} \partial_u \bX \right]
$
but it is
\begin{align}
\lim_{u\rightarrow x}
\Big(
&
\partial_u^{n-1} \left[ (u-x)^{\bar \epsilon} \partial_u  X \right]
\partial_u^{m-1} \left[ (u-x)^{\epsilon} \partial_u \bX \right]
\nonumber\\
&-
\partial_u^{n-1}  \partial_v^{m-1} 
\left[ (u-x)^{\bar \epsilon}  (v-x)^{\epsilon}
\partial_u \partial_v \Delta^{z \bz}_c(u-x, \bu -x; v-x, \bv -x;
\epsilon) |_{v=u}
\right]
\Big) 
\end{align}
in fact computing its OPE with the twist field $\sigma_{\epsilon,f}
(x)$ as in section \ref{sect:chiral_OPE} we get
\begin{align}
\Big(
\partial_u^{n-1} \left[ (u-x)^{\bar \epsilon} \partial_u  X \right]
\partial_u^{m-1} \left[ (u-x)^{\epsilon} \partial_u \bX \right]
&-
\partial_u^{n-1}  \partial_v^{m-1} 
  \left[ (u-x)^{\bar \epsilon}  (v-x)^{\epsilon} 
        \partial_u \partial_v \Delta^{z \bz}_c
  \right]
\Big)
\sigma_{\epsilon,f} (x)
\nonumber\\
&\sim
\left( 
\partial^n X \partial^m \bX
\sigma_{\epsilon,f}
\right)(x)
+O(u-x)
\end{align}
Notice that this OPE explains why we can use $u-x$ as argument of
$\Delta^{z \bz }$ and not another function which behaves as 
$u-x + O (u-x)^2$.
We conclude therefore that the generating function for the abstract
operators which give excited twists is given by
\begin{align}
\cT_{abs}(d,\bar d)
=
&
\exp\left\{
\sum_{n=1}^\infty  
\left[ 
\bar d_n 
\partial_u^{n-1} \left[ (u-x)^{\bar \epsilon} \partial_u X(u,\bu)\right]
+ d_n 
\partial_u^{n-1} \left[ (u-x)^{\epsilon} \partial_u \bX(u,\bu)\right]
\right]
\right\}
\nonumber\\
\exp
&\left\{
-
\sum_{n,m=1}^\infty  
\bar d_n~d_m ~\partial^{n-1}_u ~\partial^{m-1}_v|_{v=u}
\left[ (u-x)^{\bar \epsilon}  (v-x)^{\epsilon}
\partial_u \partial_v \Delta^{z\bz}_c(u-x,\bu-x; v-x,\bv-x; \epsilon)
\right]
\right\}
\label{T-abs}
\end{align}
Explicitly this means that
\begin{align}
\left[
\prod_{n=1}^\infty 
\left(\partial^n_u X\right)^{N_n}
\left(\partial^n_u \bar X \right)^{\bar N_n}
\sigma_{\epsilon,f}
\right](x)
=
\lim_{u\rightarrow x}
\left.
\prod_{n=1}^\infty 
\frac{\partial^{N_n}}{\partial \bar d_n^{N_n}}
\frac{\partial^{\bar N_n}}{\partial d_n^{\bar N_n}}
\cT_{abs}(d,\bar d) \right|_{\bar d=d=0}
\sigma_{\epsilon,f}(x)
\end{align}

\section{The path integral approach}

The classic method \cite{Dixon:1986qv} to compute twists correlators
is by the path integral
\begin{align}
\langle 
\sigma_{\epsilon_1,f_1}(x_1) \dots \sigma_{\epsilon_N,f_N}(x_N)
\rangle
=\int_{\cM(\{x_t,\epsilon_t, f_t\})} \cD X e^{-S_E}
\label{basic-path-integral}
\end{align}
where $\cM(\{x_t, \epsilon_t, f_t\})$ is the space of string configurations
satisfying the boundary conditions (\ref{global-boundary-upper}) and 
(\ref{global-boundary-points}).
Since the integral is quadratic we can then efficiently 
separate the classical fields from the quantum  fluctuations as
\begin{equation}
X(u,\bar u)= X_{cl}(u,\bar u;\{x_t,\epsilon_t,f_t\})+X_{q}(u,\bar u;\{x_t,\epsilon_t\})
\label{cl-q-splitting}
\end{equation}
where $X_{cl}$ satisfies the previous boundary conditions
(\ref{path-integral-global-boundary-upper}, \ref{global-boundary-points})
while $X_q$
satisfies the same  boundary conditions but with all $f_t=0$.
After this splitting we obtain
\begin{align}
\langle 
\sigma_{\epsilon_1,f_1}(x_1) \dots \sigma_{\epsilon_N,f_N}(x_N)
\rangle
=
\cN(x_t,\epsilon_t)
 e^{-S_{E,cl}(x_t,\epsilon_t, f_t)}
\label{general_corr_from_quantum_and_classical}
\end{align}
where the factor $\cN(x_t,\epsilon_t)$  is the quantum contribution.
In particular when all $f_t$ are equal, i.e. $f_t=f$ the classical action 
$S_{E,cl}(x_t,\epsilon_t, f_t=f)$ is zero since the branes are the
boundary of a zero area polygon
therefore we can identify
\begin{equation}
\cN(x_t,\epsilon_t)
=
\langle 
\sigma_{\epsilon_1,f_1=f}(x_1) \dots \sigma_{\epsilon_N,f_N=f}(x_N)
\rangle
\end{equation}
which is also true for the quantum fluctuation for which $f=0$.
Our aim is now to compute correlators with both (excited) twist field
operators  and  untwisted operators $V_{\xi_i}(x_i)$ ($i=1\dots L$)
associated with the untwisted state $\xi_i$ of which the ones in
eq. (\ref{OPEs-epsilon}) are a particular case.
We start with correlators with plain twist field operators which can
be computed as
\begin{align}
\langle 
\sigma_{\epsilon_1,f_1}(x_1) \dots \sigma_{\epsilon_N,f_N}(x_N)
\prod_{i=1}^L V_{\psi_i}(x_i)
\rangle
=\int_{\cM(\{x_t,\epsilon_t, f_t\})} \cD X~\prod_{i=1}^L V_{\psi_i}(x_i)~ e^{-S_E}
\end{align}
To do so we notice that it is by far easier not to compute the previous
correlator but to compute the generating function of all the
correlators, i.e. the Reggeon vertex, in the form of the previous path integral
(\ref{basic-path-integral}) plus linear sources
\begin{align}
V_{N+L}(J_i)
=
\int_{\cM(\{x_t,\epsilon_t, f_t\})} \cD X~ 
e^{-S_E+\sum_{i=1}^L \int d x~ J_{i I}(x) X^I(x,x)}
\label{not-reg-path-integral}
\end{align}
where $J_{i I}(x)= \sum_{n=0}^\infty c_{(i) n I} \partial^n_{x_i}\delta(x-x_i)$
since all untwisted operators can be obtained by taking derivatives
with respect to the coefficients $ c_{(i) n I} $ as explained in the
previous section.
This starting point is very similar to
(\cite{DiVecchia:1986mb},\cite{Petersen:1988cf}) where it was
recognized that the generator for all closed (super)string amplitudes 
is a quadratic path integrals.
The idea in the previous papers is that the appropriate boundary
condition for R and/or NS sector can be obtained simply by inserting
{\sl linear} sources with the desired boundary conditions.
Because of this assumption the quantum fluctuations are the same for
all the amplitudes: from the purely NS to the mixed ones.
It was later realized that this prescription misses a proper treatment
of quantum fluctuations \cite{DiVecchia:1989hf} 
and that when this part is considered the
amplitudes factorize correctly \cite{DiBartolomeo:1990fw}.

\subsection{Boundary correlators with non excited twists on $\R^2$}
\label{sect:boundary_corr}

Our strategy is therefore to compute the path integral
(\ref{not-reg-path-integral}) by properly defining it in order to
regularize the divergences
which arise as usual because of current self interactions.
Taking inspiration from what is done for deriving  the generating
function we first regularize the $\delta$ functions in the currents,
a step which corresponds to the point splitting in operatorial formalism
and then subtract the self interaction of the untwisted string.
We are therefore led to consider the path integral
\begin{align}
V_{N+L}(J_i)
=
\lim_{\{\eta_i\} \rightarrow 0}
&
\int_{\cM(\{x_t,\epsilon_t, f_t\})} 
\cD X~ 
e^{-S_E}
\nonumber\\
\times
&
\prod_{i=1}^L \left[
e^{-\oh \int d\xii a \int d\xii b J_{i I}(\xii a, \eta_i )  
   J_{i J}(\xii b,\eta_i)
  G^{I J}_{U(t_i) ,bou}(\xii a,\xii b )}
e^{\int d x~ J_{i I}(x, \eta_i) X^I(x,x) }
\right]
\label{reg-path-integral}
\end{align}
where the regularized currents are defined as
\begin{equation}
J_{i I}(x, \eta_i)= \sum_{n=0}^\infty c_{(i) n I}
\partial^n_{x_i}\delta(x-x_i; \eta_i)
\end{equation}
with $\delta(x-x_i; \eta_i)$ a regularization of the $\delta$ such
that $\lim_{\eta_i \rightarrow 0} \delta(x-x_i; \eta_i)=
\delta(x-x_i)$ and
$ G^{I J}_{U(t_i) ,bou}(\xii a,\xii b )$ is the untwisted Green function
with boundary conditions which depends on the brane on which the point
$x_i$ is. This dependence is the reason why we have written $U(t_i)$.

It is then immediate to compute the previous path integral
by using the splitting of $X$ into quantum and classical part 
(\ref{cl-q-splitting}) and get
\begin{align}
V_{N+L}(J_i)
&=
\langle 
\sigma_{\epsilon_1,f_1}(x_1) \dots \sigma_{\epsilon_N,f_N}(x_N)
\rangle
\nonumber\\
\times
&
\prod_{i=1}^L
e^{
\sum_{n=0}^\infty c_{(i) n I}
\partial^n_{x_i} X^I_{cl}(x_i,x_i)
}
\nonumber\\
\times
&
\prod_{i=1}^L
e^{
\oh
\sum_{n=0}^\infty c_{(i) n I}
\sum_{m=0}^\infty c_{(i) m J} 
\partial^n_{x_i} \partial^m_{ \hat x_i}
\Delta^{I J}_{(N,M), bou (i)}(x_i, \hat x_i ; \{ x_t, \epsilon_t\}) |_{ \hat x_i= x_i}
}
\nonumber\\
\times
&
\prod_{1\le i < j\le L}
e^{
\sum_{n=0}^\infty c_{(i) n I}
\sum_{m=0}^\infty c_{(j) m J} 
\partial^n_{x_i} \partial^m_{x_j}
G^{I J}_{(N,M), bou}(x_i, x_j ; \{ x_t, \epsilon_t\})
}
\label{V_N+M}
\end{align}
where we have introduced the boundary Green function
\begin{equation}
G^{I J}_{(N,M), bou}(x_i, x_j ; \{ x_t, \epsilon_t\})
=
G^{I J}_{(N,M)}(x_i +i 0^+, x_i -i 0^+; x_j +i 0^+, x_j -i 0^+; 
\{ x_t, \epsilon_t\})
\end{equation}
in presence of $N$ twist
fields $\sigma_{\epsilon_t, f_t=0}(x_t)$ (with $f_t=0$ since these
terms come from the quantum fluctuations) in the sector
$M=\sum_{t=1}^N \epsilon_t$ 
and the regularized Green function
\begin{equation}
\Delta^{I J}_{(N,M), bou (i)}(x_i, \hat x_i ; \{ x_t, \epsilon_t\}) 
=
G^{I J}_{(N,M), bou}(x_i, \hat x_i ; \{ x_t, \epsilon_t\})
-
G^{I J}_{U(t_i),bou}(x_i, \hat x_i )
\end{equation}
As it happened for the $N=2$ boundary Green function and its regularized
version all the components of the boundary Green function are
proportional to real symmetric functions $G_{(N,M), bou}(x_i, x_j)$
and $\Delta_{(N,M), bou (i)}(x_i, x_j)$ respectively up to phases. 
Using the result from the appendix (\ref{app:boundray_green}) where
the explicit expressions for $G_{(N,M), bou}(x_i, x_j)$
and $\Delta_{(N,M), bou}(x_i, x_j)$ are given  we can
then write
\begin{align}
V_{N+L}(J_i)
&=
\langle 
\sigma_{\epsilon_1,f_1}(x_1) \dots \sigma_{\epsilon_N,f_N}(x_N)
\rangle
\nonumber\\
\times
&
\prod_{i=1}^L
e^{
\sum_{n=0}^\infty c_{(i) n I}
\partial^n_{x_i} X^I_{cl}(x_i,x_i)
}
\nonumber\\
\times
&
\prod_{i=1}^L
e^{
\sum_{n=0}^\infty c_{(i) n \parallel D_{t_i}}
\sum_{m=0}^\infty c_{(i) m \parallel D_{t_i}} 
\partial^n_{x_i} \partial^m_{ \hat x_i}
\Delta_{(N,M), bou (i)}(x_i, \hat x_i ; \{ x_t, \epsilon_t\}) |_{ \hat x_i= x_i}
}
\nonumber\\
\times
&
\prod_{1\le i < j\le L}
e^{
2 \sum_{n=0}^\infty c_{(i) n \parallel D_{t_i}}
\sum_{m=0}^\infty c_{(j) m \parallel D_{t_i}} 
\partial^n_{x_i} \partial^m_{x_j}
G_{(N,M), bou}(x_i, x_j ; \{ x_t, \epsilon_t\})
}
\label{simplified_V_N+M}
\end{align}
where $c_{n \parallel D_{t_i}}$ are defined as in
eq. (\ref{c-parallel}) and $t_i$ is chosen by the brane on which
$x_i$ lies, i.e. $\xii {t_i+1}< x_i <\xii {t_i}$.

\subsubsection{Some consequences}
There are now two immediate consequences. The first and more trivial
is that all correlators with just one derivative vertex have
contribution only from the classical solution $X_{cl}$
\cite{Anastasopoulos:2013sta}, i.e
\begin{align}
\langle
\sigma_{\epsilon_1,f_1}(x_1) \dots \sigma_{\epsilon_N,f_N}(x_N)
\partial_x^n X^I(x,x)
\rangle
=
\partial_x^n X^I_{cl}(x,x)
\langle
\sigma_{\epsilon_1,f_1}(x_1) \dots \sigma_{\epsilon_N,f_N}(x_N)
\rangle
\end{align}
The second and more interesting is that all correlators can be
essentially computed with Wick theorem plus classical contributions
plus self interactions which are absent in Wick theorem. This implies
that a  string in presence of not excited twist fields (defects) 
is free and can be
quantized almost in the usual manner.

\subsubsection{Some examples}
\label{sect:some_bou_examples}
As a first example we want to compute the correlator of a tachyon with
$N$ twist fields
\begin{align}
&\langle 
e^{i k_I X^I(x,x)}
\prod_{t=1}^N\sigma_{\epsilon_t,f_t}(x_t)
\rangle
=
V_{N+1}|_{c_1= i k, \bar c_1= i \bar k}
\nonumber\\
&=
\langle \prod_{t=1}^N\sigma_{\epsilon_t,f_t}(x_t) \rangle
~
e^{i k \bar X_{cl}(x,x) +i \bar k  X_{cl}(x,x)}
e^{- k_{\parallel D_t}^2 \Delta_{(N,M), bou (i)}(x,x)}
\end{align}
which shows that untwisted matter sees a kind of form factor of the
interacting twisted matter in accord with the result from OPE and what
discussed in  \cite{Pesando:2011yd} for the case of a stringy instanton.
Again a priori we can take $k^I$ not parallel to the brane $D_t$ but 
the explicit form of the Green function implies a
projection in the direction parallel to $D_t$.
It is interesting to compare the previous result with what we can get
using the OPE (\ref{eikX-sigma+}) in the previous correlator.
Using the OPE at the leading order we get
\begin{align}
&\langle 
e^{i k_I X^I(x,x)}
\prod_{t=1}^N\sigma_{\epsilon_t,f_t}(x_t)
\rangle
\sim
(x-\xii t)^{ - \alpha' k^2_{\parallel D_t}} 
e^{-\oh R^2(\epsilon_t)  \alpha' k^2_{\parallel D_t} } 
e^{i ( k  \bar f_t+ \bar  k   f_t ) } 
\langle \prod_{t=1}^N\sigma_{\epsilon_t,f_t}(x_t) \rangle
\end{align}
Now using the obvious behavior
\begin{equation}
X_{cl}(u,\bu)
\sim_{u\rightarrow x_t}
f_t
+O( (u-x_t)^{\bar \epsilon_t} )
+O( (\bu-x_t)^{\epsilon_t} )
\end{equation}
and the result shown in app. \ref{app:boundray_green}
\begin{equation}
\Delta_{(N,M), bou  (i)}(x,\hat x)
\sim_{x\rightarrow x_t}
\frac{\alpha'}{2} \ln |x-x_t|^2
+R^2(\epsilon_t)
+ O( x-\hat x)
+ O(x-x_t)
\end{equation}
in the exact expression we find the expected consistency with the OPE result.

Secondly let us consider the correlator of a gauge boson with twisted matter
\begin{align}
\langle 
\epsilon_I \partial_x X^I e^{i k_I X^I(x,x)}
&
\prod_{t=1}^N\sigma_{\epsilon_t,f_t}(x_t)
\rangle
=
\langle \prod_{t=1}^N\sigma_{\epsilon_t,f_t}(x_t) \rangle
~
e^{i k \bar X_{cl}(x,x) +i \bar k  X_{cl}(x,x)}
e^{- k_{\parallel D_t}^2 \Delta_{(N,M), bou (i)}(x,x)}
\nonumber\\
&
 \sqrt{2} e^{-i \pi \alpha_t} \epsilon_{\parallel D_t}
\left( 
  \partial_x X_{cl}(x,x)
+  \sqrt{2} e^{i \pi \alpha_t} i  k_{\parallel D_t} 
\partial_{x} \Delta_{(N,M), bou (i)}(x,\hat x)|_{\hat x =x} 
\right)
\end{align}
which exhibits the same structure as the vertex
(\ref{boson-vertex-twisted-carrying-string}) and the OPE it can be
computed using it.

Next we consider the interaction of two tachyons with the twisted
matter in order to show the Wick-like expression in a simple case
\begin{align}
\langle 
&
e^{i k_{1 I} X^I(x_1,x_1)}
e^{i k_{2 I} X^I(x_2,x_1)}
\prod_{t=1}^N\sigma_{\epsilon_t,f_t}(x_t)
\rangle
=
V_{N+2}|_{c_j= i k_j, \bar c_j= i \bar k_j}
\nonumber\\
=
& 
\langle \prod_{t=1}^N\sigma_{\epsilon_t,f_t}(x_t) \rangle
\prod_{j=1}^2
\left[
e^{i k_j \bar X_{cl}(x_j,x_j) +i \bar k_j  X_{cl}(x_j,x_j)}
e^{- k_{\parallel D_\ttt j}^2 \Delta_{(N,M), bou (i)}(x_j,x_j)}
\right]
\nonumber\\
&
e^{- k_{\parallel D_\ttt 1} k_{\parallel D_\ttt 2} G_{(N,M), bou}(x_1,x_2)}
\end{align}
where $G_{(N,M), bou}(x_1,x_2)$ is the common factor of all the
components of the boundary Green function $ G_{(N,M), bou}^{I
  J}(x_1,x_2)$ and is given in eq. (\ref{G-NM-bou}) whose explicit
expression (\ref{GIJNM-bou-GNM-bou}) 
implies that only the momenta parallel to the brane on
which the vertex lies contributes.

Finally, we consider a more lengthy example
\begin{align}
\langle 
&
(\partial_x^2 X \partial_x X \partial_x \bar X ) (x_1)
(\partial_x^2 \bar X)(x_2)
\prod_{t=1}^N\sigma_{\epsilon_t,f_t}(x_t)
\rangle
=
\frac{\partial^3}{\partial \bar c_{(1) 2} \partial \bar c_{(1) 1 }
  \partial c_{(1) 1}}
\frac{\partial}{\partial c_{(2) 2} }
V_{N+2}\Big|_{ c_{(j) n}=0 }
\nonumber\\
=
&
\langle \prod_{t=1}^N\sigma_{\epsilon_t,f_t}(x_t) \rangle
\Bigg[
(\partial_x^2 X_{cl} \partial_x X_{cl} \partial_x \bar X_{cl} ) (x_1)
(\partial_x^2 \bar X_{cl})(x_2)
\nonumber\\
%
%
&+
e^{i \pi( -\alpha_{t_1} - \alpha_{t_2} ) }
~
(\partial_x^2 X_{cl} \partial_x X_{cl}) (x_1)
~
\partial_{x_1} \partial_{x_2}^2 G_{(N,M), bou}(x_1,x_2)
\nonumber\\
&+
e^{i \pi(  \alpha_{t_1} - \alpha_{t_2} ) }
~
(\partial_x^2 X_{cl} \partial_x \bar X_{cl}) (x_1)
~
\partial_{x_1} \partial_{x_2}^2 G_{(N,M), bou}(x_1,x_2)
\nonumber\\
&+
e^{i \pi( \alpha_{t_1} - \alpha_{t_2} ) }
~
(\partial_x X_{cl} \partial_x \bar X_{cl}) (x_1)
~
\partial_{x_1}^2 \partial_{x_2}^2 G_{(N,M), bou}(x_1,x_2)
\nonumber\\
%
%
&+
\oh
~
\partial_x^2 X_{cl}(x_1) \partial_x \bar X_{cl}(x_2)
~
\partial_{x_1} \partial_{\hat x_1} \Delta_{(N,M), bou}(x_1,\hat
x_1)|_{\hat x_1=x_1}
\nonumber\\
&+
\oh
~
\partial_x X_{cl}(x_1)  \partial_x \bar X_{cl}(x_2)
~
\partial_{x_1}^2 \partial_{\hat x_1} \Delta_{(N,M), bou}(x_1,\hat
x_1)|_{\hat x_1=x_1}
\nonumber\\
&+
e^{2 i \pi \alpha_{t_1}}
~
\partial_x \bar X_{cl} (x_1) \partial_x \bar X_{cl}(x_2)
~
\partial_{x_1}^2 \partial_{\hat x_1} \Delta_{(N,M), bou}(x_1,\hat
x_1)|_{\hat x_1=x_1}
\nonumber\\
%
%
&+
\oh
e^{i \pi ( \alpha_{t_1}- \alpha_{t_2})}
~
\partial_{x_1} \partial_{\hat x_1} \Delta_{(N,M), bou}(x_1,\hat x_1)|_{\hat x_1=x_1}
~
\partial_{x_1}^2 \partial_{x_2}^2 G_{(N,M), bou}(x_1,x_2)
\nonumber\\
&+
\oh
e^{i \pi ( \alpha_{t_1}- \alpha_{t_2})}
~
\partial_{x_1}^2 \partial_{\hat x_1} \Delta_{(N,M), bou}(x_1,\hat x_1)|_{\hat x_1=x_1}
~
\partial_{x_1} \partial_{x_2}^2 G_{(N,M), bou}(x_1,x_2)
\nonumber\\
&+
\oh
e^{i \pi ( \alpha_{t_1}- \alpha_{t_2})}
~
\partial_{x_1}^2 \partial_{\hat x_1} \Delta_{(N,M), bou}(x_1,\hat x_1)|_{\hat x_1=x_1}
~
\partial_{x_1} \partial_{x_2}^2 G_{(N,M), bou}(x_1,x_2)
\Bigg]
\end{align}

\subsection{Boundary correlators with non excited twists on $T^2$}
\label{sect:bou_corr_on_T2}
The wrapping contributions have been studied for the pure twist field
correlators in
\cite{Cremades:2003qj} for the N=3 case and in 
\cite{Abel:2003vv}  for
the case $M=N-2$ and there is not any difference among the different
$M$ values therefore the results obtained there are valid.
Let us anyhow quickly review them.
Given a minimal $N$-polygon in $T^2$ with vertices $\{f_t\}$,
i.e. with all vertices in the fundamental cell,  
we can consider non minimal polygons which wrap the $T^2$.
These can be easierly described
as polygons which have vertices $\{ \tilde f_t \}$ 
in the covering $\R^2$ where $T^2 \equiv R^2 / \Lambda$ with
 the lattice defined as $\Lambda=\{ n_1 e_1+ n_2 e_2 | n_1,n_2\in \Z\}$.
These configurations give an additive contribution to the classical
path integral as
\begin{align}
V_{N+M}^{(T^2)}(J_i)
&=
\sum _{\{\tilde  f_t\}}
\left[
\langle 
\sigma_{\epsilon_1,\tilde f_1}(x_1) \dots \sigma_{\epsilon_N,\tilde f_N}(x_N)
\rangle
\prod_{i=1}^M
e^{
\sum_{n=0}^\infty c_{(i) n I}
\partial^n_{x_i} X^I_{cl}(x_i,x_i;\{x_t,\epsilon_t,\tilde f_t\})
}
\right]
\nonumber\\
\times
&
\prod_{i=1}^M
e^{
\oh
\sum_{n=0}^\infty c_{(i) n I}
\sum_{m=0}^\infty c_{(i) m J} 
\partial^n_{x_i} \partial^m_{ \hat x_i}
\Delta^{I J}_{(N, M), bou (i)}(x_i, \hat x_i ; \{ x_t, \epsilon_t\}) |_{ \hat x_i= x_i}
}
\nonumber\\
\times
&
\prod_{1\le i < j\le M}
e^{
\sum_{n=0}^\infty c_{(i) n I}
\sum_{m=0}^\infty c_{(j) m J} 
\partial^n_{x_i} \partial^m_{x_j}
G^{I J}_{(N, M), bou}(x_i, x_j ; \{ x_t, \epsilon_t\})
}
\label{V_N+M-T2}
\end{align}
In order to determine the possible vertices  $\{ \tilde f_t \}$
without redundancy it is necessary to keep a vertex fixed and then
expand the polygon. For definiteness we keep fixed the vertex $\tilde f_1=f_1$ 
which lies at the intersection between $D_N$ and $D_1$.
We then move the next vertex $f_2$ along the $D_1$ brane. 
Explicitly we write $\tilde f_2= \tilde f_1+ (f_2-f_1) + n_1 t_1= 
f_2+ n_1 t_1$ with $n_1\in \Z$ and
$t_1$  the shortest tangent vector to $D_1$ which is in $\Lambda$.
We can now continue for all the other vertices for which we have
$\tilde f_t = \tilde f_{t-1} +( f_{t}-f_{t-1})+ n_{t-1} t_{t-1}
=f_t + \sum_{k=1}^{t-1} n_k t_k$.
For consistency we need requiring $\tilde f_{N+1} \equiv \tilde f_1 =
f_1$,
therefore the possible wrapped polygons are obtained from the solution
of the Diophantine equation
\begin{align}
\sum_{\htt=1}^N n_\htt t_\htt =0
\end{align}
which cannot be solved in general terms but only on a case by case
basis as discussed in \cite{Abel:2003vv}.

\subsection{Chiral correlators with non excited twists.}
\label{sect:chiral_corr}
As a warming up for the computation of excited twist fields which we
perform in the next section we consider the interaction of chiral
vertices with plain twists.

We can now follow the same strategy we used in section
\ref{sect:boundary_corr} and compute the path integral with the
insertion of an arbitrary number $L_c$ of currents which act as
generating functions for the chiral vertex operators. 
As done for the boundary correlators in section \ref{sect:boundary_corr}
we first regularize the currents
and then subtract the self interaction of the untwisted string.
We are therefore led to consider the path integral
\begin{align}
V_{N+L_c}(J_c)
=
\lim_{\{\eta_c\} \rightarrow 0}
&
\int_{\cM(\{x_t,\epsilon_t, f_t\})} 
\cD X~ 
e^{-S_E}
\nonumber\\
\times
&
\prod_{c=1}^{L_c} \Big[
e^{-\oh \int d^2 u_a \int d^2 u_ b J_{c I}(u_a, \eta_c )  J_{c J}(u_b,\eta_c)
  \partial_{u_a} \partial_{u_b} G^{I J}_{U(t_c)}(u_a,\bu_a;u_b,\bu_b )}
\nonumber\\
&
e^{\int d^2 u~ \partial_{u} J_{c I}(u, \eta_c) \partial_u X^I(u,\bu) }
\Big]
\label{reg-path-integral-chiral}
\end{align}
where in the second line we have written the regularization factor
analogous to  the one used in section \ref{sect:boundary_corr}
which regularizes the currents in the third line. 
In the previous expression the regularized currents are defined as
\begin{equation}
J_{c I}(u,\bu, \eta_c)=  \sum_{n=1}^\infty c_{(c) n I}
\partial^{n-1}_{u}\delta^2(u-u_c; \eta_c)
\end{equation}
with $\delta^2(u-u_c; \eta_c)$ a regularization of the $\delta^2()$ such
that $\lim_{\eta_c \rightarrow 0} \delta^2(u; \eta_c)= \delta^2(u)$.
Notice that we need using a $\delta^2()$ in the previous expression
since we use ``directional'' derivatives along $u$.
To subtract the self interaction of the untwisted string we 
used $G^{I J}_{U(t_c)}(u_a,\bu_a; u_b,\bu_b)$, the untwisted Green
function computed for an arbitrary $t_c$ since we chose to consider
currents with at least one derivative $\partial_u$ so that only $G^{z
  \bz}_U$ give with a non vanishing contribution which is also
independent on the brane $D_t$.

Finally performing the path integral we get
\begin{align}
V_{N+L_c}(J_c)
&=
\langle 
\sigma_{\epsilon_1,f_1}(x_1) \dots \sigma_{\epsilon_N,f_N}(x_N)
\rangle
\nonumber\\
\times
&
\prod_{c=1}^{L_c}
\Bigg\{
e^{
\sum_{n=1}^\infty c_{(c) n I }
\partial^{n-1}_{u_c} [ \partial_u X^I_{cl}(u_c, \bu_c) ]
}
\nonumber\\
\times
&
e^{
\oh
\sum_{n,m=1}^\infty c_{(c) n I } c_{(c) m J} 
\partial^{n}_{u_d} \partial^{m}_{v_{\hat c}}
\Delta_{(N,M) (c)}^{I J}(u_c,\bu_c; v_c, \bv_c; \{ x_\htt, \epsilon_\htt\}) |_{v_c=u_c}
]
}
\Bigg\}
\nonumber\\
\times
&
\prod_{1\le c < \hat c \le N}
e^{
\sum_{n,m=1}^\infty d_{(c) n I } d_{(\hat c) m J} 
\partial^{n}_{u_t} \partial^{m}_{ v_\htt}
G_{(N,M)}^{I J}(u_c, \bu_c; v_{\hat c}, \bv_{\hat c}  ; \{ x_t, \epsilon_t\})
}
\end{align}
where we have written the dependence on the complex conjugate
variables such as $\bu$ even if the derivatives are independent in
order to be consistent with the notation used in the boundary case. 
The regularized chiral Green function is defined as expected as
\begin{align}
\partial_u \partial_v \Delta_{(N,M) (c)}^{I J}(u, \bu; v, \bv \{ x_\htt,\epsilon_\htt\})
&=
\partial_u \partial_v G_{(N,M)}^{I J}(u, \bu; v, \bv; \{ x_\htt, \epsilon_\htt\})
-
\partial_u \partial_v G^{I J}_{U(t)}(u, \bu; v, \bv; \epsilon_t)
\end{align}
and we notice that the subtraction term is different from zero only
when $I J= z \bz$ or $I J= \bz z$ because of the derivatives.

\subsection{Correlators of excited twists on $\R^2$}
Finally we can compute the correlators of excited twist fields by
letting the appropriate chiral currents collide with the twist fields. 
We follow the same strategy we used in section
\ref{sect:boundary_corr} and in the previous section \ref{sect:chiral_corr}
and compute the path integral with the
insertion of one generating function (\ref{T-abs}) for each twist field.
As done for the boundary correlators in section \ref{sect:boundary_corr}
we first regularize the $\delta^2()$ functions in the currents
and then subtract the self interaction of the untwisted string.
We are therefore led to consider the path integral
\begin{align}
V_{N}(K_t)
=
&
\lim_{\{u_t\} \rightarrow \{x_t\} }
\lim_{\eta_t \rightarrow 0}
\int_{\cM(\{x_t,\epsilon_t, f_t\})} 
\cD X~ 
e^{-S_E}
\nonumber\\
\times
&
\prod_{t=1}^N \Big[
e^{-\oh \int d^2 u_a \int d^2 u_ b K_{t I}(u_a, \eta_t )  K_{t J}(u_b,\eta_t)
  (u_a-x_t)^{\epsilon_{t I}} (u_b-x_t)^{\epsilon_{t J}}
  \partial_{u_a} \partial_{u_b} G^{I J}_{U(t)}(u_a,\bu_a;u_b,\bu_b )}
\nonumber\\
&
e^{\int d^2 u~ K_{t I}(u, \eta_t) 
   (u-x_t)^{\epsilon_{t I}} \partial_u X^I(u,\bu) }
\nonumber\\
&
e^{
-
\int d^2 u_a \int d^2 u_ b K_{t z}(u_a, \eta_t )  K_{t \bz}(u_b,\eta_t)
\left[ (u_a-x_t)^{\bar \epsilon_t}  (u_b-x_t)^{\epsilon_t}
\partial_{u_a} \partial_{u_b} 
\Delta^{z\bz}_c(u_a-x_t,\bu_a-x_t; u_b-x_t,\bu_b-x_t; \epsilon_t)
\right]
}
\Big]
\label{reg-path-integral-exc-twist}
\end{align}
where in the second line we have written the regularization factor
analogous to  the one used in section \ref{sect:boundary_corr}
which regularizes the third line and finally in the last line we have
the necessary subtraction term which is in eq. (\ref{T-abs}).
In the previous expression the regularized currents are defined as
\begin{equation}
K_{t I}(u,\bu, \eta_t)= \sum_{n=1}^\infty d_{(t) n I}
\partial^{n-1}_{u}\delta^2(u-u_t; \eta_t)
\end{equation}
with $\delta^2(u-u_t; \eta_t)$ a regularization of the $\delta^2()$
as in previous section.
\COMMENTO{ 
The factor $2$ is due that $x_t$ is on the boundary and therefore the
$\delta^2()$ gives $\oh$ when integrated. 
}
For writing a more compact expression we have also used $\epsilon_{t I}$
defined as 
\begin{equation}
\epsilon_{t z}= \bar \epsilon_t,~~~
\epsilon_{t \bz}=\epsilon_t
\end{equation}

Finally performing the path integral we get
\begin{align}
V_{N}(K_t)
&=
\lim_{\{u_t\} \rightarrow \{x_t\}}
\langle 
\sigma_{\epsilon_1,f_1}(x_1) \dots \sigma_{\epsilon_N,f_N}(x_N)
\rangle
\nonumber\\
\times
&
\prod_{t=1}^N
\Bigg\{
e^{
\sum_{n=1}^\infty d_{(t) n I }
\partial^{n-1}_{u_t} [ (u_t-x_t)^{\epsilon_{t I}} \partial_u
  X^I_{cl}(u_t, \bu_t) ]
}
\nonumber\\
\times
&
e^{
\oh
\sum_{n,m=1}^\infty d_{(t) n I } d_{(t) m J} 
\partial^{n-1}_{u_t} \partial^{m-1}_{v_t}
[(u_t-x_t)^{\epsilon_{t I}} (v_t-x_t)^{\epsilon_{t J}}
\partial_u \partial_v 
\Delta_{(N,M) (t)}^{I J}(u_t, \bu_t; v_t, \bv_t; \{ x_\htt, \epsilon_\htt\}) 
]
|_{v_t=u_t}
}
\Bigg\}
\nonumber\\
\times
&
\prod_{1\le t < \htt \le N}
e^{
\sum_{n,m=1}^\infty d_{(t) n I } d_{(\htt) m J} 
\partial^{n-1}_{u_t} \partial^{m-1}_{ v_\htt}
[(u_t-x_t)^{\epsilon_{t I}} (v_\htt-x_\htt)^{\epsilon_{\htt J}}
\partial_u \partial_v
G_{(N,M)}^{I J}(u_t, \bu_t; v_\htt, \bv_\htt  ; \{ x_\bt, \epsilon_\bt\})
]
}
\end{align}
where we have defined the regularized Green function at the twist
fields $t$ to be
\begin{align}
\partial_u \partial_v &\Delta_{(N,M) (t)}^{I J}(u, \bu; v, \bv; \{ x_\htt,\epsilon_\htt\})
=
\nonumber\\
=
&
\partial_u \partial_v \Delta_{(N,M) (c)}^{I J}(u, \bu; v, \bv; \{ x_\htt, \epsilon_\htt\})
-
\partial_u \partial_v \Delta^{I J}_c(u-x_t, \bu-x_t; v-x_t, \bv-x_t; \epsilon_t)
\nonumber\\
&=
\partial_u \partial_v G_{(N,M)}^{I J}(u, \bu; v, \bv ; \{ x_\htt, \epsilon_\htt\})
-
\partial_u \partial_v G_{N=2}^{I J}(u-x_t, \bu-x_t; v-x_t, \bv-x_t; \epsilon_t)
\end{align}
and we used $G_{U(t)}^{I J}(u-x_t, v-x_t) =G_{U(t)}^{I J}(u, v)$
to write the last line
and again we have written the dependence on $\bu$ and $\bv$ even if
the derivatives are independent on  them for having a consistent notation. 
Actually because of the chiral derivatives the previous expression
simplifies in two cases to
\begin{align}
\partial_u \partial_v \Delta_{(N,M) (t)}^{z z}(u, \bu; v, \bv ; \{ x_\htt,\epsilon_\htt\})
&= 
\partial_u \partial_v G_{(N,M)}^{z z}(u, \bu; v, \bv; \{ x_\htt, \epsilon_\htt\})
\nonumber\\
\partial_u \partial_v \Delta_{(N,M) (t)}^{\bz \bz}(u, \bu; v, \bv;; \{ x_\htt,\epsilon_\htt\})
&=
\partial_u \partial_v G_{(N,M)}^{\bz \bz}(u, \bu; v, \bv; \{ x_\htt, \epsilon_\htt\})
\end{align}

It is interesting to notice that regularized Green functions are
got by subtracting the divergent part with the proper monodromy at the
point of regularization which at the points where a twist field is
located means $G_{N=2}$ while in all other points means $G_U$.
In particular both 
$(u-x_t)^{\epsilon_{t I}} (v-x_t)^{\epsilon_{t J}}
\partial_u \partial_v  \Delta_{(N,M) (t)}^{I J}$
and
$
(u-x_t)^{\epsilon_{t I}} (v-x_t)^{\epsilon_{t J}} \partial_u \partial_v  G_{(N,M)}^{I J}
$
are analytic functions at $u=x_t$ whose explicit expression is given
in appendix \ref{app:functions-for_excited}.

More explicitly the previous generating function can be written as
\begin{align}
V_{N}(K_t)
&=
\lim_{\{u_t\} \rightarrow \{x_t\}}
\langle 
\sigma_{\epsilon_1,f_1}(x_1) \dots \sigma_{\epsilon_N,f_N}(x_N)
\rangle
\nonumber\\
\times
&
\prod_{t=1}^N
\Bigg\{
e^{
\sum_{n=1}^\infty 
\left[
\bar d_{(t) n }
 \partial^{n-1}_{u_t} [ (u_t-x_t)^{\bar \epsilon_t} \partial_u X_{cl}(u_t,\bu_t) ]
+
 d_{(t) n }
 \partial^{n-1}_{u_t} [ (u_t-x_t)^{\epsilon_t} \partial_u \bX_{cl}(u_t,\bu_t) ]
\right]
}
\nonumber\\
\times
&
e^{
\oh
\sum_{n,m=1}^\infty 
\bar d_{(t) n} \bar d_{(t) m} 
\partial^{n-1}_{u_t} \partial^{m-1}_{v_t}
[(u_t-x_t)^{\bar \epsilon_{t}} (v_t-x_t)^{\bar \epsilon_{t}}
\partial_u \partial_v 
G_{(N,M) (t)}^{z z}(u_t, \bu_t; v_t, \bv_t ; \{ x_\htt, \epsilon_\htt\}) 
]
|_{v_t=u_t}
}
\nonumber\\
\times
&
e^{
\oh
\sum_{n,m=1}^\infty 
d_{(t) n} d_{(t) m} 
\partial^{n-1}_{u_t} \partial^{m-1}_{v_t}
[(u_t-x_t)^{\epsilon_{t}} (v_t-x_t)^{\epsilon_{t}}
\partial_u \partial_v 
G_{(N,M) (t)}^{\bz \bz}(u_t, \bu_t; v_t, \bv_t; \{ x_\htt, \epsilon_\htt\}) 
]
|_{v_t=u_t}
}
\nonumber\\
\times
&
e^{
\sum_{n,m=1}^\infty 
\bar d_{(t) n} d_{(t) m} 
\partial^{n-1}_{u_t} \partial^{m-1}_{v_t}
[(u_t-x_t)^{\bar \epsilon_{t}} (v_t-x_t)^{\epsilon_{t}}
\partial_u \partial_v 
\Delta_{(N,M) (t)}^{z \bz}(u_t, \bu_t; v_t, \bv_t; \{ x_\htt, \epsilon_\htt\}) 
]
|_{v_t=u_t}
}
\Bigg\}
\nonumber\\
\times
&
\prod_{1\le t < \htt \le N}
\Bigg\{
e^{
\sum_{n,m=1}^\infty \bar d_{(t) n } \bar d_{(\htt) m } 
\partial^{n-1}_{u_t} \partial^{m-1}_{ v_\htt}
[(u_t-x_t)^{\bar \epsilon_{t}} (v_\htt-x_\htt)^{\bar \epsilon_{\htt}}
\partial_u \partial_v
G_{(N,M)}^{z z}(u_t, \bu_t; v_\htt, \bv_\htt  ; \{ x_\bt, \epsilon_\bt\})
]
}
\nonumber\\
&
\times
e^{
\sum_{n,m=1}^\infty d_{(t) n } d_{(\htt) m } 
\partial^{n-1}_{u_t} \partial^{m-1}_{ v_\htt}
[(u_t-x_t)^{\epsilon_{t}} (v_\htt-x_\htt)^{\epsilon_{\htt}}
\partial_u \partial_v
G_{(N,M)}^{\bz \bz}(u_t, \bu_t; v_\htt, \bv_\htt  ; \{ x_\bt, \epsilon_\bt\})
]
}
\nonumber\\
&
\times
e^{
\sum_{n,m=1}^\infty \bar d_{(t) n } d_{(\htt) m } 
\partial^{n-1}_{u_t} \partial^{m-1}_{ v_\htt}
[(u_t-x_t)^{\bar \epsilon_{t}} (v_\htt-x_\htt)^{\epsilon_{\htt}}
\partial_u \partial_v
G_{(N,M)}^{z \bz}(u_t, \bu_t; v_\htt, \bv_\htt  ; \{ x_\bt, \epsilon_\bt\})
]
}
\nonumber\\
&
\times
e^{
\sum_{n,m=1}^\infty d_{(t) n } \bar d_{(\htt) m } 
\partial^{n-1}_{u_t} \partial^{m-1}_{ v_\htt}
[(u_t-x_t)^{\epsilon_{t}} (v_\htt-x_\htt)^{\bar \epsilon_{\htt}}
\partial_u \partial_v
G_{(N,M)}^{\bz z}(u_t, \bu_t; v_\htt, \bv_\htt  ; \{ x_\bt, \epsilon_\bt\})
]
}
\Bigg\}
\label{reggeon-excited-twists}
\end{align}

\subsubsection{Some examples}
Let us consider the simplest non trivial correlator with one excited
twist field
\begin{align}
&
\langle 
\left(\partial^n X \sigma_{\epsilon_t, f_t} \right)
\prod_{\bt\ne t}  \sigma_{\epsilon_\bt, f_\bt}
\rangle
=
\frac{\partial}{ \partial \bar d_{(t) n z} } V_N|_{d=\bar d=0}
\nonumber\\
&=
\langle \prod_{\bt}  \sigma_{\epsilon_\bt, f_\bt} \rangle
\lim_{u_t\rightarrow x_t} \partial^{n-1}_{u_t} 
[ (u_t-x_t)^{\bar\epsilon_{t}} \partial_u X_{cl}(u_t) ]
\end{align}
where the limit is not strictly necessary since the expression 
 $(u_t-x_t)^{\bar \epsilon_{t }} \partial_u X_{cl}(u_t)$ is
regular in $x_t$. The explicit expression can be easily computed and
it is given in appendix \ref{app:explicit_expressions_N=3} for some cases.
Nevertheless correlators with only one excited twist can be less
trivial as 
\begin{align}
\langle 
&
\left( \partial^n X \partial^m \bar X \sigma_{\epsilon_t, f_t} \right)
\prod_{\bt\ne t}  \sigma_{\epsilon_\bt, f_\bt}
\rangle
=
\frac{\partial^2}{ \partial \bar d_{(t) n z}  \partial d_{(t) m z}} V_N|_{d=\bar d=0}
\nonumber\\
&=
\langle \prod_{\bt}  \sigma_{\epsilon_\bt, f_\bt} \rangle
\lim_{u_t\rightarrow x_t} 
\Big[
\partial^{n-1}_{u_t} [ (u_t-x_t)^{\bar\epsilon_{t}} \partial_u X_{cl}(u_t) ]
\partial^{m-1}_{u_t} [ (u_t-x_t)^{\bar\epsilon_{t}} \partial_u X_{cl}(u_t) ]
\nonumber\\
&
+
\partial^{n-1}_{u_t} \partial^{m-1}_{v_t}
[(u_t-x_t)^{\bar \epsilon_{t}} (v_t-x_t)^{\epsilon_{t}}
\partial_u \partial_v 
\Delta_{(N,M) (t)}^{z \bz}(u_t, \bu_t; v_t, \bv_t; \{ x_\htt, \epsilon_\htt\}) 
]
|_{v_t=u_t}
\Big]
\end{align}

Finally we can also consider
\begin{align}
\langle 
&
\left( \partial^n X \sigma_{\epsilon_t, f_t} \right)
\left( \partial^m \bar X \sigma_{\epsilon_\htt, f_\htt} \right)
\prod_{\bt\ne t,\htt}  \sigma_{\epsilon_\bt, f_\bt}
\rangle
=
\frac{\partial^2}{ \partial \bar d_{(t) n z}  \partial d_{(\htt) m z}} V_N|_{d=\bar d=0}
\nonumber\\
&=
\langle \prod_{\bt}  \sigma_{\epsilon_\bt, f_\bt} \rangle
\lim_{u_{t,\htt}\rightarrow x_{t,\htt}} 
\Big[
\partial^{n-1}_{u_t} [ (u_t-x_t)^{\bar\epsilon_{t}} \partial_u X_{cl}(u_t) ]
\partial^{m-1}_{u_\htt} [ (u_\htt-x_\htt)^{\bar\epsilon_{\htt}} \partial_u X_{cl}(u_\htt) ]
\nonumber\\
&
+
\partial^{n-1}_{u_t} \partial^{m-1}_{u_\htt}
[(u_t-x_t)^{\bar \epsilon_{t}} (u_\htt-x_\htt)^{\epsilon_{\htt}}
\partial_{u_t} \partial_{u_\htt} 
G_{(N,M)}^{z \bz}(u_t, \bu_t; u_\htt, \bu_\htt; \{ x_\htt, \epsilon_\htt\}) 
]
\Big]
\end{align}
This correlator is the correlator of eq. (4.83) of
\cite{Anastasopoulos:2013sta}, i.e. 
$ 
\langle \tau_\alpha(x_1) \tau_{1-\alpha}(x_2) \sigma_\beta(x_3)
\sigma_{1-\beta}(x_4)\rangle 
$ when we set $N=4$, $n=m=1$ and 
$\epsilon_1=\alpha$, $\epsilon_2=1-\alpha$,
$\epsilon_3=\beta$ and $\epsilon_4=1-\beta$.

\subsection{Correlators of boundary operators and excited twists on $\R^2$}
Finally we can assemble the results from previous section to write the
generating function for correlators of boundary operators and excited
twists on $R^2$ to be
\begin{align}
V_{N+L}(K_t,J_i)
&=
\lim_{\{u_t\} \rightarrow \{x_t\}}
\langle 
\sigma_{\epsilon_1,f_1}(x_1) \dots \sigma_{\epsilon_N,f_N}(x_N)
\rangle
\nonumber\\
\times
&
\prod_{t=1}^N
\Bigg\{
e^{
\sum_{n=1}^\infty d_{(t) n I }
\partial^{n-1}_{u_t} [ (u_t-x_t)^{\epsilon_{t I}} \partial_u
  X^I_{cl}(u_t, \bu_t) ]
}
\nonumber\\
\times
&
e^{
\oh
\sum_{n,m=1}^\infty d_{(t) n I } d_{(t) m J} 
\partial^{n-1}_{u_t} \partial^{m-1}_{v_t}
[(u_t-x_t)^{\epsilon_{t I}} (v_t-x_t)^{\epsilon_{t J}}
\partial_u \partial_v 
\Delta_{(N,M) (t)}^{I J}(u_t, \bu_t; v_t, \bv_t; \{ x_\bt, \epsilon_\bt\}) ]
|_{v_t=u_t}
}
\Bigg\}
\nonumber\\
\times
&
\prod_{i=1}^L
\Bigg\{
e^{
\sum_{n=0}^\infty c_{(i) n I}
\partial^n_{x_i} X^I_{cl}(x_i,x_i)
}
\nonumber\\
\times
&
e^{
\oh
\sum_{n=0}^\infty c_{(i) n I}
\sum_{m=0}^\infty c_{(i) m J} 
\partial^n_{x_i} \partial^m_{ \hat x_i}
\Delta^{I J}_{(N,M), bou (i)}(x_i, \hat x_i ; \{ x_t, \epsilon_t\}) |_{ \hat x_i= x_i}
}
\Bigg\}
\nonumber\\
\times
&
\prod_{1\le t < \htt \le N}
e^{
\sum_{n,m=1}^\infty d_{(t) n I } d_{(\htt) m J} 
\partial^{n-1}_{u_t} \partial^{m-1}_{ v_\htt}
[(u_t-x_t)^{\epsilon_{t I}} (v_\htt-x_\htt)^{\epsilon_{\htt J}}
\partial_u \partial_v
G_{(N,M)}^{I J}(u_t, \bu_t; v_\htt, \bv_\htt  ; \{ x_\bt, \epsilon_\bt\})
]
}
\nonumber\\
\times
&
\prod_{1\le i < j\le L}
e^{
\sum_{n=0}^\infty c_{(i) n I}
\sum_{m=0}^\infty c_{(j) m J} 
\partial^n_{x_i} \partial^m_{x_j}
G^{I J}_{(N,M), bou}(x_i, x_j ; \{ x_t, \epsilon_t\})
}
\nonumber\\
\times
&
\prod_{1\le t \le N}
\prod_{1\le j \le L}
e^{
\sum_{n=1}^\infty d_{(t) n I } c_{(j) m J} 
\partial^{n-1}_{u_t} \partial^{m}_{x_j}
[(u_t-x_t)^{\epsilon_{t I}} \partial_u 
G_{(N,M)}^{I J}(u_t, \bu_t; x_j, x_j  ; \{ x_\bt, \epsilon_\bt\})
]
}
\label{reggeon-excited-twists+bou}
\end{align}
where the last line is the interaction between the twist fields and
the boundary operators.
The generating function for correlators on $T^2$ can be formally
easily obtained as done in section \ref{sect:bou_corr_on_T2} by
summing over all possible wrapping contributions as in eq. \ref{V_N+M-T2}.

\COMMENTOO{
Notice that  if we fix the excited twist operators, i.e. we perform
the necessary derivative wrt $d_{I}$, the resulting expression cannot
be interpreted as coming from a modified Green function since the term
quadratic in $c$ is not modified.
}

\subsection{Correlators of bulk operators and excited twists on
  $\R^2$}
We can now make an educated guess of the generating function of the
correlators of bulk operators and excited twists.
As long as the bulk vertex operators  do not involve momenta there is
no doubt on the result since the bulk field can be written as
 the product of a chiral vertex times an antichiral vertex therefore
 the generating function is nothing else but the product of the
 generating function for the chiral part times 
the generating function for the antichiral part times 
the obvious interaction of the chiral current with the antichiral.
What requires an educated guess is when the bulk vertex operators
involve momenta since in this case we know that the local description
requires to separate the right moving from the left moving part and
normal order them separately, i.e. to the abstract vertex $e^{i k_I X^I(u,\bu)}$
corresponds the local version $:e^{i k_I X^I_{(loc) L}(\ul)}: :e^{i
  k_I X^I_{(loc) R}(\bul)}:$ up to cocycles (\cite{Polchinski:1998rq},
\cite{Pesando:2011yd} but see also \cite{DiVecchia:2007dh})
Hence we guess that in the presence of momenta the twisted Green
function must be split in its chiral-chiral, chiral-antichiral and so
on pieces, i.e. $G=G_{(L L)}+ G_{(L R)}+ G_{(R L)}+ G_{(R R)}$. 
This is obviously consistent with the case where
derivatives are present since applying a  derivative like $\partial_u$
of the Green function is actually projecting it on the chiral piece.
We therefore guess that the generating function of the
correlators of $L_c$ bulk operators and $N$ excited twists read up to
phases due to cocycles
\begin{align}
V_{N+L_c}(K_t,J_c)
&=
\lim_{\{u_t\} \rightarrow \{x_t\}}
\langle 
\sigma_{\epsilon_1,f_1}(x_1) \dots \sigma_{\epsilon_N,f_N}(x_N)
\rangle
\nonumber\\
\times
&
\prod_{t=1}^N
\Bigg\{
e^{
\sum_{n=1}^\infty d_{(t) n I }
\partial^{n-1}_{u_t} [ (u_t-x_t)^{\epsilon_{t I}} \partial_u
  X^I_{cl}(u_t, \bu_t) ]
}
\nonumber\\
\times
&
e^{
\oh
\sum_{n,m=1}^\infty d_{(t) n I } d_{(t) m J} 
\partial^{n-1}_{u_t} \partial^{m-1}_{v_t}
[(u_t-x_t)^{\epsilon_{t I}} (v_t-x_t)^{\epsilon_{t J}}
\partial_u \partial_v 
\Delta_{(N,M) (t)}^{I J}(u_t, \bu_t; v_t, \bv_t; \{ x_\bt, \epsilon_\bt\}) ]
|_{v_t=u_t}
}
\Bigg\}
\nonumber\\
%
\times
&
\prod_{c=1}^{L_c}
\Bigg\{
e^{
\sum_{n=0}^\infty c_{(L, c) n I}
\partial^n_{u_c} X^I_{(L) cl }(u_c)
}
\nonumber\\
\times
&
e^{
\oh
\sum_{n=0}^\infty c_{(L, c) n I}
\sum_{m=0}^\infty c_{(L, c) m J} 
\partial^n_{u_c} \partial^m_{ \hat u_c}
\Delta^{I J}_{(N,M), (L L, c)}(u_c, \hat u_c ; \{ x_t, \epsilon_t\}) |_{ \hat u_c= u_c}
}
\Bigg\}
\nonumber\\
%
\times
&
\prod_{c=1}^{L_c}
\Bigg\{
e^{
\sum_{n=0}^\infty c_{(R, c) n I}
\partial^n_{\bu_c} X^I_{(R) cl }(\bu_c)
}
\nonumber\\
\times
&
e^{
\oh
\sum_{n=0}^\infty c_{(R, c) n I}
\sum_{m=0}^\infty c_{(R, c) m J} 
\partial^n_{\bu_c} \partial^m_{ \hat \bu_i}
\Delta^{I J}_{(N,M), (R R, c)}(u_c, \hat {\bu}_c ; 
\{ x_t, \epsilon_t\}) |_{ \hat \bu_c= \bu_c}
}
\Bigg\}
\nonumber
\end{align}
\begin{align}
%
\times
&
\prod_{1\le t < \htt \le N}
e^{
\sum_{n,m=1}^\infty d_{(t) n I } d_{(\htt) m J} 
\partial^{n-1}_{u_t} \partial^{m-1}_{ v_\htt}
[(u_t-x_t)^{\epsilon_{t I}} (v_\htt-x_\htt)^{\epsilon_{\htt J}}
\partial_u \partial_v
G_{(N,M)}^{I J}(u_t, \bu_t; v_\htt, \bv_\htt  ; \{ x_\bt, \epsilon_\bt\})
]
}
\nonumber\\
%
\times
&
\prod_{1\le c < d\le L_c}
e^{
\sum_{n=0}^\infty c_{(L, c) n I}
\sum_{m=0}^\infty c_{(L, d) m J} 
\partial^n_{u_c} \partial^m_{u_d}
G^{I J}_{(N,M) (L L)}(u_c, u_d ; \{ x_t, \epsilon_t\})
}
\nonumber\\
%
\times
&
\prod_{1\le c,d\le L_c}
e^{
\sum_{n=0}^\infty c_{(L, c) n I}
\sum_{m=0}^\infty c_{(R, d) m J} 
\partial^n_{u_c} \partial^m_{\bu_d}
\left[
G^{I J}_{(N,M) (L L)}(u_c, \bu_d ; \{ x_t, \epsilon_t\})
+
G^{J I}_{(N,M) (L L)}(\bu_d, u_c ; \{ x_t, \epsilon_t\})
\right]
}
\nonumber\\
%
\times
&
\prod_{1\le c < d\le L_c}
e^{
\sum_{n=0}^\infty c_{(R, c) n I}
\sum_{m=0}^\infty c_{(R, d) m J} 
\partial^n_{\bu_c} \partial^m_{\bu_d}
G^{I J}_{(N,M) (R R)}(\bu_c, \bu_d ; \{ x_t, \epsilon_t\})
}
\nonumber\\
%
\times
&
\prod_{1\le t \le N}
\prod_{1\le c \le L_c}
e^{
\sum_{n=1}^\infty d_{(t) n I } c_{(L, c) m J} 
\partial^{n-1}_{u_t} \partial^{m}_{u_c}
[(u_t-x_t)^{\epsilon_{t I}} \partial_u 
\left[
G_{(N,M) (L L) }^{I J}(u_t; u_c  ; \{ x_\bt, \epsilon_\bt\})
+G_{(N,M) (R L) }^{I J}(\bu_t; u_c  ; \{ x_\bt, \epsilon_\bt\})
\right]
]
}
\nonumber\\
%
\times
&
\prod_{1\le t \le N}
\prod_{1\le c \le L_c}
e^{
\sum_{n=1}^\infty d_{(t) n I } c_{(R, c) m J} 
\partial^{n-1}_{u_t} \partial^{m}_{\bu_c}
[(u_t-x_t)^{\epsilon_{t I}} \partial_u 
\left[
G_{(N,M) (L L) }^{I J}(u_t; \bu_c  ; \{ x_\bt, \epsilon_\bt\})
+G_{(N,M) (R L) }^{I J}(\bu_t; \bu_c  ; \{ x_\bt, \epsilon_\bt\})
\right]
]
}
\label{reggeon-excited-twists+closed}
\end{align}
Using the previous generating function it would be interesting deriving
the boundary state with $N$ twist fields. This could be done
as in  \cite{Pesando:2009tt} and would be an interesting
generalization of the boundary state with open string interactions
derived in \cite{Pesando:2003ww}. This boundary state could be used as
in \cite{Di Vecchia:1997pr} to derive useful information about the
long distance spacetime geometry generated by branes at angles.

\subsection{Rewriting the Reggeon vertex using auxiliary Fock spaces}
In the previous section we have given the explicit form of the
generating function for correlators of boundary operators and excited
twists. Traditionally and for sewing a different expression is used
where to any operator insertion, i.e. external leg 
is associated an auxiliary Fock space.
If we associate to any twist operator an auxiliary Fock space with
vacuum $|T_t\rangle$ and we identify\footnote{
The normalization is chosen in the usual way such that applying the map $d$ to
auxiliary operators on eq. (\ref{T_chiral})  we have
$\langle T_{aux} |
\cT(d\rightarrow \bar \alpha_{aux} ,\bar d \rightarrow \alpha_{aux})
~ \lim_{u_{aux}\rightarrow 0} \partial_{u_{aux}}^{n} \left[
  {u_{aux}}^{\bar \epsilon} \partial_{u_{aux}} X\loc({u_{aux}},{\bar u_{aux}})\right]
| T_{aux} \rangle
=
\partial_\ul^{n} \left[ \ul^{\bar \epsilon} \partial_\ul X\loc(\ul,\bul)\right]
$
.
}
\begin{equation}
(n-1)! \bar d_{(t) n} \leftrightarrow 
\frac{-2}{\alpha'} 
\kei t
\frac{\alpha_{(t,aux) n- \bar\epsilon_t}}{n-\bar\epsilon_t}
,~~~~
(n-1)! d_{(t) n} \leftrightarrow 
\frac{-2}{\alpha'} 
\kbei t
\frac{\bar \alpha_{(t,aux) n-\epsilon_t}}{n-\epsilon_t},~~~~
\end{equation} 
and we  do the same for any boundary vertex operator to which we
associate an auxiliary Fock space with vacuum $| 0_a, p_{(j)}=0\rangle$
and we identify ($n \ge 0$)
 \begin{equation}
n! \bar c_{(j) n} \leftrightarrow 
\frac{-2}{\alpha'} 
\kei {t_j}
\alpha_{(j, aux) n }
,~~~~
n! c_{(j) n} \leftrightarrow 
\frac{-2}{\alpha'} 
\kbei {t_j}
\bar \alpha_{(j, aux) n}
\end{equation} 
we can write the generating function as a usual Reggeon vertex as
\begin{align*}
\langle V_{N+L} |=
&
\nonumber\\ 
\prod_{t=1}^N \langle T_t|
&
\prod_{i=1}^L \langle 0_a, x_{(i)}=0|
\nonumber\\
&\prod_{t=1}^N 
e^{
\frac{-2}{\alpha'}
 \oint_{z=x_t} \frac{d z}{ 2\pi i}
\left[ 
\bar \chi_{(t,aux)}^{(+)}(z-x_t) ~\partial \chi_{cl}(z) 
+
\chi_{(t,aux)}^{(+)}(z-x_t) ~\partial \bar \chi_{cl}(z) 
\right]
}
\nonumber\\
&\prod_{i=1}^L 
e^{
\frac{-2}{\alpha'}
 \oint_{z=x_t} \frac{d z}{ 2\pi i}
\left[ 
\bar \chi_{(i,aux)}^{(+)}(z-x_t) ~\partial \chi_{cl}(z) 
+
\chi_{(i,aux)}^{(+)}(z-x_t) ~\partial \bar \chi_{cl}(z) 
\right]
}
\nonumber\\
\prod_{t,\htt=1}^N
\Bigg[
&
e^{
\frac{2}{\alpha'^2}
\oint_{z=x_t} \frac{d z}{ 2\pi i}
\oint_{w=x_\htt} \frac{d w}{ 2\pi i}
\bar \chi_{(t, aux)}(z-x_t)  ~\bar \chi_{(\htt, aux)}(w-x_\htt) 
~\partial_z \partial_w G^{z z}_{(N,M)}(z, \bz; w , \bar w)
}
\nonumber\\
&
e^{
\frac{2}{\alpha'^2}
\oint_{z=x_t} \frac{d z}{ 2\pi i}
\oint_{w=x_\htt} \frac{d w}{ 2\pi i}
\chi_{(t, aux)}(z-x_t)  ~\chi_{(\htt, aux)}(w-x_\htt) 
~\partial_z \partial_w G^{\bz \bz}_{(N,M)}(z, \bz; w , \bar w)
}
\nonumber\\
&
e^{
\frac{4}{\alpha'^2}
\oint_{z=x_t} \frac{d z}{ 2\pi i}
\oint_{w=x_\htt} \frac{d w}{ 2\pi i}
\bar \chi_{(t, aux)}(z-x_t)  ~\bar \chi_{(\htt, aux)}(w-x_\htt) 
~\partial_z \partial_w G^{z \bz}_{(N,M)}(z, \bz; w , \bar w)
}
\Bigg]
\end{align*}
\begin{align*}
\phantom{\langle V_{N+L} |=}&
\nonumber\\
\prod_{i,j=1}^L
\Bigg[
&
e^{
\frac{1}{\alpha'^2}
\oint_{z=x_i} \frac{d z}{ 2\pi i}
\oint_{w=x_j} \frac{d w}{ 2\pi i}
\partial_z \bar \chi_{(i, aux)}(z-x_t)  ~\partial_w \bar \chi_{(j, aux)}(w-x_j) 
  ~G^{z z}_{(N,M)}(z, \bz; w , \bar w)
}
\nonumber\\
&
e^{
\frac{1}{\alpha'^2}
\oint_{z=x_i} \frac{d z}{ 2\pi i}
\oint_{w=x_j} \frac{d w}{ 2\pi i}
\partial_z \chi_{(i, aux)}(z-x_t)   ~\partial_w \chi_{(j, aux)}(w-x_j) 
 ~G^{\bz \bz}_{(N,M)}(z, \bz; w , \bar w)
}
\nonumber\\
&
e^{
\frac{2}{\alpha'^2}
\oint_{z=x_i} \frac{d z}{ 2\pi i}
\oint_{w=x_j} \frac{d w}{ 2\pi i}
 \partial_z \bar \chi_{(i, aux)}(z-x_t)   ~\partial_w \bar \chi_{(j, aux)}(w-x_j) 
 ~G^{z \bz}_{(N,M)}(z, \bz; w , \bar w)
}
\Bigg]
\end{align*}
\begin{align}
\phantom{\langle V_{N+L} |=}&
\nonumber\\
\prod_{i=1}^L \prod_{t=1}^N
\Bigg[
&
e^{
\frac{1}{\alpha'^2}
\oint_{z=x_i} \frac{d z}{ 2\pi i}
\oint_{w=x_t} \frac{d w}{ 2\pi i}
\partial_z \bar \chi_{(i, aux)}(z-x_t)  ~\bar \chi_{(t, aux)}(w-x_t) 
~\partial_w  G^{z z}_{(N,M)}(z, \bz; w , \bar w)
}
\nonumber\\
&
e^{
\frac{1}{\alpha'^2}
\oint_{z=x_i} \frac{d z}{ 2\pi i}
\oint_{w=x_t} \frac{d w}{ 2\pi i}
\partial_z \chi_{(i, aux)}(z-x_t)  ~\chi_{(t, aux)}(w-x_t) 
~ \partial_w  G^{\bz \bz}_{(N,M)}(z, \bz; w , \bar w)
}
\nonumber\\
&
e^{
\frac{1}{\alpha'^2}
\oint_{z=x_i} \frac{d z}{ 2\pi i}
\oint_{w=x_t} \frac{d w}{ 2\pi i}
 \partial_z \bar \chi_{(i, aux)}(z-x_t)  ~\chi_{(t,  aux)}(w-x_t) 
~\partial_w G^{z \bz}_{(N,M)}(z, \bz; w , \bar w)
}
\nonumber\\
&
e^{
\frac{1}{\alpha'^2}
\oint_{z=x_i} \frac{d z}{ 2\pi i}
\oint_{w=x_t} \frac{d w}{ 2\pi i}
 \partial_z \chi_{(i, aux)}(z-x_t)  ~\bar \chi_{(t,  aux)}(w-x_t) 
~\partial_w G^{\bz z}_{(N,M)}(z, \bz; w , \bar w)
}
\Bigg]
\label{traditional-reggeon-excited-twists+bou}
\end{align}
\COMMENTOO{How to define the Green function on all the complex plane?}
where we have used the doubled fields $\chi_I(z)$ defined as in
eq. \ref{loc-calX-calXbar} for the twisted case and in the usual way
for the untwisted one.
Notice that the terms $t=\htt$ and $i=j$ 
must be regularized by properly computing the
two contour integrals as discussed in \cite{Petersen:1988cf}

\noindent {\large {\bf Acknowledgments}}

The author thanks R. Richter for useful comments on a preliminary draft.

This work is supported in part by the Compagnia di San Paolo contract
"Modern Application in String Theory" (MAST) TO-Call3-2012-0088


\appendix


\section{Details and useful formula for the untwisted string and $N=2$
case}
\label{app:Delta}

We start considering the untwisted string
associated to the $D_t$ brane, i.e. the string with both ends on $D_t$. 
This has boundary conditions
\begin{align}
Re( e^{-i \pi \alpha_{t}} \dy X_{loc} |_{y=0} ) 
=
Im( e^{-i \pi \alpha_{t}} X_{loc} |_{y=0}) -g_{t} 
=0
\end{align}
in the upper half plane $H$ and has  the expansion 
\begin{align}
X_{loc}(\ul,\bul)
&=
e^{i \pi \alpha_{t}}
\Bigg[
x^{\hat 1} + i g_t
-i~2\alpha'  p^{\hat 1} ~\ln |\ul|
&+
i\oh \sqrt{2\alpha'}
\sum_{n=1}^\infty
\left[ 
\frac{\bar \alpha_{(t)n} }{ n } \ul^{-n}
-
\frac{\alpha_{(t)n}^\dagger}{ n } \ul^{n}
\right]
\nonumber\\
&
&+
i \oh \sqrt{2\alpha'}
\sum_{n=1}^\infty 
\left[
-
\frac{\bar \alpha_{(t)n}^\dagger }{ n } \bul^{n}
+
\frac{\alpha_{(t)n} }{  n} \bul^{-n}
\right]
\Bigg]
\nonumber\\
\bar X_{loc}(\ul,\bul)
&=
e^{-i \pi \alpha_{t}}\Bigg[
x^{\hat 1} - i g_t
-2\alpha' i p^{\hat 1} ~\ln |\ul|
&+
i \oh \sqrt{2\alpha'}
\sum_{n=1}^\infty 
\left[
-
\frac{\bar \alpha_{(t)n}^\dagger }{ n } \ul^{n}
+
\frac{\alpha_{(t)n} }{ n } \ul^{-n}
\right]
\nonumber\\
&
&+
i \oh \sqrt{2\alpha'}
\sum_{n=1}^\infty
\left[ 
\frac{\bar \alpha_{(t)n} }{ n } \bul^{-n)}
-
\frac{\alpha_{(t)n}^\dagger }{ n } \bul^{n}
\right]
\Bigg]
\label{loc-exp-UX}
\end{align}
We have the non trivial commutators
\begin{equation}
[ \alpha_{(t) n}, \alpha_{(t) m}^\dagger]
= n \delta_{m,n}
,~~~~
[ \bar\alpha_{(t) n}, \bar\alpha_{(t) m}^\dagger]
=n \delta_{m,n}
,~~~~
[x^{\hat 1}, p^{\hat 1}] = i
\end{equation}
where $x^{\hat 1}, p^{\hat 1}$ are the zero mode position and momentum
of string $X^{\hat 1}=\frac{1}{\sqrt{2}}(e^{-i \pi \alpha_{t}} X+
e^{+i \pi \alpha_{t}} \bar X)$ with $N N$ boundary condition.
The vacuum is defined in the usual way by
\begin{equation}
\alpha_{(t) n} |0_t\rangle 
=\bar \alpha_{(t) n}|0_t\rangle 
= p^{\hat 1} |0_t\rangle
=0
~~~~ n\ge 1
\label{untwisted-vacuum}
\end{equation}
even if care must be taken in order to deal with the $D D$ zero modes ***\cite{}.
Then we can compute the untwisted Green functions
\begin{align}
G^{z z}_{U(t)}(\ul,\bul; \vl,\bvl)
&= 
[X\loc^{(+)}(\ul,\bul), X\loc^{(-)}(\vl,\bvl)]
=\ke^2 \ln |\ul-\bvl|^2
\nonumber\\
G^{\bz \bz}_{U(t)}(\ul,\bul; \vl,\bvl)
&= 
[\bar X\loc^{(+)}(\ul,\bu), \bar X\loc^{(-)}(\vl,\bvl)]
=\kbe^2 \ln |\ul-\bvl|^2
\nonumber\\
G^{z \bz}_{U(t)}(\ul,\bul; \vl,\bvl)
&= 
[X\loc^{(+)}(\ul,\bul), \bar X\loc^{(-)}(\vl,\bvl)]
=\ke \kbe \ln |\ul-\vl|^2
\label{untwisted-green-function}
\end{align}
where 
$\kei{t}= -i\oh\sqrt{2\alpha'} e^{i \pi \alpha_{t}}$ and
$\kbei{t}= -i\oh\sqrt{2\alpha'} e^{-i \pi \alpha_{t}}$ as in the main text.
Notice that $G^{z \bz}_{U(t)}$ does not feel that the brane is rotated
while both $G^{z z}_{U(t)}$ and $G^{\bz \bz}_{U(t)}$ do because of the
phase in $\kei{t}^2$ and $\kbei{t}^2$.

In a similar way we can compute the $N=2$ twisted Green functions
\begin{align}
%
%
G^{z z}_{N=2}(u,\bu; v,\bv;\{ 0,\epsilon; \infty, \bar\epsilon\})
&= 
[X\loc^{(+)}(\ul,\bul), X\loc^{(-)}(\vl,\bvl)]
\nonumber\\
&=
-\kei{t}^2\left[
\frac{1}{\epsilon} \left(\frac{\vl }{ \bul}\right)^\epsilon
{}_2F_1(1,\epsilon;1+\epsilon; \frac{\vl }{ \bul})
+
\frac{1}{\bep} \left(\frac{\bvl }{ \ul}\right)^\bep
{}_2F_1(1,\bep;1+\bep; \frac{\bvl }{ \ul})
\right]
\nonumber\\
%
%
G^{\bz \bz}_{N=2}(u,\bu; v,\bv; \{ 0,\epsilon; \infty, \bar\epsilon\})
&= 
[\bar X\loc^{(+)}(\ul,\bul), \bar X\loc^{(-)}(\vl,\bvl)]
\nonumber\\
&=
-\kbei{t}^2\left[
\frac{1}{\epsilon} \left(\frac{\bvl }{ \ul}\right)^\epsilon
{}_2F_1(1,\epsilon;1+\epsilon; \frac{\bvl }{ \ul})
+
\frac{1}{\bep} \left(\frac{\vl }{ \bul}\right)^\bep
{}_2F_1(1,\bep;1+\bep; \frac{\vl }{ \bul})
\right]
\nonumber\\
G^{z \bz}_{N=2}(u,\bu; v,\bv; \{ 0,\epsilon; \infty, \bar\epsilon\})
&= 
[ X\loc^{(+)}(\ul,\bul), \bar X\loc^{(-)}(\vl,\bvl)]
\nonumber\\
&=
G^{\bz z}_{N=2}(v,\bv; u,\bu;\{ 0,\epsilon; \infty, \bar\epsilon\})
=
G^{\bz z}_{N=2}(u,\bu; v,\bv;\{ 0,\epsilon; \infty, \bar\epsilon\})
\nonumber\\
&=
-\kei{t}\kbei{t}\left[
\frac{1}{\epsilon} \left(\frac{\bvl }{ \bul}\right)^\epsilon
{}_2F_1(1,\epsilon;1+\epsilon; \frac{\bvl }{ \bul})
+
\frac{1}{\bep} \left(\frac{\vl }{ \ul}\right)^\bep
{}_2F_1(1,\bep;1+\bep; \frac{\vl }{ \ul})
\right]
\label{Green-N=2}
\end{align}
where we have used
\begin{equation}
{}_2F_1(1,\epsilon;1+\epsilon; x)
=
\sum_{n=0}^\infty \frac{\epsilon}{n+\epsilon} x^n
~~~~
|x|<1
\end{equation}
as follows from the general expression for the hypergeometric function
${}_2F_1(a, b; c; x)= \sum_{n=0}^\infty \frac{ (a)_n (b)_n}{ n! (c)_n}x^n$ 
with $(a)_n= \Gamma(a+n) /\Gamma(a)$ the Pochhammer symbol.
They have the following symmetry properties
\begin{align}
G^{I J}_{N=2}(u,\bu; v,\bv; \{ 0,\epsilon; \infty, \bar \epsilon\})
=
G^{J I}_{N=2}( v,\bv; u,\bu;\{ 0,\epsilon; \infty, \bar \epsilon\})
=
G^{I J}_{N=2}( v,\bv; u,\bu;\{ 0,\bar\epsilon; \infty, \epsilon\})
\label{Green-N=2-symmetry}
\end{align}
which follow from the hypergeometric transformation properties in
particular
${}_2 F_1(1, \epsilon; 1+\epsilon; x) =  
\frac{\epsilon }{\bar \epsilon x}
{}_2F_1(1, \bar\epsilon; 1+\bar\epsilon; 1/x)$.

For future use f.x. in eq. (\ref{more-explicit-Delta-t}) we notice that
\begin{align}
g(u,v; \epsilon)
&= 
(\kei{t}\kbei{t})^{-1}
\partial_u \partial_v G^{z \bz}=
 z^{-\bar \epsilon } w ^{-\epsilon}
 \frac{\bar \epsilon z + \epsilon w}{(u-v)^2}
\nonumber\\
l(u,v; \epsilon)&= \partial_u \partial_v G^{z z}= 0
\nonumber\\
h(u,v; \epsilon)&= \partial_u \partial_v G^{\bz \bz}= 0
\end{align}

In order to write $\Delta^{ I J}$, the regularized Green function,  
in a more compact and transparent way we introduce the quantity
\begin{equation}
\DDelta(x; \epsilon)
= 
\sum_{n=0}^\infty \frac{x^{n+\epsilon}}{n+\epsilon}
-\sum_{n=1}^\infty \frac{x^{n}}{n}
=
\frac{x^{\epsilon}}{\epsilon} {}_2F_1(1,\epsilon;1+\epsilon; x)
+\log(1-x)
~~
|arg(x)|<\pi
\end{equation}
which can be expanded around $x=1$ as
\begin{equation}
\DDelta(x; \epsilon)
= 
\psi(1)-\psi(\epsilon)
-\sum_{n=1}^\infty \comb{\epsilon-1}{n} \frac{ (x-1)^n}{n}
\end{equation}
where $\psi(x)= d \log \Gamma(x) / d x $ is the digamma function.
All of this expansion but the constant term can be easily obtained by
computing $\DDelta'(x; \epsilon)$ and the integrating on $x$.
To get the constant term is necessary to use the  
\begin{align}
{}_2F_1(a, b; a+b; x)=
&
\frac{\Gamma(a+b)}{\Gamma(a) \Gamma(b)}
\sum_{n=0}^\infty
 \frac{ (a)_n (b)_n}{ (n!)^2 }
\left[2 \psi(n+1) -\psi(a+n) - \psi(b+n) - \log(1-x) \right]
(1-x)^n
\nonumber\\
&~~~~
|arg(1-x)|<\pi, ~~|1-x|<1
\end{align}

We can now write the boundary $\Delta\bou^{ I J}$  defined as
\begin{equation}
\Delta\bou^{I J}(\xii1; \xii2; \epsilon)
=
\left\{\begin{array}{c l}
\scriptstyle
G^{I J}_{N=2}(\xii1+i0^+,\xii1-i0^+; \xii2+i0^+,\xii2-i0^+; 
\{ 0,\epsilon; \infty, \bar\epsilon\}) 
&
\scriptstyle
- G^{I J}_{U(t)}(\xii1+i0^+,\xii1-i0^+; \xii2+i0^+,\xii2-i0^+)
\\
& \xii1,\xii2>0
\\
\scriptstyle
G^{I J}_{N=2}(\xii1+i0^+,\xii1-i0^+; \xii2+i0^+,\xii2-i0^+; 
\{ 0,\epsilon; \infty, \bar\epsilon\}) 
&
\scriptstyle
- G^{I J}_{U(t+1)}(\xii1+i0^+,\xii1-i0^+; \xii2+i0^+,\xii2-i0^+)
\\
& \xii1,\xii2<0
\end{array}\right. 
\label{Delta-N=2-bou}
\end{equation}
as
\begin{align}
\Delta\bou^{z z}(\xii1; \xii2; \epsilon)
&=
\left\{\begin{array}{c c}
e^{i 2\pi \alpha_t}  ~\Delta_{bou}(x_1,x_2) 
& \xii1,\xii2>0
\\
e^{i 2\pi \alpha_{t+1}} ~\Delta_{bou}(x_1,x_2) 
& \xii1,\xii2<0
\end{array}\right. 
\nonumber\\
\Delta\bou^{\bz \bz}(\xii1; \xii2; \epsilon)
&=
\left\{\begin{array}{c c}
e^{-i 2\pi \alpha_t} ~ \Delta_{bou}(x_1,x_2) 
& \xii1,\xii2>0
\\
e^{-i 2\pi \alpha_{t+1}} ~\Delta_{bou}(x_1,x_2) 
& \xii1,\xii2<0
\end{array}\right. 
\nonumber\\
\Delta\bou^{z \bz}(\xii1; \xii2; \epsilon)
&=
 ~\Delta_{bou}(x_1,x_2)
~~~
\label{expl-Delta-expr}
\end{align}
where we have defined the common factor
\begin{equation}
\mdda
\Delta_{bou}(x_1,x_2)
=
\DDelta\left(\frac{\xii1}{\xii2}; \epsilon \right)
+\DDelta\left(\frac{\xii1}{\xii2}; \bar\epsilon \right)
+ \log (\xii1 ^2)
\sim
\log (\xii1 ^2)
+ R^2(\epsilon_t)
\label{Delta-N=2}
\end{equation}
We have to consider the cases $\xii1,\xii2>0$ and
$\xii1,\xii2<0$  because, for example,
$\frac{\vl }{ \bul}= \frac{\xii2 }{\xii1}$ when $\xii1,\xii2>0$ and
$\frac{\vl }{ \bul}= \frac{|\xii2| }{|\xii1|} e^{i 2\pi}$ when $\xii1,\xii2<0$
and this gives raise to different phases.
This is issue is not present for $\Delta\bou^{z \bz}$ because
$\frac{\vl }{ \ul}=\frac{\xii2 }{\xii1}$ independently on $\xii1,\xii2>0$ or
$\xii1,\xii2<0$.
Notice that these phases are fundamental for projecting an arbitrary
momentum $(k,\bar k)$ in the direction parallel to the $D_t$ brane as
shown explicitly in section \ref{sect:boundary_corr}.

In a similar way we define the regularized Green function with the
twist $\sigma_{\epsilon,f}$  at $\xl=0$ and the
anti-twist $ \sigma_{\bar \epsilon,f}$ in $\xl=\infty$ used in chiral
operators correlators. In particular because of the fact that there
are at least $\partial_u\partial_v$ the only piece which contributes is
\begin{align}
\Delta_c^{z \bz}(\ul,\bul; \vl,\bvl; \epsilon)
&=
G^{z \bz}_{N=2}(\ul,\bul; \vl,\bvl; \{ 0,\epsilon; \infty, \bar\epsilon\} ) 
-
G^{z \bz}_{U(t)}((\ul,\bul; \vl,\bvl)
\nonumber\\
&=
\madd \left[ 
\DDelta\left(\frac{\bvl}{\bul}; \epsilon \right)
+\DDelta\left(\frac{\vl}{\ul}; \bar\epsilon \right)
+ \log |\ul|^2\right]
\end{align}
which is again independent on the phase of $\ke_t$.

\section{Classical solutions}
\label{app:classical_soultions}
In this appendix we would like to summarize the results of the
previous work \cite{Pesando:2012cx} (see also 
\cite{Cvetic:2003ch} and \cite{Abel:2003vv}).
Defined the anharmonic ratio for a complex variable $z\in\C$ to be
\begin{equation}
\omega_z = \frac{z-x_2}{z-x_N} \frac{x_1-x_N}{x_1-x_2}
\label{anh-ratio}
\end{equation}
so that $\omega_1=1$, $\omega_2=0$ and $\omega_N=-\infty$, 
a basis of the derivatives of zero modes of the two dimensional
laplacian satisfying the boundary conditions (\ref{path-integral-global-boundary-upper})
is
\begin{align}
\dz_\omega\cX^{(n)}(\omega_z)
&=
~\prod_{t=1}^{N-1} (\omega_z-\omega_t)^{-(1-\epsilon_t)} ~\omega_z^n
,~~~~ 0\le n \le N-M-2
\nonumber\\
\dz_\omega\cbX^{(r)}(\omega_z)
&=
~\prod_{t=1}^{N-1} (\omega_z-\omega_t)^{-\epsilon_t}~\omega_z^r
,~~~~ 0\le r \le M-2
\label{basis-class-sols}
\end{align}
so that we can write the classical solution 
\begin{align}
X_{cl}(u,\bu;\{x_t,\epsilon_t,f_t\})
=
f_N
&+
\sum_{n=0}^{N-M-2} a_n(\omega_t) 
\int_{-\infty; \omega\in H^+}^{\omega_u} d \omega~
\dz_\omega\cX^{(n)}(\omega)
\nonumber\\
&+
\sum_{r=0}^{M-2} b_r(\omega_t) 
\left[
\int_{-\infty; \omega\in H^+}^{\omega_u} d \omega~
\dz_\omega\cbX^{(r)}(\omega)
\right]^*
\end{align}
which satisfies also the global constraints (\ref{global-boundary-points}).
The real coefficients $e^{-i\pi \alpha_1} a_n(\omega_t)$ and 
$e^{-i\pi \alpha_1} b_r(\omega_t)$ are fixed by 
the constraints
\begin{equation}
X_{cl}(x_{t+1}, \bar x_{t+1})- X_{cl}(x_{t}, \bar x_{t})
= f_{t+1}-f_{t}
~~~~
t=2, \dots N-1
\end{equation}
The ``reality'' of the coefficients can be easily seen once we introduce 
the following functions which are real on the real axis
\begin{align}
\dz_\omega\cXX^{(n)}(\omega)
&=
~\prod_{t=1}^{N-1} |\omega-\omega_t|^{-\bar \epsilon_t} ~\omega^n
,~~~~ 0\le n \le N-M-2
\nonumber\\
\dz_\omega\bcXX^{(r)}(\omega)
&=
~\prod_{t=1}^{N-1} |\omega-\omega_t|^{-\epsilon_t}~\omega^r
,~~~~ 0\le r \le M-2
\label{basis-real-class-sols}
\end{align}
and their integrals (which can be expressed using the type D 
Lauricella functions)
\begin{align}
I^{(N)}_{t,n}(\bar \epsilon)
&=
\int_{\omega_{t+1}}^{\omega_t} d \omega~ \dz_\omega\cXX^{(n)}(\omega)
\nonumber\\
I^{(N)}_{t,r}(\epsilon)
&=
\int_{\omega_{t+1}}^{\omega_t} d \omega~ \dz_\omega\bcXX^{(r)}(\omega)
\end{align}
so that we can write the constraints as
\begin{align}
(-1)^{t-1}
\sum_{n=0}^{N-M-2} a_n I^{(N)}_{t,n}(\bar \epsilon)
+
\sum_{r=0}^{M-2} b_r I^{(N)}_{t,r}(\epsilon)
= 
e^{-i \pi \alpha_1} \left[ e^{-i \pi \alpha_{t}} (f_t - f_{t+1}) \right]
~~~~
\end{align}
where the quantity between square brackets on the right hand side is real.
Finally the classical action can be written as
\begin{align}
S_{cl}
=&
\frac{1}{8 \pi \alpha'}\Bigg[
\sum_{n,m=0}^{N-M-2}  (e^{-i\pi\alpha_1}a_n)~ (e^{-i\pi\alpha_1}a_m)
\sum_{t=1}^{N-2}
\sum_{\htt=t+1}^{N-1}
\sin\left(\pi \sum_{u=t+1}^\htt \bar \epsilon_u \right)
 I^{(N)}_{t,n}(\bar\epsilon) 
I^{(N)}_{\htt,m}(\bar\epsilon)
\nonumber\\
+&
\sum_{r,s=0}^{M-2}  (e^{-i\pi\alpha_1}b_r) (e^{-i\pi\alpha_1}b_s)
\sum_{t=1}^{N-2}
\sum_{\htt=t+1}^{N-1}
\sin\left(\pi \sum_{u=t+1}^\htt \epsilon_u \right)
 I^{(N)}_{t,r}(\epsilon) 
I^{(N)}_{\htt,s}(\epsilon)
\Bigg]
\end{align}
\section{Green function}
\label{app:Green_functions}
In (\cite{Pesando:2011ce}) following previous works on the subjects we
defined the derivatives of the Green function on the whole complex
plane $\C$ using the doubling trick as
\begin{align}
g_{(N,M)}(z, w; \{x_t\} )
&=
\mdda
\frac{
\langle 
\dz \cX_q(z) \dw \cbX_q(w) \sigma_{\epsilon_1,f}(x_1) \dots \sigma_{\epsilon_N,f}(x_N)
\rangle
}{
\langle 
\sigma_{\epsilon_1,f}(x_1) \dots \sigma_{\epsilon_N,f}(x_N)
\rangle
}
\nonumber\\
h_{(N,M)}(z, w; \{x_t\} )
&=
\mdda
\frac{
\langle 
\dz \cbX_q(z) \dw \cbX_q(w) \sigma_{\epsilon_1,f}(x_1) \dots \sigma_{\epsilon_N,f}(x_N)
\rangle
}{
\langle 
\sigma_{\epsilon_1,f}(x_1) \dots \sigma_{\epsilon_N,f}(x_N)
\rangle
}
\nonumber\\
l_{(N,M)}(z, w; \{x_t\} )
&=
\mdda
\frac{
\langle 
\dz \cX_q(z) \dw \cX_q(w) \sigma_{\epsilon_1,f}(x_1) \dots \sigma_{\epsilon_N,f}(x_N)
\rangle
}{
\langle 
\sigma_{\epsilon_1,f}(x_1) \dots \sigma_{\epsilon_N,f}(x_N)
\rangle
}
\end{align}
with expansions
\begin{align}
g_{(N,M)}(z, w; \{x_t\} )
&=
\frac{\partial \omega_z}{\partial z}
\frac{\partial \omega_w}{\partial w}
\frac{1}{(\omega_z-\omega_w)^2}
\sum_{\hat n=0}^{N-M}\sum_{\hat s=0}^{M} a_{\hat n \hat s}(\omega_{t\ne 1,2, N})
~\dz_\omega \cX^{(\hat n)}(\omega_z)
~\dz_\omega \cbX^{(\hat s)}(\omega_w)
\nonumber\\
&=
\frac{1}{(z-w)^2}
\sum_{\hat n=0}^{N-M}\sum_{\hat s=0}^{M} a_{\hat n \hat s}(\omega_{t\ne 1,2, N})
~\dz_\omega \cX^{(\hat n)}(\omega_z)
~\dz_\omega \cbX^{(\hat s)}(\omega_w)
\nonumber\\
h_{(N,M)}(z, w; \{x_t\} )
&=
e^{-i 2\pi \alpha_1}
\frac{\partial \omega_z}{\partial z}
\frac{\partial \omega_w}{\partial w}
\sum_{r,s=0}^{M-2} b_{r s}(\omega_{t\ne 1,2,N}) 
~\dz_\omega \cbX^{(r)}(\omega_z)
~\dz_\omega \cbX^{(s)}(\omega_w)
\nonumber\\
l_{(N,M)}(z, w; \{x_t\} )
&=
e^{i 2\pi \alpha_1}
\frac{\partial \omega_z}{\partial z}
\frac{\partial \omega_w}{\partial w}
\sum_{n,m=0}^{N-M-2} c_{n m}(\omega_{t\ne 1,2, N}) 
~\dz_\omega \cX^{(n)}(\omega_z)
~\dz_\omega \cX^{(m)}(\omega_w)
\end{align}
where we have used the anharmonic ratio as defined in eq. (\ref{anh-ratio})
and
we have extended the range of definition from $N-M-2$ to $N-M$
for $\cX^{(n)}$
and from $M-2$ to $M$ for $ \cbX^{(s)}$ in order to write in a more compact
way $g_{(N,M)}$ and have therefore used hatted indexes.
The previous quantities are subject to the constraints
\begin{align}
&
\left[
\prod_{t= 1}^{N-1}
\frac{(\omega_z-\omega_t)^{\epsilon_t-1}}{(\omega_w-\omega_t)^{\epsilon_t}}
\sum_{\hat n=0}^{N-M}\sum_{\hat  s=0}^{M} a_{\hat n \hat  s}(\omega_{t\ne1,2,N}) 
\omega_z^{\hat n} \omega_w^{\hat s}
\right]
\Big |_{\omega_w=\omega_z}=1
\nonumber\\
&\frac{\partial}{\partial \omega_w}\Big|_{\omega_w=\omega_z}
\left[
\prod_{t= 1}^{N-1}
\frac{(\omega_z-\omega_t)^{\epsilon_t-1}}{(\omega_w-\omega_t)^{\epsilon_t}}
\sum_{\hat n=0}^{N-M}\sum_{\hat s=0}^{M} a_{\hat n \hat s}(\omega_{t\ne1,2,N}) 
\omega_z^{\hat n} \omega_w^{\hat s}
\right]
=0
\label{constr-on-g}
\end{align}
because $g$ must have only a double pole with coefficient $1$
and
\begin{align}
&\int_{x_{t+1}}^{x_{t}} d x~ g_{(N,M)}(x+i 0^+,w)
+ e^{i 2\pi \alpha_1} h_{(N,M)}(x-i 0^+,w)=0
\nonumber\\
&\int_{x_{t+1}}^{x_{t}} d x~   l_{(N,M)}(z,x-i 0^+) 
+ e^{i 2\pi  \alpha_1} g_{(N,M)}(z,x+i 0^+)=0
\label{glob-consts-gh-gl}
\end{align}
due to the boundary conditions.
This set of equations is an overdetermined but consistent one as discussed
in \cite{Pesando:2012cx}.
Using the previous quantities we wrote that the Green function in
presence of $N$ twists is given by
\begin{align}
\mdda G^{ z \bz}_{(N,M)}(u, \bu; v, \bv; \{x_t\})
&=
\int_{x_\ttt1; u'\in H}^u d u' \int_{x_\ttt2; v'\in H}^v d v' ~g_{(N,M)}(u', v';  \{x_t\}))
\nonumber\\
&+
e^{-i 2\pi  \alpha_1}
\int_{x_\ttt1; u\in H}^u d u' \int_{x_\ttt2; \bv' \in H^-}^\bv d \bv' 
~l_{(N,M)}(u', \bv';  \{x_t\}))
\nonumber\\
&+
e^{i 2\pi  \alpha_1}
\int_{x_\ttt1; \bu'\in H^-}^\bu d \bu' \int_{x_\ttt2; v'\in H}^v d v' 
~h_{(N,M)}(\bu', v';  \{x_t\}))
\nonumber\\
&+
\int_{x_\ttt1; \bu\in H^-}^\bu d \bu' \int_{x_\ttt2; \bv' \in H^-}^\bv d \bv' ~g_{(N,M)}(\bv', \bu';  \{x_t\}))
\end{align}
and
\begin{align}
\mdda G^{z z}_{(N,M)}(u, \bu; v, \bv; \{x_t\})
&=
\int_{x_\ttt1; u'\in H}^u d u' \int_{x_\ttt2; v'\in H}^v d v' ~l_{(N,M)}(u', v';  \{x_t\}))
\nonumber\\
&+
e^{i 2\pi  \alpha_1}
\int_{x_\ttt1; u\in H}^u d u' \int_{x_\ttt2; \bv' \in H^-}^\bv d \bv' ~g_{(N,M)}(u', \bv';  \{x_t\}))
\nonumber\\
&+
e^{i 2\pi  \alpha_1}
\int_{x_\ttt1; \bu'\in H^-}^\bu d \bu' \int_{x_\ttt2; v'\in H}^v d v' ~g_{(N,M)}(v', \bu';  \{x_t\}))
\nonumber\\
&+
e^{i 4\pi  \alpha_1}
\int_{x_\ttt1; \bu\in H^-}^\bu d \bu' \int_{x_\ttt2; \bv' \in H^-}^\bv d \bv' ~h_{(N,M)}(\bu', \bv';  \{x_t\}))
\end{align}
and
\begin{align}
\mdda G^{ \bz \bz}_{(N,M)}(u, \bu; v, \bv; \{x_t\})
&=
\int_{x_\ttt1; u'\in H}^u d u' \int_{x_\ttt2; v'\in H}^v d v' ~h_{(N,M)}(u', v';  \{x_t\}))
\nonumber\\
&+
e^{-i 2\pi  \alpha_1}
\int_{x_\ttt1; u\in H}^u d u' \int_{x_\ttt2; \bv' \in H^-}^\bv d \bv'
~g_{(N,M)}(\bv', u';  \{x_t\}))
\nonumber\\
&+
e^{-i 2\pi  \alpha_1}
\int_{x_\ttt1; \bu'\in H^-}^\bu d \bu' \int_{x_\ttt2; v'\in H}^v d v'
~g_{(N,M)}(\bu', v';  \{x_t\}))
\nonumber\\
&+
e^{-i 4\pi  \alpha_1}
\int_{x_\ttt1; \bu\in H^-}^\bu d \bu' \int_{x_\ttt2; \bv' \in H^-}^\bv d \bv' 
~l_{(N,M)}(\bu', \bv';  \{x_t\}))
\end{align}
where the normalization is needed in order to match the
singularity of the untwisted Green function (\ref{untwisted-green-function})
\footnote{
This is obvious for $G^{ z \bz}_{(N,M)}$ but for $G^{z z}_{(N,M)}$
there would seem to be a mismatch of phases since 
$\partial_u \partial_\bv G^{z z}_{U(t)}= e^{i 2\pi \alpha_t}
\mdda \frac{1}{(u-\bv)^2}$ has a phase which depends on
the brane while naively $\partial_u \partial_\bv G^{z z}_{(N,M)}= 
e^{i 2\pi \alpha_1} \mdda g_{(N,M)}(u,\bv)
\sim
e^{i 2\pi \alpha_1} \mdda \frac{1}{(u-\bv)^2}$.
The issue is solved by noticing that the singularity is only there
when $u,\bv \rightarrow x\in\R$ and therefore 
$ g_{(N,M)}(x+i\delta_1,x-i\delta_2)$ ($\delta_{1,2}>0$) has the
singularity with the required phase, explicitly for $x_t< x < x_{t-1}$
$g_{(N,M)}(x+i\delta_1,x-i\delta_2) = e^{i 2\pi (\alpha_t-\alpha_1)}
g_{(N,M)}(x+i\delta_1,x+i\delta_2)$
since
$
\prod_{u= 1}^{N-1} (\omega_{x-i\delta}-\omega_u)^{-\epsilon_u}
=  e^{i 2\pi (\alpha_t-\alpha_1)}
\prod_{u= 1}^{N-1} (\omega_{x+i\delta}-\omega_u)^{-\epsilon_u}
$.
}.
The arbitrariness of the lower integration limit is due to the
constraints (\ref{glob-consts-gh-gl}) 
which allow to change $x_\ttt1\rightarrow x_\ttt3$ for whichever
$\ttt3$ ans similarly for $x_\ttt2$.
We would now justify this result since it is important in computing
correlators involving momenta.
The reason why we fixed the lower integration limit to one of the
twist location is because we want 
\begin{equation}
G^{I J}(x_\ttt1, x_\ttt1; v, \bv; \{x_t\})
=
G^{I J}(u, \bu; x_\ttt1, x_\ttt1; \{x_t\})=0
\end{equation}
as follows from the boundary condition (\ref{global-boundary-points})
in the case of the quantum fluctuation where $f_t\rightarrow 0$.

\section{Boundary Green functions and their regularized version
  $\Delta_{(N,M), bou}$}
\label{app:boundray_green}
For the computation of the boundary correlators it is interesting and
useful to notice that all the components of the Green function are
proportional, analogously to eq.s (\ref{expl-Delta-expr}) we have
\begin{align}
G^{z z}_{(N,M), bou}(x_1,x_2) &= 
e^{i \pi(\alpha_{t_1}+ \alpha_{t_2})} G_{(N,M), bou}(x_1,x_2)
\nonumber\\
G^{\bz \bz}_{(N,M), bou}(x_1,x_2) &= 
e^{-i \pi(\alpha_{t_1}+ \alpha_{t_2})} G_{(N,M), bou}(x_1,x_2)
\nonumber\\
G^{z \bz}_{(N,M), bou}(x_1,x_2) &= 
e^{i \pi(\alpha_{t_1}- \alpha_{t_2})} G_{(N,M), bou}(x_1,x_2)
\nonumber\\
G^{\bz z}_{(N,M), bou}(x_1,x_2) &= 
e^{i \pi(-\alpha_{t_1}+ \alpha_{t_2})} G_{(N,M), bou}(x_1,x_2)
\label{GIJNM-bou-GNM-bou}
\end{align}
where the point $x_1$ is on the brane $D_{t_1}$, i.e. $x_{t_1}<
x_1<x_{t_1-1}$ and similarly for
$x_2$ and we have defined the common real 
symmetric function $G_{(N,M),  bou}(x_1,x_2)=G_{(N,M), bou}(x_2,x_1)$ 
\footnote{The easiest way to verify that it is symmetric is to notice
  that $G^{z z}_{(N,M), bou}(x_1,x_2)$ is symmetric.
} to be
\begin{align}
\mdda&G_{(N,M), bou}(x_1,x_2)
=
\nonumber\\
&
e^{i \pi (\bar N_\ttt1 + \bar N_\ttt2) }
\sum_{n,m=0}^{N-M-2} c_{n m}(\omega_{t\ne 1,2,N}) 
~ \int _{\omega_{t_{1}}}^{\omega_{x_1}} d \omega~ \dz_\omega
  \cXX^{(n)}(\omega)
~ \int _{\omega_{t_{2}}}^{\omega_{x_2}} d \omega~ \dz_\omega \cXX^{(m)}(\omega)
\nonumber\\
+&
e^{i \pi ( N_\ttt1 +  N_\ttt2) }
\sum_{r,s=0}^{M-2} b_{r s}(\omega_{t\ne 1,2,N}) 
~ \int _{\omega_{t_{1}}}^{\omega_{x_1}} d \omega~ \dz_\omega \bcXX^{(r)}(\omega)
~ \int _{\omega_{t_{2}}}^{\omega_{x_2}} d \omega~ \dz_\omega \bcXX^{(s)}(\omega)
\nonumber\\
+&
e^{i \pi (\bar N_\ttt1 +  N_\ttt2) }
\sum_{\hat n=0}^{N-M} \sum_{\hat s=0}^{M} a_{\hat n \hat s}(\omega_{t\ne 1,2,N}) 
~ \int _{\omega_{t_{1}}}^{\omega_{x_1}} d \omega_z~
~ \int _{\omega_{t_{2}}}^{\omega_{x_2}}  d \omega_w 
\frac{  
\dz_\omega \cXX^{(\hat n)}(\omega_z)
\dz_\omega \bcXX^{(\hat s)}(\omega_w)
}{(\omega_z-\omega_w)^2} 
\nonumber\\
+&
e^{i \pi (N_\ttt1 +\bar  N_\ttt2) }
\sum_{\hat n=0}^{N-M} \sum_{\hat s=0}^{M} a_{n s}(\omega_{t\ne 1,2,N}) 
~ \int _{\omega_{t_{1}}}^{\omega_{x_1}} d \omega_z~
~ \int _{\omega_{t_{2}}}^{\omega_{x_2}}  d \omega_w 
\frac{  
\dz_\omega \cXX^{(\hat n)}(\omega_w)
\dz_\omega \bcXX^{(\hat s)}(\omega_z)
}{(\omega_z-\omega_w)^2} 
\label{G-NM-bou}
\end{align}
where $\dz_\omega \cXX^{(n)}(\omega)$ and $\dz_\omega \bcXX^{(s)}(\omega) $
are the functions (\ref{basis-real-class-sols}) which are real 
when $\omega\in\R$
and we have introduced the integers
\begin{align}
N_t &= \sum_{u=1}^{t-1} \theta( \alpha_u- \alpha_{u+1})
\nonumber\\
\bar N_t &= N_t + (t-1)
\label{def-N_t}
\end{align}
which enter the game because of the way $\epsilon$ is defined in
eq. (\ref{eps-alf-alf}). 

Another interesting point is to study the behavior of 
$\Delta_{(N,M), bou}(x_1,x_2)$ with $x_1,x_2\in (x_t, x_{t-1})$,
i.e. they are on $D_t$ when  $x_1, x_2 \rightarrow x_t$. 
This is an important check since we should recover
both the singularity and the form factor $R^2(\epsilon_t)$ of the
function $\Delta\bou$ given in eq. (\ref{Delta-N=2}).
In the limit $x_1, x_2 \rightarrow x_t$ only the last two lines of
eq. (\ref{G-NM-bou}) contributes, if we change variables as
$\omega_z=\omega_t+(\omega_{x_1}-\omega_t) y_1$ and
$\omega_w=\omega_t+(\omega_{x_1}-\omega_t) y_2$ in the third line and
in a similar way in the forth one we get
\begin{align}
\mdda&
G_{(N,M), bou}(x_1,x_2)
\sim_{x_1,x_2 \rightarrow x_t}
\sum_{\hat n=0}^{N-M} \sum_{\hat s=0}^{M} a_{\hat n \hat s}(\omega_{t\ne 1,2,N}) 
\Big[
\nonumber\\
(-)^{t-1}
&
\int_0^1 d y_1 \int_0^{\hat y} d y_2
\frac{  
\dz_\omega \cXX^{(\hat n)}( \omega_t+(\omega_{x_1}-\omega_t) y_1 )
\dz_\omega \bcXX^{(\hat s)}(\omega_t+(\omega_{x_2}-\omega_t) y_2 )
+ (y_1 \leftrightarrow y_2)
}{(y_1-y_2)^2}
\Big]
\nonumber\\
&=
\int_0^1 d y_1 \int_0^{\hat y} d y_2
\frac{  
y_1^{-\bar \epsilon_t}y_2^{-\epsilon_t} 
\left(\bar\epsilon_t y_1 + \epsilon_t y_2 +O(y^2) \right)
+ (y_1 \leftrightarrow y_2)
}{(y_1-y_2)^2}
\end{align}
where we used eq.s (\ref{constr-on-g}) which imply 
$\sum_{\hat n=0}^{N-M}\sum_{\hat s=0}^{M} a_{\hat n \hat s}(\omega_{u\ne1,2,N}) 
\omega_t^{\hat n+\hat s}=0 $,
$
\sum_{\hat n=0}^{N-M}\sum_{\hat s=0}^{M} a_{\hat n \hat s}(\omega_{u\ne1,2,N}) 
\hat n
\omega_t^{\hat n+\hat s-1}=
\bar \epsilon_t  \prod_{u\ne t,N} (\omega_t-\omega_u)
$ and the analogous equation
$
\sum_{\hat n=0}^{N-M}\sum_{\hat s=0}^{M} a_{\hat n \hat s}(\omega_{u\ne1,2,N}) s
\omega_t^{\hat n+\hat s-1}=
\epsilon_t  \prod_{u\ne t,N} (\omega_t-\omega_u)
$.
In the previous equation we have introduced also
$\hat y
= \frac{\omega_2-\omega_t}{\omega_1-\omega_t}
= \frac{x_2-x_t}{x_2-x_N} / \frac{x_1-x_t}{x_1-x_N}
$
where $x_1, x_2$ are not the location of the twist fields but the
points where the Green function is evaluated.
This expression has to be compared with the analogous for $N=2$ which
can be written as
\begin{align}
\mdda G_{N=2}(\bar x_1, \bar x_2)
&=
\int_0^1 d y_1 \int_0^{\bar x_2 / \bar x_1} d y_2
\frac{  
y_1^{-\bar \epsilon_t}y_2^{-\epsilon_t} 
\left(\bar\epsilon_t y_1 + \epsilon_t y_2 \right)
+ (y_1 \leftrightarrow y_2)
}{(y_1-y_2)^2}
\end{align}
then we can write
\begin{equation}
G_{(N,M), bou}(x_1,x_2)
\sim
G_{N=2}(1,\hat y)=
G_{N=2}( \frac{x_1-x_t}{x_1-x_N},  \frac{x_2-x_t}{x_2-x_N})
\end{equation}
In the limit $x_1, x_2 \rightarrow x_t$ we notice that 
$\hat y = \frac{x_1-x_t}{x_2-x_t}+ O(x_1-x_2)$ then we can use the
previous result (\ref{Delta-N=2}) to write
\begin{align}
\Delta_{(N,M), bou}(x_1,x_2)
\sim
\log (\xii1 ^2)
+ R^2(\epsilon_t)
+O(x_1-x_2)
\label{Delta-N}
\end{align}

\section{Functions entering excited twist fields correlators}
\label{app:functions-for_excited}
If we look at eq. (\ref{reggeon-excited-twists}) and the more general
eq. (\ref{reggeon-excited-twists+bou})  we see immediately that
the key quantities are
\begin{align}
& 
(u_t-x_t)^{\epsilon_{t I}} \partial_u X^I_{cl}(u_t)
\nonumber\\
& 
(u_t-x_t)^{\epsilon_{t I}} (v_t-x_t)^{\epsilon_{t J}}
\partial_u \partial_v 
\Delta_{(N,M) (t)}^{I J}(u_t, v_t; \{ x_\htt, \epsilon_\htt\})
\nonumber\\
&
(u_t-x_t)^{\epsilon_{t I}} (v_\htt-x_\htt)^{\epsilon_{\htt J}}
\partial_u \partial_v
G_{(N,M)}^{I J}(u_t, \bu_t; v_\htt, \bv_\htt  ; \{ x_t, \epsilon_t\})
\nonumber\\
&
(u_t-x_t)^{\epsilon_{t I}} 
\partial_u 
G_{(N,M)}^{I J}(u_t, \bu_t; x_j, x_j  ; \{ x_t, \epsilon_t\})
\end{align}
We would now like to give a more explicit expression for these
quantities while a completely explicit expression is given in the next
section for few cases.
Actually the previous expressions except the cases with only one derivative
can be written as
\begin{align}
&
(u-x_t)^{\bar\epsilon_{t}} \partial_u X_{cl}(u)
= 
\sum_{n=0}^{N-M-2} a_n(\omega_\htt)~
 (u-x_t)^{\bar\epsilon_{t}} \dz_u\cX^{(n)}(\omega_u)
\nonumber\\
&
(u-x_t)^{\epsilon_{t}} \partial_u \bar X_{cl}(u)
=
\sum_{r=0}^{M-2} b_r(\omega_\htt)
~(u-x_t)^{\epsilon_{t}} \dz_u\cX^{(r)}(\omega_u)
\end{align}
and
%
%
\begin{align}
%
%
\mdda
(u-x_t)^{\bar \epsilon_{t}} 
&
(v-x_t)^{\epsilon_{t}}
\partial_u \partial_v 
\Delta_{(N,M) (t)}^{z \bz}(u, v; \{ x_\htt, \epsilon_\htt\})
=
\nonumber\\
\frac{1}{(u-v)^2}
\Bigg[
&
\sum_{\hat n=0}^{N-M}\sum_{\hat s=0}^{M} a_{\hat n \hat s}(\omega_{\htt\ne 1,2, N})  
~~ (u-x_t)^{\bar\epsilon_{t}} \dz_\omega \cX^{(\hat n)}(\omega_u)
~~ (v-x_t)^{\epsilon_{t}} \dz_\omega \cbX^{(\hat s)}(\omega_v)
\nonumber\\
&
-
\left( \bar\epsilon_t (u-x_t)+ \epsilon_t (v-x_t) \right)
\Bigg] 
%
\nonumber\\
%
%
%
\mdda
(u-x_t)^{\bar \epsilon_{t}} 
&
(v-x_t)^{\bar \epsilon_{t}}
\partial_u \partial_v 
\Delta_{(N,M) (t)}^{z z}(u, v; \{ x_\htt, \epsilon_\htt\})
=
\nonumber\\
\mdda
(u-x_t)^{\bar \epsilon_{t}} 
&
(v-x_t)^{\bar \epsilon_{t}}
\partial_u \partial_v 
G_{(N,M) (t)}^{z z}(u, v; \{ x_\htt, \epsilon_\htt\})
=
\nonumber\\
=&
e^{i 2\pi \alpha_1}
\frac{\partial \omega_u}{\partial u}
\frac{\partial \omega_v}{\partial v}
\sum_{n,m=0}^{N-M-2} 
c_{n m}(\omega_{\htt\ne 1,2, N})  
~~ (u-x_t)^{\bar\epsilon_{t}} \dz_\omega \cX^{(n)}(\omega_u)
~~ (v-x_t)^{\bar\epsilon_{t}} \dz_\omega \cX^{(m)}(\omega_v)
\nonumber\\
%
%
%
\mdda
(u-x_t)^{\epsilon_{t}} 
&
(v-x_t)^{\epsilon_{t}}
\partial_u \partial_v 
\Delta_{(N,M) (t)}^{\bz \bz}(u, v; \{ x_\htt, \epsilon_\htt\})
=
\nonumber\\
\mdda
(u-x_t)^{\epsilon_{t}} 
&
(v-x_t)^{\epsilon_{t}}
\partial_u \partial_v 
G_{(N,M) (t)}^{\bz \bz}(u, v; \{ x_\htt, \epsilon_\htt\})
=
\nonumber\\
=&
e^{-i 2\pi \alpha_1}
\frac{\partial \omega_u}{\partial u}
\frac{\partial \omega_v}{\partial v}
\sum_{r,s=0}^{M-2} 
b_{r s}(\omega_{\htt\ne 1,2, N})  
~~ (u-x_t)^{\epsilon_{t}} \dz_\omega \cbX^{(r)}(\omega_u)
~~ (v-x_t)^{\epsilon_{t}} \dz_\omega \cbX^{(s)}(\omega_v)
\label{more-explicit-Delta-t}
\end{align}
and
\begin{align}
%
%
\mdda
(u-x_t)^{\bar \epsilon_{t}} 
&
(v-x_\bt)^{\epsilon_{\bt}}
\partial_u \partial_v 
G_{(N,M)}^{z \bz}(u, \bu; v,\bv ; \{ x_\htt, \epsilon_\htt\})
=
\nonumber\\
\frac{1}{(u-v)^2}
&
\sum_{\hat n=0}^{N-M}\sum_{\hat s=0}^{M} a_{\hat n \hat s}(\omega_{\htt\ne 1,2, N})  
~~ (u-x_t)^{\bar\epsilon_{t}} \dz_\omega \cX^{(\hat n)}(\omega_u)
~~ (v-x_\bt)^{\epsilon_{\bt}} \dz_\omega \cbX^{(\hat s)}(\omega_v)
\label{more-explicit-G-t}
\end{align}

The remaining cases which involve only one derivative and are needed
for computing the interaction among  untwisted vertices with momenta
and excited twists require a little more work.
 The easiest cases can be written as
\begin{align}
%
%
%
\mdda
(u-x_t)^{\bar \epsilon_{t}} 
&
\partial_u
G_{(N,M) (t)}^{z z}(u, \bu; x_j, x_j; \{ x_\htt, \epsilon_\htt\})
=
\nonumber\\
=&
e^{i 2\pi \alpha_1}
\frac{\partial \omega_u}{\partial u}
\sum_{n,m=0}^{N-M-2} 
c_{n m}(\omega_{\htt\ne 1,2, N})  
~~ (u-x_t)^{\bar\epsilon_{t}} \dz_\omega \cX^{(n)}(\omega_u)
~~ \int_{0; \omega\in H}^{\omega_j}  d\omega
\dz_\omega \cX^{(m)}(\omega)
\nonumber\\
&
+
e^{i 2\pi \alpha_1}
\frac{\partial \omega_u}{\partial u}
\sum_{\hat n=0}^{N-M} \sum_{\hat s=0}^{M} 
a_{\hat n \hat s}(\omega_{\htt\ne 1,2, N})  
~~ (u-x_t)^{\bar\epsilon_{t}} \dz_\omega \cX^{(\hat n)}(\omega_u)
~~ \int_{0; \bar \omega\in H^-}^{\omega_j} d \bar\omega 
\frac{ 
\dz_\omega \cbX^{(\hat s)}(\bar \omega)
}{
(\omega_u -\bar \omega)^2
}
\nonumber\\
%
%
%
\mdda
(u-x_t)^{\epsilon_{t}} 
&
\partial_u 
G_{(N,M) (t)}^{\bz \bz}(u, \bu; x_j, x_j; \{ x_\htt, \epsilon_\htt\})
=
\nonumber\\
=&
e^{-i 2\pi \alpha_1}
\frac{\partial \omega_u}{\partial u}
\sum_{r,s=0}^{M-2} 
b_{r s}(\omega_{\htt\ne 1,2, N})  
~~ (u-x_t)^{\epsilon_{t}} \dz_\omega \cbX^{(r)}(\omega_u)
~~  \int_{0; \omega\in H}^{\omega_j} \dz_\omega \dz_\omega \cbX^{(s)}(\omega)
\nonumber\\
&
+
e^{-i 2\pi \alpha_1}
\frac{\partial \omega_u}{\partial u}
\sum_{\hat n=0}^{N-M} \sum_{\hat s=0}^{M} 
a_{\hat n \hat s}(\omega_{\htt\ne 1,2, N})  
~~ (u-x_t)^{\epsilon_{t}} \dz_\omega \cbX^{(\hat s)}(\omega_u)
~~ \int_{0; \bar \omega\in H^-}^{\omega_j} d \bar\omega 
\frac{
\dz_\omega \cX^{(\hat n)}(\bar \omega)
}{
(\omega_u -\bar \omega)^2
}
\label{more-explicit-der-Green-zz+bzbz}
\end{align}
then the $I J = z \bz, \bz z $ cases are
\begin{align}
%
%
%
\mdda
(u-x_t)^{\bar \epsilon_{t}} 
&
\partial_u
G_{(N,M) (t)}^{z \bz}(u, \bu; x_j, x_j; \{ x_\htt, \epsilon_\htt\})
=
\nonumber\\
=&
\frac{\partial \omega_u}{\partial u}
\sum_{\hat n=0}^{N-M} \sum_{\hat s=0}^{M} 
a_{\hat n \hat s}(\omega_{\htt\ne 1,2, N})  
~~ (u-x_t)^{\bar\epsilon_{t}} \dz_\omega \cX^{(\hat n)}(\omega_u)
~~ \int_{0; \omega\in H}^{\omega_j} d \omega 
\frac{ 
\dz_\omega \cbX^{(\hat s)}(\omega)
}{
(\omega_u -\omega)^2
}
\nonumber\\
&+
e^{-i 2\pi \alpha_1}
\frac{\partial \omega_u}{\partial u}
\sum_{n,m=0}^{N-M-2} 
c_{n m}(\omega_{\htt\ne 1,2, N})  
~~ (u-x_t)^{\bar\epsilon_{t}} \dz_\omega \cX^{(n)}(\omega_u)
~~ \int_{0; \bar\omega\in H^-}^{\omega_j}  d\bar\omega
\dz_\omega \cX^{(m)}(\bar\omega)
\nonumber\\
%
%
%
\mdda
(u-x_t)^{\epsilon_{t}} 
&
\partial_u 
G_{(N,M) (t)}^{\bz z}(u, \bu; x_j, x_j; \{ x_\htt, \epsilon_\htt\})
=
\nonumber\\
=&
\frac{\partial \omega_u}{\partial u}
\sum_{\hat n=0}^{N-M} \sum_{\hat s=0}^{M} 
a_{\hat n \hat s}(\omega_{\htt\ne 1,2, N})  
~~ (u-x_t)^{\epsilon_{t}} \dz_\omega \cbX^{(\hat s)}(\omega_u)
~~ \int_{0; \omega\in H}^{\omega_j} d \omega 
\frac{
\dz_\omega \cX^{(\hat n)}(\omega)
}{
(\omega_u -\omega)^2
}
\nonumber\\
&
+
e^{i 2\pi \alpha_1}
\frac{\partial \omega_u}{\partial u}
\sum_{r,s=0}^{M-2} 
b_{r s}(\omega_{\htt\ne 1,2, N})  
~~ (u-x_t)^{\epsilon_{t}} \dz_\omega \cbX^{(r)}(\omega_u)
~~  \int_{0; \bar \omega\in H^-}^{\omega_j}  d \bar \omega
\dz_\omega \cbX^{(s)}(\bar \omega)
\label{more-explicit-der-Green-zbz+bzz}
\end{align}

We see therefore that all the previous cases boil down to computing
the blocks
\begin{align}
(u-x_t)^{\bar\epsilon_{t}} \dz_u \cX^{(\hat n)}(\omega_u)
&=
(u-x_t)^{\bar\epsilon_{t}}
\frac{\partial \omega_u}{\partial u}
 \dz_\omega \cX^{(\hat n)}(\omega_u)
\nonumber\\
(u-x_t)^{\epsilon_{t}} \dz_\omega \cbX^{(\hat r)}(\omega_u)
&=
(u-x_t)^{\epsilon_{t}}
\frac{\partial \omega_u}{\partial u}
 \dz_\omega \cbX^{(\hat r)}(\omega_u)
\label{basic-block-twisted}
\end{align}

\section{Explicit formula for the $N=3$, $M=1$ case}
\label{app:explicit_expressions_N=3}
In this appendix we collect all formula from \cite{Pesando:2012cx} in
order to allow a quick computation of $N=3$ amplitudes.
We start with the basis of derivatives of  zero modes  on the
laplacian operator (\ref{basis-class-sols}) which in this case is
\begin{align}
\dz_\omega\cX^{(0)}(\omega_z)
&=
(\omega_z-1)^{- \bar \epsilon_1}
\omega_z^{- \bar \epsilon_2}
\nonumber\\
\dz_\omega\cbX^{(r)}(\omega_z)
&=
0
\end{align}
since as usual $\omega_1=1$, $\omega_2=0$.
We find then classical solution
\begin{align}
X_{cl}^{(3,1_{cw})}(u,\bar u)
=
f_2
+ %
\frac{ 
e^{i\pi (1-\epsilon_2)} (f_1-f_2) 
}{
B(1, \epsilon_2) B(\epsilon_1, \epsilon_2)}
~_2F_1(\epsilon_2,~ 1-\epsilon_1;~ 1+\epsilon_2;~ \omega_u)
~\omega_u^{\epsilon_2}
\end{align}
and its classical action
\begin{align}
S_{cl}^{(3,1_{cw})}(\{\epsilon_t,f_t\})
&=
\frac{1}{4\pi \alpha'}
\oh |f_1-f_2|^2 
\frac{ \sin(\pi \epsilon_1) \sin(\pi\epsilon_2)
}{
  \sin( \pi \epsilon_3)
}
\end{align}
which is nothing else but the area of the triangle delimited by the branes.

We have the twist fields correlator on $\R^2$
\begin{align}
\langle \prod_{t=1}^3 \sigma_{\epsilon_t, f_t}(x_t) \rangle
&=
\frac{
k
\left[ 
\prod_{t=1}^3 {\Gamma(\bar\epsilon_t)} / {\Gamma(\epsilon_t) }
\right]^{1/4}
e^{-S_{cl}(\{\epsilon_t,f_t\}) }
}{          
x_{12}^{ \oh ( \epsilon_1 \bar\epsilon_1 
                + \epsilon_2 \bar\epsilon_2 - \epsilon_3 \bar\epsilon_3)
          }
x_{13}^{ \oh ( \epsilon_1 \bar\epsilon_1 
                + \epsilon_3 \bar\epsilon_3 - \epsilon_2 \bar\epsilon_2)
          }
x_{23}^{ \oh ( \epsilon_2 \bar\epsilon_2 
                + \epsilon_3 \bar\epsilon_3 - \epsilon_1 \bar\epsilon_1)
          }
}
\end{align}
The derivatives of the Green function when defined on the whole
complex plane minus cuts are
\begin{align}
g_{(3,1)}(z,w;\{x_t\})
&=
\frac{
\frac{\partial \omega_z}{\partial z}
\frac{\partial \omega_w}{\partial w}
}{(\omega_z-\omega_w)^2}
\frac{1 }{ (\omega_z-1)^{\bar \epsilon_1}  \omega_z^{\bar \epsilon_2}  }
\frac{1 }{ (\omega_w-1)^{\epsilon_1} \omega_w^{\epsilon_2} }
\nonumber\\
&
\Big[
(1-\epsilon_1-\epsilon_2) \omega_z^2 
+(\epsilon_1+\epsilon_2) \omega_z \omega_w
-(1-\epsilon_2) \omega_z -\epsilon_2 \omega_w
\Big]
\COMMENTO{
\frac{ (\omega_z-1)^{\epsilon_2-1} }{ (\omega_w-1)^{\epsilon_2} }
\frac{ \omega_z^{\epsilon_3-1} }{ \omega_w^{\epsilon_3} }
\nonumber\\
&
\Big[
(1-\epsilon_2-\epsilon_3) \omega_z^2 
+(\epsilon_2+\epsilon_3) \omega_z \omega_w
-(1-\epsilon_3) \omega_z -\epsilon_3 \omega_w
\Big]
}
\nonumber\\
h_{(3,1)}(z,w;\{x_t\})
&=
0
\nonumber\\
l_{(3,1)}(z,w;\{x_t\})
&=
e ^{i 2 \pi \alpha_1}
c_{0 0} 
\frac{
\frac{\partial \omega_z}{\partial z}
 }{ (\omega_z-1)^{\bar \epsilon_1}  \omega_z^{\bar \epsilon_2}  }
\frac{ 
\frac{\partial \omega_w}{\partial w}
}{ (\omega_w-1)^{\bar\epsilon_1} \omega_w^{\bar\epsilon_2} }
\nonumber\\
&=
e ^{i 2 \pi \alpha_1}
c_{0 0} 
x_{12} x_{1N} x_{2N}
\prod_{t=1}^{N=3} (z-x_t)^{-\bar\epsilon_t}
\prod_{t=1}^{N=3} (w-x_t)^{-\bar\epsilon_t}
\end{align}
where 
\begin{align}
c_{0 0}
=
- 
\epsilon_1
\frac{ B(\bar\epsilon_2, \bar\epsilon_3) }{ B(\epsilon_2, \epsilon_3)}
\end{align}
Then we can write the Green function
%
%
\begin{align}
\mdda
G_{(3,1)}^{z \bz}
&
(u, \bu; v,\bv ; \{ x_\htt, \epsilon_\htt\})
=
\nonumber\\
&
\omega_u^{\epsilon_2} (\omega_u-1)^{\epsilon_1}
\cdot
\int_{0; \omega \in H}^{\omega_v} d \omega
(\omega-\omega_u)^{-1}
\omega^{-\epsilon_2} (\omega-1)^{-\epsilon_1}
\nonumber\\
&
+
\bar \omega_v^{\epsilon_2} (\bar\omega_v-1)^{\epsilon_1}
\cdot
\int_{0; \bar \omega \in H^-}^{\bar\omega_v} d\bar \omega~
(\bar\omega-\bar\omega_v)^{-1} 
\bar\omega^{-\epsilon_2} (\bar\omega-1)^{-\epsilon_1}
\nonumber\\
&
+
c_{00}
\int_{0; \omega \in H}^{\omega_u} d \omega~
\omega^{-\bar\epsilon_2} (\omega-1)^{-\bar\epsilon_1}
\cdot
\int_{0; \bar\omega \in H^-}^{\bar \omega_v} d \omega~
\bar\omega^{-\bar\epsilon_2} (\bar\omega-1)^{-\bar\epsilon_1}
\end{align}
and
%
%
\begin{align}
\mdda
G_{(3,1)}^{z z}
&
(u, \bu; v,\bv ; \{ x_\htt, \epsilon_\htt\})
=
\nonumber\\
&
e^{i 2\pi \alpha_1}
\omega_u^{\epsilon_2} (\omega_u-1)^{\epsilon_1}
\cdot
\int_{0; \bar\omega \in H^-}^{\bar\omega_v} d \bar\omega~
(\bar\omega-\omega_u)^{-1}
\bar\omega^{-\epsilon_2} (\bar\omega-1)^{-\epsilon_1}
\nonumber\\
&
+
e^{i 2\pi \alpha_1}
\omega_v^{\epsilon_2} (\omega_v-1)^{\epsilon_1}
\cdot
\int_{0; \bar \omega \in H^-}^{\bar\omega_u} d\bar \omega~
(\bar\omega-\omega_v)^{-1} 
\bar\omega^{-\epsilon_2} (\bar\omega-1)^{-\epsilon_1}
\nonumber\\
&
+
e^{i 2\pi \alpha_1}
c_{00}
\int_{0; \omega \in H}^{\omega_u} d \omega~
\omega^{-\bar\epsilon_2} (\omega-1)^{-\bar\epsilon_1}
\cdot
\int_{0; \omega \in H}^{\omega_v} d \omega~
\omega^{-\bar\epsilon_2} (\omega-1)^{-\bar\epsilon_1}
\end{align}
and
%
%
\begin{align}
\mdda
G_{(3,1)}^{\bz \bz}
&
(u, \bu; v,\bv ; \{ x_\htt, \epsilon_\htt\})
=
\nonumber\\
&
e^{-i 2\pi \alpha_1}
\bar\omega_u^{\epsilon_2} (\bar\omega_u-1)^{\epsilon_1}
\cdot
\int_{0; \omega \in H}^{\omega_v} d \omega~
(\omega-\bar \omega_u)^{-1}
\omega^{-\epsilon_2} (\omega-1)^{-\epsilon_1}
\nonumber\\
&
+
e^{-i 2\pi \alpha_1}
\bar \omega_v^{\epsilon_2} (\bar \omega_v-1)^{\epsilon_1}
\cdot
\int_{0;  \omega \in H}^{\omega_u} d \omega~
(\omega-\bar \omega_v)^{-1} 
\omega^{-\epsilon_2} (\omega-1)^{-\epsilon_1}
\nonumber\\
&
+
e^{-i 2\pi \alpha_1}
c_{00}
\int_{0; \bar\omega \in H^-}^{\bar\omega_u} d \bar\omega~
\bar\omega^{-\bar\epsilon_2} (\bar\omega-1)^{-\bar\epsilon_1}
\cdot
\int_{0; \bar\omega \in H^-}^{\bar\omega_v} d \bar\omega~
\bar\omega^{-\bar\epsilon_2} (\bar\omega-1)^{-\bar\epsilon_1}
\end{align}
which clearly shows the logarithmic singularity as $u \rightarrow v$.
The double integral of $g_{(3,1)}$ can be expressed as the product of
two single integrals by using an
integration by part and then rewriting the new resulting integral.
If we would not use this procedure we would obtained an integral 
of Lauricella function which is by far more complex.
The idea is simple and amounts to write the new integral as
\begin{align}
\frac{ 1 }{
\omega_z-\omega_w
}
\frac{\partial}{\partial \omega_w}
&
\left[
\prod_{t= 1}^{N-1} (\omega_w-\omega_t)^{-\epsilon_t}
\sum_{\hat n=0}^{N-M}\sum_{\hat s=0}^{M} a_{\hat n \hat s}(\omega_{t\ne1,2,N}) 
\omega_z^{\hat n} \omega_w^{\hat s}
\right]
\nonumber\\
&
=
\frac{\partial}{\partial \omega_w}
\left[
\prod_{t= 1}^{N-1} (\omega_w-\omega_t)^{-\epsilon_t}
Polynomial_1(\omega_z, \omega_w)
\right]
\nonumber\\
&+
\prod_{t= 1}^{N-1} (\omega_w-\omega_t)^{-\epsilon_t}
Polynomial_2(\omega_w)
\end{align}
where the key point is that $Polynomial_2(\omega_w)$ depends only on $\omega_w$.
The previous step always possible because the function $g_{(N,M)}$ has only a
double pole.
Moreover a choice of which integral to do first either $\omega_z$ or
$\omega_w$ can drastically simplify the final result.
The final and simplest result is then in our case 
\begin{align}
\int_0^{\omega_w} d \hat\omega
\int_0^{\omega_z} d \omega
g_{(3,1)}(\omega, \hat \omega)
&=
\int_0^{\omega_w} d \hat\omega
\frac{1}{\hat \omega - \omega_z}
\hat \omega^{-\epsilon_2} (\hat \omega-1)^{-\epsilon_1}
\omega_z^{\epsilon_2} (\omega_z-1)^{\epsilon_1}
\end{align}

The boundary Green function reads
\COMMENTOOK{and the signs??}
%
%
\begin{align}
\mdda
&
(-1)^{N_{t_1}+N_{t_2}}
G_{(3,1), bou}
(x_1,x_2 ; \{ x_\htt, \epsilon_\htt\})
=
\nonumber\\
&
|\omega_{x_1}|^{\epsilon_2} |\omega_{x_1}-1|^{\epsilon_1}
\cdot
\int_{0}^{\omega_{x_2}} d \omega
(\omega-\omega_{x_1})^{-1}
|\omega|^{-\epsilon_2} |\omega-1|^{-\epsilon_1}
\nonumber\\
&
+
| \omega_{x_2}|^{\epsilon_2} |\omega_{x_2}-1|^{\epsilon_1}
\cdot
\int_{0 }^{\omega_{x_1}} d\omega~
(\omega-\omega_{x_2})^{-1} 
|\omega|^{-\epsilon_2} |\omega-1|^{-\epsilon_1}
\nonumber\\
&
+
e^{i \pi (N_{t_1} + N_{t_2})}
c_{00}
\int_{0 }^{\omega_{x_1}} d \omega~
|\omega|^{-\bar\epsilon_2} |\omega-1|^{-\bar\epsilon_1}
\cdot
\int_{0}^{\omega_{x_2}} d \omega~
|\omega|^{-\bar\epsilon_2} |\omega-1|^{-\bar\epsilon_1}
\end{align}
where the integer $N_t$ is defined in eq. (\ref{def-N_t}) and the sign
entering the definition of the Green boundary function is chosen
consistently with eq.s \ref{GIJNM-bou-GNM-bou}.
The regularized version of the boundary Green function at $x_1$
(remember that in this case both  $x_1$ and $x_2$ are on the same brane) is
\COMMENTOOK{and the signs??}
%
%
\begin{align}
\mdda
&
\Delta_{(3,1), bou (1)}
(x_1,x_2 ; \{ x_\htt, \epsilon_\htt\})
=
\nonumber\\
&
|\omega_{x_1}|^{\epsilon_2} |\omega_{x_1}-1|^{\epsilon_1}
\cdot
\int_{0}^{\omega_{x_2}} d \omega
\frac{
|\omega|^{-\epsilon_2} |\omega-1|^{-\epsilon_1}
-
|\omega_{x_1}|^{-\epsilon_2} |\omega_{x_1}-1|^{-\epsilon_1}
}
{\omega-\omega_{x_1}}
\nonumber\\
&
+
| \omega_{x_2}|^{\epsilon_2} |\omega_{x_2}-1|^{\epsilon_1}
\cdot
\int_{0 }^{\omega_{x_1}} d\omega~
\frac{
|\omega|^{-\epsilon_2} |\omega-1|^{-\epsilon_1}
-
|\omega_{x_2}|^{-\epsilon_2} |\omega_{x_2}-1|^{-\epsilon_1}
}
{\omega-\omega_{x_2}}
\nonumber\\
&
+
c_{00}
\int_{0 }^{\omega_{x_1}} d \omega~
|\omega|^{-\bar\epsilon_2} |\omega-1|^{-\bar\epsilon_1}
\cdot
\int_{0}^{\omega_{x_2}} d \omega~
|\omega|^{-\bar\epsilon_2} |\omega-1|^{-\bar\epsilon_1}
\end{align}
which is nothing else but the unregularized boundary Green function to
which we have subtracted the logarithm.

We also have the basic blocks for the twisted computations which
correspond to eq.s \ref{basic-block-twisted}
\begin{align}
(u-x_1)^{\bar\epsilon_{1}} \dz_u \cX^{(\hat n)}(\omega_u)
&
=
e^{i \pi \epsilon_1} 
x_{12}^{\epsilon_N -\hat n} x_{1N}^{\epsilon_2 +\hat n} x_{2N}^{\epsilon_1}
(u-x_2)^{-\bar \epsilon_2+\hat n} (u-x_N)^{-\bar \epsilon_N-\hat n}
\nonumber\\
(u-x_2)^{\bar\epsilon_{2}} \dz_u \cX^{(\hat n)}(\omega_u)
&
=
e^{i \pi \epsilon_1} 
x_{12}^{\epsilon_N - \hat n} x_{1N}^{\epsilon_2 +\hat n} x_{2N}^{\epsilon_1}
(u-x_1)^{-\bar \epsilon_1} (u-x_2)^{-\bar \epsilon_2+\hat n}
(u-x_N)^{-\bar \epsilon_N-\hat n}
\nonumber\\
(u-x_3)^{\bar\epsilon_{3}} \dz_u \cX^{(\hat n)}(\omega_u)
&
=
e^{i \pi \epsilon_1} 
x_{12}^{\epsilon_N -\hat n} x_{1N}^{\epsilon_2+\hat n} x_{2N}^{\epsilon_1}
(u-x_1)^{-\bar \epsilon_1} (u-x_2)^{-\bar \epsilon_2+\hat n}
(u-x_N)^{-\hat n}
\end{align}
with $x_{t \htt}= x_t -x_\htt$ and $\hat n=0,1,2$.
Similarly for $ \dz_u \cX^{(\hat r)}(\omega_u)$ which can be obtained
fro the previous ones with $\epsilon \leftrightarrow \bar \epsilon$
and $\hat n \leftrightarrow \hat r$.

It is also a good check that in the limit 
$u,v\rightarrow x_t$ the Green function gives the expected
singularities.
This can be obtained with a simple change of variable or, essentially
in the same way, rewriting the Green function using the Lauricella
functions, for example 
\begin{align}
\mdda
G_{(3,1)}^{z \bz}
&
(u, \bu; v,\bv ; \{ x_\htt, \epsilon_\htt\})
=
\nonumber\\
&
- \bar \epsilon_2 (1- \omega_u)^{\epsilon_1}
\left( \frac{\omega_v}{\omega_u} \right)^{\epsilon_2}
F^{(2)}_D(\bar \epsilon_2; 1, \epsilon_1;  
1+\bar \epsilon_2;
\frac{\omega_v}{\omega_u}; \omega_v)
\nonumber\\
&
- \bar \epsilon_2 (1- \bar\omega_u)^{\epsilon_1}
\left( \frac{\bar\omega_v}{\bar\omega_u} \right)^{\epsilon_2}
F^{(2)}_D(\bar \epsilon_2; 1, \epsilon_1;  
1+\bar \epsilon_2;
\frac{\bar\omega_v}{\bar\omega_u}; \bar\omega_v)
\nonumber\\
&+
c_{0 0}
\bar \epsilon_2^2
\omega_u^{\epsilon_2}  \bar \omega_v^{\epsilon_2}
~{}_2 F_1(\bar \epsilon_2; \bar\epsilon_1;  
1+\bar \epsilon_2;
\omega_u)
~{}_2 F_1(\bar \epsilon_2; \bar\epsilon_1;  
1+\bar \epsilon_2;
\bar\omega_v)
\label{Green-N=3-special-function}
\end{align}
where for $Re \,c > Re \,a > 0 ~$
\begin{equation}
F_D^{(n)}(a, b_1,\ldots,b_n, c; x_1,\ldots,x_n) = 
\frac{\Gamma(c)} {\Gamma(a) \Gamma(c-a)} 
\int_0^1 t^{a-1} (1-t)^{c-a-1} (1-x_1t)^{-b_1} \cdots (1-x_nt)^{-b_n} 
\,\mathrm{d}t
\end{equation}
and then using the obvious relation
\begin{equation}
F_D^{(n)}(a, b_1,\ldots,b_n, c; x_1,\ldots,x_n=0)
=
F_D^{(n-1)}(a, b_1,\ldots,b_{n-1}, c; x_1,\ldots,x_{n-1})
\end{equation}
to get in the limit $u,v\rightarrow x_2$, $\omega_v / \omega_u$ constant
the desired behavior of the Green function  for $N=2$ given in eq.s
(\ref{Green-N=2}) upon the use of 
$ 
\frac{\omega_v}{\omega_u} \rightarrow \frac{v-x_2}{u-x_2}
$. For the limit $u,v\rightarrow x_{1,3}$ we can proceed in the same
way but we have to use the relation which connects the hypergeometric
computed at $x$ to that computed in $1/x$ or to use the symmetries 
(\ref{Green-N=2-symmetry}).



\end{document}